\documentclass[]{IEEEtran}
\usepackage{cuted}
\usepackage{epsf}
\usepackage{subfigure} 
\usepackage{bbm}
\usepackage{dsfont}
\usepackage{algorithm}
\usepackage{algorithmic}
\usepackage{textcomp }
\usepackage{amsthm}
\usepackage{amsmath,amssymb}
\usepackage[hidelinks]{hyperref}
\usepackage{lipsum}
\usepackage{caption}
\usepackage{stackrel}
\usepackage{float}
\usepackage{graphicx}
\usepackage{epstopdf}
\usepackage{color}
\usepackage{soul}
\usepackage{listings}
\makeatletter
\def\@seccntformatinl#1{\csname the#1dis\endcsname\hskip 1em\relax}
\makeatother
\usepackage{cite}
\usepackage{multirow,tabularx}
\usepackage{ifthen}
\usepackage{times}
\usepackage{array,url,kantlipsum}
\usepackage{amsmath}

\usepackage{multicol}
\usepackage[english]{babel}
\usepackage{blindtext}

\usepackage{lipsum, babel}

\usepackage{amsfonts}

\usepackage{amsxtra}

\usepackage{latexsym}

\makeatletter
\newcommand{\removelatexerror}{\let\@latex@error\@gobble}
\makeatother

\usepackage{lipsum}

\usepackage{mathtools}

\DeclarePairedDelimiter\floor{\lfloor}{\rfloor}

\begin{document}

\bstctlcite{IEEEexample:BSTcontrol}

\title{ A Comprehensive Survey of Machine Learning Based Localization with Wireless Signals 
 \thanks{D. Burghal, A. T. Ravi, A. F. Molisch are at the Ming Hsieh Department of Electrical Engineering, University of Southern California, Los Angeles, CA 90089, USA. (e-mail:\{burghal,telagima,molisch\}@usc.edu.}
 \thanks{ V. Rao  was with the University of Southern California, Los Angeles, CA 90089 USA. He is now with Amazon.com Services, Inc., Seattle, 98109 USA (e-mail: varunrao.vr41@gmail.com).}
 \thanks{ A. A. Alghafis is with the Communication and Information Technology Research Institute, King Abdulaziz City for Science and Technology, Riyadh 11442, Saudi Arabia (e-mail: alghafis@kacst.edu.sa).}}

\author{
Daoud Burghal, Ashwin T. Ravi, Varun Rao, Abdullah A. Alghafis, Andreas F. Molisch, { \it Fellow, IEEE} }

\maketitle
\allowdisplaybreaks
\allowbreak

\begin{abstract}
The last few decades have witnessed a growing interest in location-based services. Using localization systems based on Radio Frequency (RF) signals has proven its efficacy for both indoor and outdoor applications. However, challenges remain with respect to both complexity and accuracy of such systems. Machine Learning (ML) is one of the most promising methods for mitigating these problems, as ML (especially deep learning) offers powerful practical data-driven tools that can be integrated into localization systems. In this paper, we provide a comprehensive survey of ML-based localization solutions that use RF signals. The survey spans different aspects, ranging from the system architectures, to the input features, the ML methods, and the datasets.

A main point of the paper is the interaction between the domain knowledge arising from the physics of localization systems, and the various ML approaches. Besides the ML methods, the utilized input features play a major role in shaping the localization solution; we present a detailed discussion of the different features and what could influence them, be it the underlying wireless technology or standards or the preprocessing techniques. A detailed discussion is dedicated to the different ML methods that have been applied to localization problems, discussing the underlying problem and the solution structure. Furthermore, we summarize the different ways the datasets were acquired, and then list the publicly available ones. Overall, the survey categorizes and partly summarizes insights from almost 400 papers in this field.

This survey is self-contained, as we provide a concise review of the main ML and wireless propagation concepts, which shall help the researchers in either field navigate through the surveyed solutions, and suggested open problems.
\end{abstract}

\section{Introduction}
Location information is becoming a cornerstone for many applications, e.g., smart transportation systems, assisted driving services, public safety, location-based advertising, individualized location-based tourist information, and many others. Part of this success can be attributed to the wide availability of efficient localization with Global Navigation Satellite Systems (GNSSs), in particular GPS (Global Positioning System). As a matter of fact, in environments where GNSSs are available, the localization problem can be considered as mostly solved. However, localization with a GNSS is based on trilateration, using radio frequency (RF) signals, which requires free line of sight (LOS) to at least 3 satellites. As a result, in indoor environments, dense urban environments (e.g., street canyons), underground (e.g., mines), and other environments or scenarios, localization is challenging and has many open problems, even after many years of research.  

In the last two decades, efforts were made to use light, magnetic field, acoustic signals, images, and inertial sensor readings for localization. However, RF-based localization systems have continued to attract most of the research interest due to the wide availability of wireless systems and technologies, and the attractive properties of RF signals \cite{zekavat2013handbook}. 
\subsection{Localization Methods and Challenges}
RF based localization methods can be roughly categorised into four main classes:
 \begin{itemize}

     \item {\em Trilateration and triangulation methods}. The system uses one or more observed quantities to infer the distance or the angles between the {\em target} (i.e., a wireless node whose position we wish to determine), and {\em anchors} (i.e., nodes with known location). It then applies simple geometric calculation to infer the coordinates of the target. GNSS, and, more generally, time-of-arrival (ToA) -based trilateration \cite{gezici2005localization} are the most important example of this category.
     
     \item  {\em Proximity}. In these solutions, the goal is to identify whether a target is within a certain range from an infrastructure node (e.g., a sensing devices). 
     
     \item {\em Fingerprint matching}. In this type of solutions, the system relies on the availability of a database that contains the fingerprints, i.e., observations of characteristics of signals at a set of locations; when localization of a target is required, the system uses either statistical or deterministic methods to match the observed signal to one of the priori stored fingerprints to infer the approximate user coordinates. The construction of the database is usually done "offline", e.g., when the system is initially deployed.
     
     \item   {\em Direct methods}. In these solutions, the system estimates the {\em coordinates} directly without estimating latent quantities such as distances and angles. A possible way is through a set of assumptions about the joint relation between the coordinates and wireless channels, e.g., statistical models.

 \end{itemize}
 
None of these approaches can be considered as universally "the best", as the success of any method is contingent on the availability of the required data and technology, level of complexity, and the properties of the environment or the setup. For instance, ToA-based trilateration works best for wideband systems that have precise synchronization between the anchors, and operating in environments where pure LOS propagation to at least three anchors is present; triangulation requires antenna arrays and also environments with LOS or quasi-LOS propagation. Proximity-based methods, in applications where the relative range is enough, may require additional infrastructure. Finger printing techniques are mainly constrained by the difficulty and expense of building up the database (also known as radio-map); furthermore establishing a framework for robust and accurate identification of the similarity between the observed channel and stored fingerprint is not trivial.

The attempt to use model-based direct methods relies on the accuracy of the underlying assumptions, e.g., the statistical distribution of the signal, which are usually constrained by requirements for analytical tractability and thus could deviate from actual environmental properties. Note that triangulation/trilateration are based on a physical model of the wireless propagation. This model relates the position of the device to some intermediate characteristics, such as signal runtime, that in turn can be derived from the received signals. The advantage of such an approach is a clear physical relationship between the measurements and the solutions. However, model-based methods become less and less reliable in the presence of effects that are not included in the underlying models - be it because they cannot be measured realistically (such as a calibrated antenna pattern for every single handset), because their inclusion in the model would be too complicated (such as non-Gaussian noise statistics), or because they would make computations too time-consuming for real-time implementation. Important information might be lost due to the necessity of simplified models.  
\subsection{A Case for Machine Learning} 
To address some of the aforementioned limitations, recent years have seen increased interest in utilizing Machine Learning (ML) based localization. ML has emerged as a powerful framework to solve many challenging practical problems in a wide variety of areas, ranging from image recognition, to translation between different languages, to scheduling of wireless transmissions, to autonomous driving. Fundamentally, ML uses real-world data to train an ML solution to capture the complex relations between the input data (features) and the output values (labels).   
 
The labels can be limited to discrete values that represent different classes, such as classifying an observed image to be one of $N$ objects, e.g., dog vs. cat, in which case the problem is referred to as {\em classification problem}. Alternatively, the labels could be a range of continuous values, such as the price of a stock; this problem is referred to as {\em regression problem}. Note that identifying a location can be formulated as either a classification problem (e.g., finding a point on a finite grid that best fits the observed input data) or a regression problem (finding the  - continuous  - coordinates of the device).
\begin{figure*}[h]
\centering
\vspace{-18 mm}
 \includegraphics[width=1\textwidth]{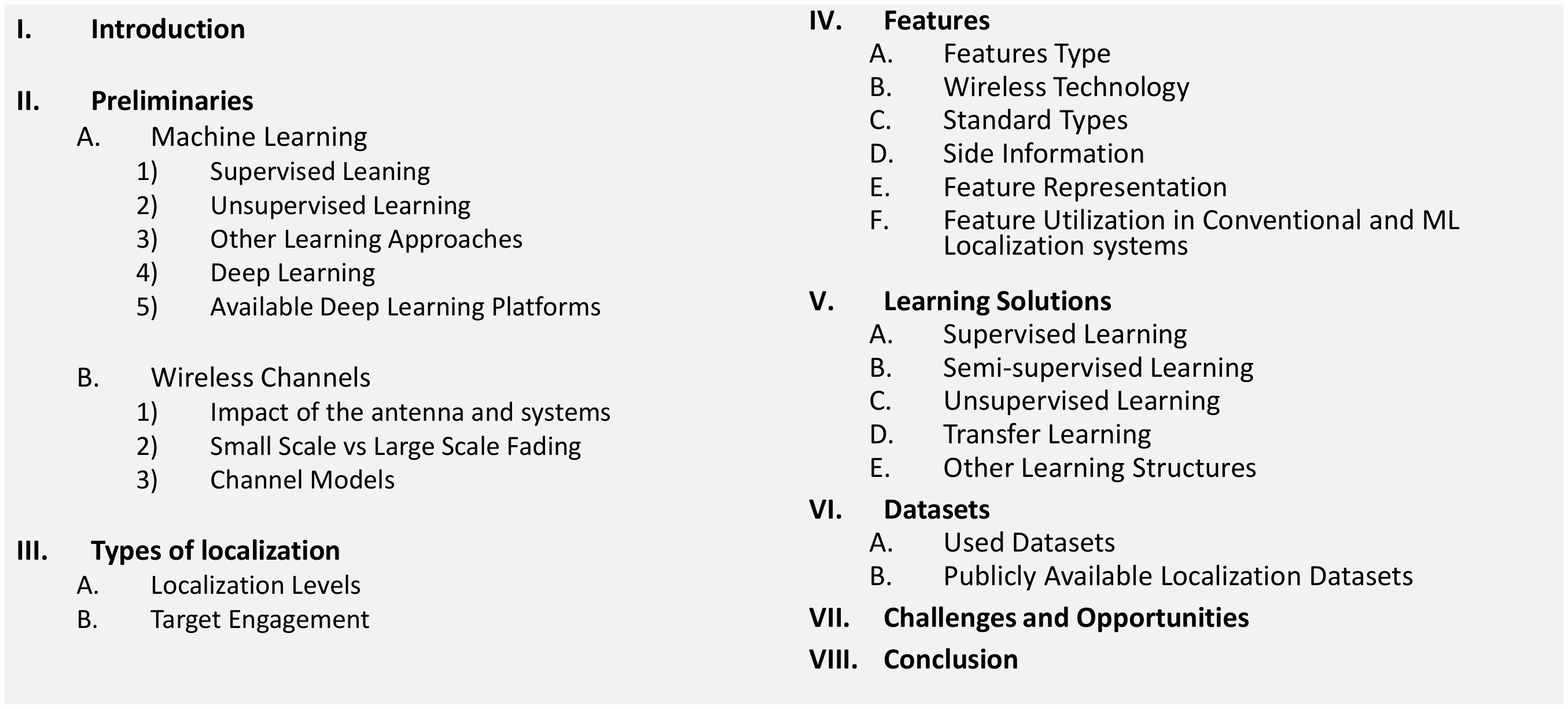}
\vspace{-45  mm}
 \caption{ \small Sections and main subsections of the paper.}
 \label{fig:ToC}
\vspace{-3 mm}
\end{figure*}

ML-based localization can be used in any of the introduced localization methods. Direct methods are a natural choice for an end-to-end ML-based solution, where observations are fed to the ML solution to produce the location. However, ML can also be integrated with any of the other localization methods. For instance, ML can be used to improve the range estimates in trilateration and proximity methods. They can be used as robust matching techniques in fingerprinting methods. The range of applications extends to more complicated systems, e.g., when the available data are constrained or when observations are complex-valued. This has ignited rapidly growing interest in ML-based localization, which can be attributed to the following.
 \begin{itemize}
     \item Recent success of ML in many fields, in particular image recognition and speech and language processing, proving that highly complex problems that have long stymied deterministic approaches can be successfully tacked with ML, and allowing the solutions to be not limited by underlying model assumption. 
    \item The ability of ML to utilize large sets of available data; the data can also be from heterogeneous sources, which are usually difficult to utilize with deterministic solutions.
    \item A number of localization problems can be naturally formulated as ML problems.
     \item The availability of efficient software and hardware solutions that are customized to accelerate ML algorithms. For instance, it is anticipated that by 2022 about 80 percent of the shipped smartphones have on-device Artificial Intelligence (AI) capabilities \cite{Gartner2020AI}. 
 \end{itemize}
 At the same time, ML approaches to localization face major challenges
 \begin{itemize}
     \item Availability of training data: large sets of training data are sometimes required for number of ML algorithms.
     \item Robustness: it is uncertain how robust ML solutions are to both small and large changes in the measurement devices and the environment, and how robust to interference. For example, can the presence/absence of a truck near a receiver completely throw off the location estimate? What is the impact of RF signal interference in cognitive and ad-hoc networks?\footnote{While it is possible that localization systems may face malign interference, such as jamming, this is rare in most civil applications. Such an issue could occur in other applications such as radars \cite{lang2020comprehensive}, which is beyond the scope of this survey.} Such interference could lead to corruption of (part of) the training data or the feature measurements of the device. 
     \item Training and computational complexity: are the algorithms simple enough to be trained and performed in real-time? This is particularly important if the localization should be done by the user device (an approach that is desirable for privacy reasons).
     \item Feature selection: what are the signal characteristics that should be used as the input for the ML algorithm. 
 \end{itemize}
\subsection{Contributions}
The anticipated accelerated research in ML-based localization calls for a comprehensive review of the fundamental aspects of ML-based localization solutions and relevant problems, as well as a summary of what has been done, and what still remains to do. The current paper aims to provide exactly such a survey.

In the literature, there have been excellent surveys of localization techniques. However, most of the available surveys are presented with the focus on solutions that are specific to certain environments \cite{zafari2019survey,liu2007survey,deak2012survey,gu2009survey}, technologies or/and standards \cite{wen2019survey,del2017survey,he2015wi,gu2009survey,lin2017positioning} , applications\cite{davidson2016survey,ferreira2017localization,tahat2016look}, localization methods \cite{he2015wi,shit2019ubiquitous}, type of used signal \cite{yang2013rssi,liu2019survey,yang2015mobility,harle2013survey}. Furthermore, most of these surveys, as well as the extensive handbook \cite{zekavat2013handbook}, concentrate on conventional (i.e., non-ML) approaches. A few provide explicit yet very limited discussion of some ML-based solutions \cite{davidson2016survey,he2015wi,liu2007survey,liu2019survey,zafari2019survey}. In one of the recent papers \cite{saeed2019state}, the authors provide a survey of a dimensionality reduction based localization, which is usually considered an ML technique (see Sec. \ref{sec:MLPrelm}). There are recent papers that survey ML-based solutions for non RF based localization, e.g., \cite{chen2020survey}. Different from all the above, in this paper, we provide a self-contained and comprehensive survey of ML-based localization solutions that utilize RF signals. 

 
 In particular, the contributions of this paper are as follows.
 \begin{itemize}
     \item We provide comprehensive review of the ML-based localization systems based on different aspects:
     \begin{itemize}
         \item Localization goal. 
         \item System/target engagement.
         \item The used RF features.
         \item The used wireless technologies and standards along with their impact on the features.
         \item Details of the ML solutions based on the ML framework and system setup. The presentation is carried out from the ML perspective, where different ML solutions and methods may differ in the underlying assumption about the type and the availability of the data (e.g., supervised, unsupervised, semi-supervised, transfer learning, etc., see Sec. \ref{sec:MLPrelm} for definitions).
     \end{itemize}
     \item A survey of key papers in various categories. We describe salient features that occur as a common theme in multiple papers, and provide tables that categorize related papers. 
     \item We highlight the current challenges that face ML-based localization solutions, and propose a number of interesting directions for future research. 
     \item Summaries of publicly available datasets and how data have been collected, are provided.
     \item We provide concise reviews of fundamental ML techniques and wireless channels and systems, to make the paper self-contained and accessible to researchers in both the wireless and the ML communities.
 \end{itemize}
 
The majority of the surveyed papers are published after 2010, with the vast majority in the last couple of years, when the ML-based localization witnessed an increased interest due to the success of Deep Learning (DL) -based methods. Nevertheless, for completeness, the survey also includes earlier works, as the interest in ML-based localization started about two decades ago (see Sec. \ref{sec:ConvVsMLSystems} for relevant discussions).

\subsection{Paper Structure}

The structure of this survey is as follows. In the first subsection of Sec. \ref{Sec:Prem}, we provide a concise yet comprehensive overview of ML, which covers many important concepts and aspect of ML and DL, in the second subsection we review relevant concepts of wireless channels and systems. This section is designed to provide an introduction for the researchers who might appreciate a tutorial or refresher in any of the two subjects, and also serves to establish some notation. We review the ultimate goals of the localization solution in Sec. \ref{sec:LocGaol}; we also discuss the role of the target object in the localization system. Sec. \ref{sec:Feat} is dedicated to wireless features, where we present the basic features used, their relation to wireless technologies, and the common wireless standards; we also discuss some of the side information that has been used to improve the localization. Next, in Sec. \ref{sec:LearnStrcut}, we summarize the framework based on the availability of the data during the training phase and the learning framework. In each category, we present a number of the proposed ML structures, where we review the used features, input-output relations, and solution details. A major part of Sec. \ref{sec:LearnStrcut} is dedicated to DL solutions, as those are the subject of many recent advances in ML-based localization. Since the ML solutions depend on the used data, in Sec. \ref{sec:Data}, we review the different types of datasets that were used to train and evaluate the models, and provide a list of the publicly available datasets for localization. In Sec. \ref{sec:Challenges}, we review different challenges that could face ML-based localization solutions. We also provide some suggested open and interesting research directions. Throughout the paper we provide extensive citations. Finally, in Sec. \ref{sec:conc}, we provide some concluding remarks followed by a table of the used acronyms (table \ref{tab:acry}). Fig. \ref{fig:ToC} displays the main subsections of the survey.

 \section{Preliminaries}\label{Sec:Prem}
 
 In this section we review some fundamental concepts of ML and of wireless communication, in particular wireless propagation channels. Since this review aims at people with different backgrounds, this covers some introductory concepts. It also helps the reader in identifying the novelty of some of the reviewed papers in the later sections, and assess the challenges of using RF signals for localization and ML techniques. The reader may skim or skip the subsections already familiar to them without losing readability of the later sections. 
 
\subsection{Machine Learning}\label{sec:MLPrelm}
ML provides a set of tools and frameworks that enable the computer to utilize available data $\mathcal{D}$ to perform a certain task, or detect patterns, without programming it with task- specific procedures. The data usually include observable attributes that the system uses to make decisions. ML can be classified into different categories depending on what and when data is available, what task we try to achieve, and the complexity constraints on the models. 
We can {\em initially} distinguish two main classes:
\begin{itemize}
    \item In {\em supervised learning}, the system has a training dataset $\mathcal{D}$ to learn the mapping between the right decisions (labels or tags) $y$ and the observed features $\boldsymbol{x}\in\mathbf{R}^d$.  $\mathcal{D}$ consists of $N$ examples (or data points) with features-label pairs, i.e., $\mathcal{D} = \{\boldsymbol{x}^{i},y^{i}\}_{i=1}^N$. 
    \item In contrast, in {\em unsupervised learning}, the collected dataset has no labels, i.e., $\mathcal{D} = \{\boldsymbol{x}^i\}_{i=1}^N$.
\end{itemize}  
The two approaches usually differ in the task. In the following we start with standard supervised learning to introduce many ML concepts. We then discuss unsupervised learning, followed by a brief introduction to other learning classes. Next, we discuss DL and briefly review some of the available frameworks. Other details and categories are provided later in the survey as needed.
\subsubsection{Supervised Learning}
Supervised learning uses labeled training data, i.e., examples annotated with correct decisions, to teach the system how to make decision when it observes data outside the set of examples. More formally, assuming that the input output relation is given as follows:
\begin{align}\label{eq:model}
 y = f(\boldsymbol{x}) + \epsilon   ~~,
\end{align} 
where the vector of $d$ features $\boldsymbol{x}\in \mathbf{R}^d$ is the observable attributes, $y$ is the associated decision (usually referred to as the label), $f$ is an unknown mapping function, and $\epsilon$ is the noise that encompasses any unobserved factors or intrinsic randomness. When $y$ takes categorical values, we refer to the problem as a {\em classification problem}, e.g., $y$ is one of four directions $\{\rm east, west, north, south\}$. When $y$ takes continuous values, the problem is referred to as a {\em regression problem}, e.g, $y \in \mathbf{R}$ is the distance to an object. Then the goal of a {\em discriminative} ML is to learn the underlying mapping $f$, such that $\hat{y} \triangleq \hat{f}(\boldsymbol{x};\Omega) \approx y$, where $\hat{f}$ is the learned mapping, and $\Omega$ are the mapping parameters. Alternatively it can be viewed as learning the conditional distribution $p(y|\boldsymbol{x})$ using the model $p(y|\boldsymbol{x}; \Omega)$, here $\boldsymbol{x}$ values can be assumed to be generated from a joint probability distribution $p(\boldsymbol{x})$. The probability notation is useful as typically the features are random, and/or the mapping is uncertain due to the noise and other hidden factors. The relation between the two, the learned model $\hat{f}$ and distribution $p(y|\boldsymbol{x};\Omega)$, may be given by \cite{murphy2012machine}
$$ \hat{y} = \hat{f}(\boldsymbol{x};\Omega) = {\rm argmax}_y p(y|\boldsymbol{x};\Omega)~~.$$

 {\bf Loss function:}
To train the model, the ML algorithm relies on the dataset $\mathcal{D}$ and a loss metric $\mathcal{L}$. The loss metric measures the deviation of the learned mapping $\hat{y}$ from the true labels $y$, i.e., $\mathcal{L}(y, \hat{y})$. A popular loss function is mean squared error (MSE):

$$\mathcal{L}_{\rm MSE}(y, \hat{y}) = \mathbb{E} [ |y - \hat{f}(\boldsymbol{x};\Omega)|^2] \approx \frac{1}{N} \sum^N_i |y^i - \hat{f}(\boldsymbol{x}^i;\Omega)|^2 $$
where $\mathbb{E}[]$ is the expectation operator over the ensemble of noise and $\boldsymbol{x}$ realizations, which may be approximated using the empirical mean of the square error over the $N$ examples. From a probabilistic perspective, we can use Kullback–Leibler (KL) divergence to measure the dissimilarity between  $p(y|\boldsymbol{x})$ (from of the dataset) and the model $p(y|\boldsymbol{x}; \Omega)$, in this case
\begin{align}
\mathcal{L}_{KL}(p(y|\boldsymbol{x}),p(y|\boldsymbol{x}; \Omega)) & = \mathbb{E}[ \log p(y|\boldsymbol{x}) - \log p(y|\boldsymbol{x};\Omega)]
\end{align}
Under certain conditions, a number of loss functions are equivalent. For instance, minimizing the loss function above means that we have to select $\Omega$ such that $\mathcal{L}_{KL}$ is small, which can be shown to be equivalent to minimizing $- \mathbb{E}[ \log p(y|\boldsymbol{x};\Omega)] $, and this is equivalent to maximizing the log-likelihood of the distribution over $\Omega$ \cite{goodfellow2016deep}. Furthermore, when the noise in (\ref{eq:model}) has Gaussian distribution, maximizing the likelihood function (i.e., maximizing $\log p(y|\boldsymbol{x}; \Omega)$) results in a metric equivalent to the MSE \cite{goodfellow2016deep}. Finally, we note that minimizing the KL divergence is equivalent to the popular cross-entropy loss function that is widely used in classification problems.

Finally, the structure of $\hat{f}$ (or $p(;\Omega)$) depends on our chosen model; it is usually controlled by the set of parameters $\Omega$. The model and the possible set of parameters define the hypothesis set $\mathcal{H}$ that we hope that $f$ belongs to. The choice of the model is usually made a priori based on domain knowledge, i.e., experience or insights into the physics of the underlying problem. When the size of the model's parameters is fixed, the model is referred to as a {\em parametric} model, otherwise it is {\em non-parametric}.

{\bf Examples:}
To demonstrate the above, let us consider the popular linear regression problem. In such problem we use the data to approximate the scalar output $y$ as follow:
\begin{align}\label{eq:LR}
    \hat{y} = \sum_{i=1}^d \omega_i x_i + \omega_0  = \boldsymbol{\omega}^\top \boldsymbol{x} + \omega_0
\end{align}
Here the goal of the ML problem is to find the best estimates of the parameters $\Omega = \{ \boldsymbol{\omega}, \omega_0\}$. When the loss function is the MSE, methods such as least square can be used to estimate the parameters, i.e., "train" the model.

Although the output can take any real number, it can be used for classification, e.g., binary classification based on the sign of $y$. However, in that case logistic regression is more appropriate, as it estimates the probability of being in one of two classes. It has the simple formulation
\begin{align}
p(y|\boldsymbol{x}; \Omega)  = \sigma( \boldsymbol{\omega}^\top \boldsymbol{x}+ \omega_0),   
\end{align}

$\sigma(.)$ is a non-linear function that modifies the output such that it lies between [0,1], thus naturally fits into the probabilistic view; the logistic (sometimes referred to as sigmoid) function is usually used, see Fig. \ref{fig:activation}. In this case it is more appropriate to maximize the likelihood function, i.e., $p(y|\boldsymbol{x}, \Omega)$, with respect to $\Omega$, which can be shown to be equivalent to minimizing the cross-entropy loss. Different from linear regression, training a logistic regression requires iterative methods such as gradient descent. Both the linear regression and logistic regression are examples of parametric models. An important point to note here is that we can incorporate non-linear features in the {\em linear} regression framework (and the logistic regression) by replacing $\boldsymbol{x}$ in (\ref{eq:LR}) with $\phi(\boldsymbol{x})$, where $\phi(.)$ is, e.g., a polynomial function. This will increase the dimensionality of the features and possibly increase the separability of the data.  

Another popular model is the K-nearest Neighbor (KNN) model. Different from above, no explicit functional form is assumed, rather, the predicted label of a newly observed data point $o$ with observations $x_o$ is based on the labels of the K-nearest neighbors. In a classification problem, with $M$ different classes, i.e., $y \in \{c_1,..,c_M\}$, the label of the observed point can be given by:
$$\hat{y}_o = \max_c \sum_{i \in \mathcal{K}_o } \mathbbm{1} (y_i=c)$$
where $\mathcal{K}_o$ is the set of $K$ neighbors of $o$,  $\mathbbm{1} (y_i=c)$ is equal to one if the label of the $i^{\rm th}$ neighbor is $c$. The choice of $K$ and what defines a "neighbor", i.e., members of $\mathcal{K}_o$, are design parameter choices that can be set using the data and the prior knowledge. For instance, Euclidean distance can be used to set the distance to points and thus select the $K$ nearest neighbors. Note that KNN is an example of a non-parametric model. It also belongs to the class to  {\em instance-based} learning as it derives the labels of the new observations by comparing them with the training instances.

 {\bf Generalization and over-fitting:} In supervised learning the goal is to label an unobserved data point, meaning we are interested in minimizing the loss function of the unobserved data. To do so we usually use the observed dataset to estimate the performance over the unobserved data, and thus rely on the generalization capability of the model.\footnote{This statement is a key difference that differentiates ML from conventional optimization problems \cite{goodfellow2016deep}.},\footnote{ Note that for the system to generalize well beyond the set of the observed examples, smoothness is assumed, such that we can anticipate similar decisions for similar inputs.} This poses challenges when we train the model. As per the discussion above, the goal is to minimize the loss function. Yet this may result in {\em overfitting}, where we use all the degrees of freedom in the model to fit all the variations in observed values (even the noise). In this case the trained model will have large error values when applied to unobserved data, i.e., results in a large variance. This usually occurs when the model has large degrees of freedom or the dataset is very small. On the other hand, when the model is relatively simple, i.e., cannot capture the true $f$, the estimate $\hat{f}$ will be biased, as we will always have non zero error with any choice $\Omega$ for $\hat{f}(. ;\Omega)$. This results in what is known as "variance-bias trade" off, which can be mapped to the size of the hypothesis set above. Note that ideally we would like to have $f \in \mathcal{H}$, however, since it is unknown it is quite likely to fall into one of the two extremes above.\footnote{In learning theory, the structure of the hypothesis set $\mathcal{H}$ can be used to assess generalization error by bounding the deviation of the generalization error from the training error.} 

Typically, over-fitting is more challenging than under-fitting, since the training error could be misleading. There are different ways to combat over-fitting, including weight regularization and validation. Many training problems can be viewed as parameter optimization. Weight regularization is a powerful technique that prevents the optimizer from over-fitting through penalizing model parameters, for instance, the objective function can be 
\begin{align}\label{eq:cost}
    {\rm argmin}_\Omega  \mathcal{L}(y, \hat{f}(x;\Omega)) + \lambda r(\Omega)
\end{align} 
 where $r()$ is a function of the parameters $\Omega$ that enforces certain properties on the parameters. A common choice is  $r(\Omega) = ||\Omega||_p $, where sub-script $p$ refers to the $p$-norm (usually referred to as $L_p$ norm) that can prevent the parameters from taking arbitrarily large values to fit outliers (or noise). $\lambda$ is a weight value that controls the importance of the regularization. The choice of $p$ also impacts the structure of the solution, e.g., $p=0$ or $p=1$ enforce sparseness of the model, and thus plays a role in the model selection.

 Validation is a powerful technique that can be used to combat overfitting and improve the overall performance of the model. Typically we divide the available dataset into disjoint training and testing subsets, where the testing subset is not used in any way to tune or select the model but used to evaluate the performance of the final model. When validation is used, we divide the training dataset into two (or more) subsets: one for training (parameter tuning) and the other for {\em hyperparameter} and model selection. For instance, we can generalize $\hat{f}$ to a polynomial with degree $n$; the choice of the degree can be selected with validation. The proper choice of the hyperparameter $\lambda$ in eq. (\ref{eq:cost}) can be done also through validation. The concept is simple: since noise and outliers are likely to be different in  unobserved realizations in the validation dataset, the model is expected to have larger error values, which could reflect the true performance. As a result, during training, when the model starts to fit noise values to drive down the training error, the validation error goes up. There are different validation techniques, a popular one is the {\em K-fold validation} \cite{james2013introduction,abu2012learning}, where the dataset is divided into $K$ different subsets. At each time, $K-1$ subsets are used for training and the other one is used for validation. The model structure and the hyper-parameters are chosen based on the validation error.  

In practice, preprocessing is usually used to improve the training and model robustness. This includes data formatting, normalization and filtering. Each has different impact on the quality of the model. Noise reduction and outlier filtration could reduce the overfitting. Dimensionality reduction and reducing co-linearity make the learning easier \cite{james2013introduction, goodfellow2016deep}. Also proper formatting of the output values impact the choice of the algorithm (e.g., representation of the different classes). Feature normalization also speeds up training and allows the model to capture the impact of feature variation correctly. 

{\bf Probabilistic models:}
 A supervised learning solution can also aim at learning the joint distribution of the features and the labels, $p(y,\boldsymbol{x})$, or the conditional distribution $p(\boldsymbol{x}|y)$. In this case the model is referred to as {\em generative} model. One can then use Bayes rule on these distributions to get to discriminative approach, i.e., $p(y|\boldsymbol{x})$, where the goal of the latter is to find the best decision boundary given the data. An example of generative models is a Naive Bayes classifier that assumes conditional independence of the features given the class (the label). The training for generative models can be done using Maximum Likelihood or Maximum A Posteriori (MAP). For Naive Bayes the estimates of $p(\boldsymbol{x}|y)$ can be simply done by empirically calculating the frequency of a certain feature given class type. The class prior $p(y)$ is also estimated empirically from the given labeled dataset. Generative models have many advantages, such as their ability to handle missing data, utilize unlabeled data, and impose prior knowledge. However, they are sensitive to pre-processing and make assumptions that may not be valid in reality. 

In probabilistic models we have the features $\boldsymbol{x}$, the labels $y$ and other {\em latent} (unobserved) variables $\boldsymbol{z}$. The inclusion of latent variables adds power to the model as it could capture the true structure of the real world. In general, we can view $y$ as one of the latent variables; a generative model would then capture $p(\boldsymbol{x},\boldsymbol{z})$. It is typically difficult to deal with high dimensional distributions as the number of parameters will grow fast with the number of variables, making both the inference and learning process very difficult \cite{murphy2012machine}. To alleviate some of these issues, certain restrictions on the variable relations can be used, such as conditional independence (CI) (e.g., two variables are CI when a third variable is given), or limiting the dependency by, e.g., Markov models. With hidden variables, Hidden Markov Models (HMM) can be used. One way to represent models with hidden variables is through Mixture Models, where the distribution can be (roughly) viewed as a mixture of a number of base distributions \cite{murphy2012machine}. A popular example is the Gaussian Mixture Model (GMM), where the base models are assumed to be Gaussian. With hidden variables, the problem of Maximum Likelihood (and MAP) is usually not convex. In that case, an iterative method such as Expectation Maximization (EM) can be used to arrive at good solutions. 


An alternative way to capture the variable dependency is through {\em graphical models}: Using graphs, the variables are represented by vertices; an edge between two vertices indicates a direct relation between the two variables, a missing link indicates CI \cite{koller2009probabilistic,murphy2012machine}. The edges can be either directed or undirected, both have their pros and cons that are beyond the scope of this overview. Using either representation allow the use of many graph theoretic algorithms to simplify the learning and inference. Note that, in general, training such models requires sampling from the underlining distribution; methods such as Markov Chain Monte Carlo Methods and Gibbs sampling are usually used.

One method to simplify the training of the probabilistic models is to approximate $p(\boldsymbol{z}|\boldsymbol{x};\Omega)$ with a $q(\boldsymbol{z}|\boldsymbol{x};\Phi)$ that has a tractable form, then maximize a lower bound that captures the relation between the two (along with the marginal distribution over $\boldsymbol{x}$). This is usually referred to as {\em variational inference}, and the lower bound is called Evidence Lower Bound (ELBO). Relevant examples are covered in the DL sections below.

\subsubsection{ Unsupervised Learning  }
\begin{figure}
 \centering 
   \vspace{-15mm}
 \includegraphics[width=1.1\linewidth]{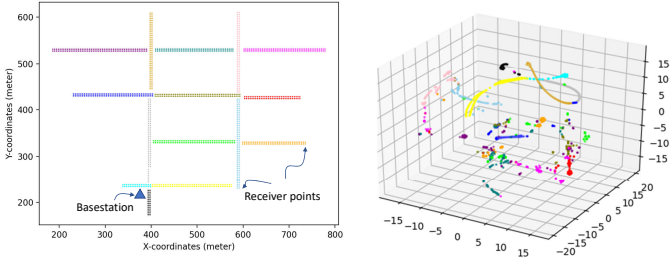}
 \vspace{-25mm}
  \caption{\small A visualization for high dimensional wireless data, revealing a possible structure of the underlying manifold. Data generated using ray-tracing software at $2.4$GHz. (a) The transmitter (Basestation) and receiving points locations, with point color coded to differentiate the different streets. (b) Dimensionality reduction using t-SNE (t-Distributed Stochastic Neighbor Embedding) \cite{maaten2008visualizing}. For each point we used 30 different features including angles of arrival, delay and power of the 5 best MPCs (see sec. \ref{sec:prelWireless}). Notice there is a reasonable agreement between the locations (streets) and the points after reducing the dimension, this is more prominent for the streets close the basestation. More about the setups can be found in \cite{burghal2018band}.}
  \vspace{-5 mm}
\label{fig:MPCmanifold}
\end{figure}

In unsupervised learning the dataset has only unlabeled data,  $\mathcal{D}= \{\boldsymbol{x}^{i}\}_{i=1}^N$. The goal of the ML in this case is to identify patterns in the data, including:
\begin{itemize}
    \item {\em Clustering}, one of the widely popular use cases, where the goal is to identify $K$ different clusters in the data. Since there is no unique metric to guide the clustering process, the choice of $K$ and final clusters may not represent the real world. The K-mean clustering is one of the simplest and most popular examples of clustering techniques; we discuss it below in the examples subsection.
    
    \item {\em Dimensionality reduction}. Usually, real-world data occur in high dimensional spaces, e.g., number of pixels in an image. This large dimensional space results in what is usually referred to as {\em curse of dimensionality}, where different phenomena start to show up in large dimensional spaces that are not important in low dimensional ones. An intuitive example is related to the number of examples needed to cover all the possible cases. For grayscale images, $2^d$ is the number of possible values for the $d$ features, which grows exponentially with the number of features \cite{geron2019hands}. Projecting the data onto lower-dimensional spaces is one method to reduce the size of the needed data. The Principle Component Analysis is one of the oldest and most effective techniques. We discuss it further in the examples section below.
    
    This problem can also be viewed as manifold learning problem. A manifold can be described, roughly, as a constraint shape in $d$ dimensions, but can be captured by $k<d$ dimensions. For instance, the points on the surface of a sphere, in a 3-Dimensional (3D) space, can be viewed as points on a bent 2D surface. In many practical problems, the observed data are on a manifold with an intrinsic dimensionality $k$ embedded in a $d$ dimensional space, see Fig. \ref{fig:MPCmanifold} . The dimension of the observed data can be reduced by unfolding the manifold in $k$ dimensional space. Learning the underlying manifold or some of its properties, such as its tangents, can be very helpful in understanding the relation and evolution of features. There is a number of algorithms to learn the manifold of the observed data; we provide examples in the following subsection.

    \item {\em Density estimation.} As discussed earlier, there is a number of advantages in knowing the underlying structure of the data, e.g., for constructing the graphical model, or regenerating data.  
    
    \item {\em Denoising.} Here, the goal is to filter out the noise and reconstruct the original signal, i.e., the goal is to recover $\boldsymbol{x}$ from a corrupted version of it $\tilde{\boldsymbol{x}}$. Knowledge of the data (or noise) structure and density can thus help in this process. 
    
    \item A number of the points above can be boiled down to the issue of data representation. The proper data representation depends on the problem. For instance, algorithms may handle different representations differently, e.g., some algorithms can utilize a sparse representation to improve performance, while others are oblivious to available sparsity and thus struggle with high dimensional data.
    
\end{itemize}
We emphasize here that the above are not the only goals and are not mutually exclusive. For instance, clustering can be viewed as sparse representation of the data, where a point in a high dimensional features space can be represented by only one of $K$ clusters. Other examples are given below. Furthermore, although the dataset $\mathcal{D}$ is assumed to have no labels, unsupervised learning can be viewed as prior step in a supervised learning problem, where the goal is to provide the right representation of the data.

{\bf Examples:}
\begin{itemize}
    \item {\em K-means clustering} is the most popular clustering algorithm. It assumes the number of clusters is known a priori, say $K$, then it proceeds as follows. It randomly initializes the location of the clusters' centroids $\boldsymbol{c}_k  \in \mathbb{R}^d$, then assign the points $\boldsymbol{x}^i$ to the closest cluster based on some distance measure to the centroid. After all the data points have been assigned to clusters, it then updates the location of the centroid based on the data present in the current clusters, e.g., by giving it the value of the mean of the data in that cluster, i.e., $\boldsymbol{c}_k = \sum_{i \in \mathcal{C}_k} \boldsymbol{x}^i$, where  the set $\mathcal{C}_k$ includes the indices of data points in cluster $k$. The procedure continues until no further changes of the cluster assignment occurs. 
    
    \item {\em Principal Component Analysis (PCA)} is a linear mapping of the data of dimension $d$ to a subspace of dimension $k$. The goal is to find a subspace of dimension $k$ that contains the largest variation of the data. In other words, PCA is a projection to the subspace (of dimension $k$) with the largest variation; by doing so it preserves most of the information but with a smaller dimensional space. When $k=1$, PCA finds the first "principal component" $\boldsymbol{u} \in \mathbb{R}^d$, such that ${\rm var}(\boldsymbol{u}^\top \boldsymbol{x})$ is the largest, where $\rm var(.)$ refers to the variance over the distribution of $\boldsymbol{x}$. $\boldsymbol{u}$ can be found to be the eigenvector that corresponds to the maximum eigenvalue of $\boldsymbol{x}$'s $d \times d$ covariance matrix. In general, for $k \leq d$, and datasets in $X = [\boldsymbol{x}^1, ...,\boldsymbol{x}^n]$ with dimension $d \times n$, i.e., $n$ realizations of $\boldsymbol{x}$, the PCA can be found to be
    $$U = W^\top X$$
    where $U$ is the transformed data with size $k\times n$, $W$ is the $ d \times k$ projection matrix, with the columns of $W$ being the $k$ eigenvectors of the covariance matrix of $\boldsymbol{x}$ that correspond to the $k$ largest eigenvalues. 
    Assuming $\boldsymbol{x}$ values have mean zero, one can empirically approximate the covariance of $\boldsymbol{x}$ by $XX^\top$, and then take the first $k$ eigenvectors of $XX^\top$ that correspond to the largest $k$ eigenvalues of $XX^\top$, or equivalently take the first $k$ columns of left singular vectors of $X$, found through Singular Value Decomposition (SVD).
    Note there are other forms of PCA, such as kernel PCA, that can use a non-linear kernel to replace the inner product between data vectors, which allows to use the possible linear separability in higher dimension before mapping to lower dimension.

    \item {\em Locally Linear Embedding (LLE) \cite{roweis2000nonlinear}} is a nonlinear dimensionality reduction algorithm. It tries to discover the structure of the manifold using linear local relations between the data points. This is done by assuming that it is possible to reconstruct a point $\boldsymbol{x}^i$ by a convex linear combination of its neighbors  $\boldsymbol{x}^j \in \mathcal{K}(i)$. This relation should be preserved when it is mapped to a lower dimension, i.e., between the lower dimensional point $ \boldsymbol{c}^i \in \mathbb{R}^k$ and $\boldsymbol{c}^j \in \mathcal{K}(i)$, where $\mathcal{K}(i)$ is preserved between the two dimensions. Formally, for given neighbor relations between the points $\mathcal{K}_i$, the solution can be found by two optimization problems:
    \begin{align}\label{eq:LLE1}
\min_w \sum_i |\boldsymbol{x}^i -\sum_{j \in \mathcal{K}_i}  w_{ij} \boldsymbol{x}^j|^2 , ~~~~~ s.t. \sum_{j \in \mathcal{K}_i} w_{ij} = 1 ~~~~~\forall i, \end{align}
    which could be solved in closed form solution. Then for given weights $w_{ij}$, we need to solve for $\boldsymbol{c}^i$:
    \begin{align}\label{eq:LLE2}
    \min_{\boldsymbol{c}} \sum_i |\boldsymbol{c}^i -\sum_{j \in \mathcal{K}_i}  w_{ij} \boldsymbol{c}^j|^2
    \end{align}
    The $\mathcal{K}_i$ can be simply the K-nearest neighbors. Alternatively, $\mathcal{K}_i$ can be found using domain knowledge. An alternative representation of  (\ref{eq:LLE2}) is,
     $$\min_{\boldsymbol{C}} |{{\bf C} ({\bf I}_N-{\bf \Omega})}|^2 $$
     where ${\bf C} = [\boldsymbol{c}^1,...,\boldsymbol{c}^N]$, is a $k \times N$ matrix, ${\bf I}_N$ is an identity matrix of size $N\times N$, and ${\bf \Omega} =[\tilde{\boldsymbol{\omega}}^1,...,\tilde{\boldsymbol{\omega}}^N]$ is a $N \times N$ matrix, such that a vector $\tilde{\boldsymbol{\omega}}^i$ has the $j^{\rm th}$ element equal to $\omega_{ij}$ and the rest are zero. We can define $L = {\bf I}_N-{\bf \Omega}$ as the graph {\em Laplacian}. This is an important matrix that is the central component in  Laplacian eigenmap dimensionality reduction algorithms that are used widely in localization, where the goal is to directly optimize
     \begin{align}\label{eq:lapEg}
         \min_{\boldsymbol{c}} \sum_{i,j} \omega_{ij} (\boldsymbol{c}^i-\boldsymbol{c}^j)^2 \equiv \min_{\bf C} {\bf C^\top L C}
     \end{align}
 Here the weights can be set (or identified) from the constructed weighted graph that capture the correlation of the vertices (i.e., $\boldsymbol{x}^i$'s).
\end{itemize}

\subsubsection{Other Learning Approaches}
As highlighted above, supervised and unsupervised learning are not the only classes of machine learning approaches. Other include
\begin{itemize}
    \item {\em Semi-supervised learning}. In such approach the task is usually similar to supervised learning, however, the dataset includes both labeled and unlabeled data, i.e., the dataset  $\mathcal{D} = \{\boldsymbol{x}^{i},y^{i}\}_{i=1}^N \cup \{\boldsymbol{x}^{j}\}_{j=1}^M$, has $N$ and $M$ labeled and unlabeled data points, respectively. 
     \item {\em Reinforcement learning}. The system observes a stream of data with partially observable labels or sometimes only feedback values that indicate the quality of the system's decision (actions).  
    \item {\em Online Learning}. The data is only available during system operation.
    \item {\em Transfer learning (TL).} The available data include a large number of data examples from a {\em source} domain that is "close" to the domain of interest ({\em target}  domain), and a few examples from the target domain. By domain we refer here to the observable features and the task of the ML problem. In TL problems at least characteristics of the data (e.g., distribution $p(\boldsymbol{x}_d)$ or $p(y_d)$), the mapping (e.g., $p(y_d|\boldsymbol{x}_d)$) could be different between the two domains, we used the subscript $d$ to refer to the domain specific feature and labels.
\end{itemize}


\subsubsection{Deep Learning}
There is no unique definition of DL but it usually refers to ML techniques that use hierarchical learning structures. DL has recently shown remarkable performance in a number of challenging fields, such as computer vision (CV) and natural language processing (NLP). Being a subset of ML techniques, DL has been incorporated in number of the different classes of ML, such as supervised, unsupervised etc. Many of the recent successful architectures are based on Neural Networks (NNs), thus in the following we describe NNs, their basic architectures, and how they are trained. Next, we summarize a number of popular architectures that have been used in localization problems. Each is a representative of certain interesting properties in DL. We then summarize novel solutions to a number of challenges in DL. 
\begin{figure*}
 \centering
   \vspace{-7mm}
 \includegraphics[width=.95\textwidth]{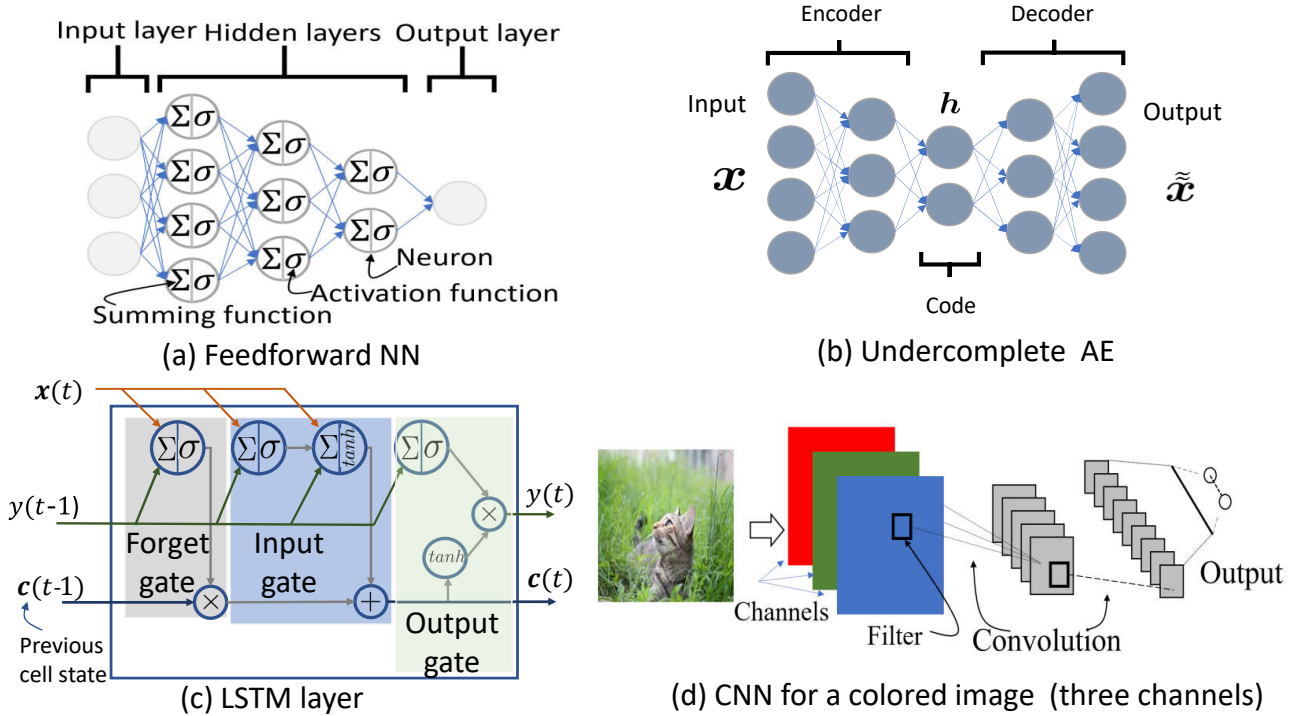}
 \vspace{-18mm}
  \caption{\small Examples of different deep NN architectures.}
  \vspace{0 mm}
 \label{fig:comb}
\end{figure*}

\begin{figure}
\centering
\vspace{- 5 mm}
 \includegraphics[width=0.75\columnwidth]{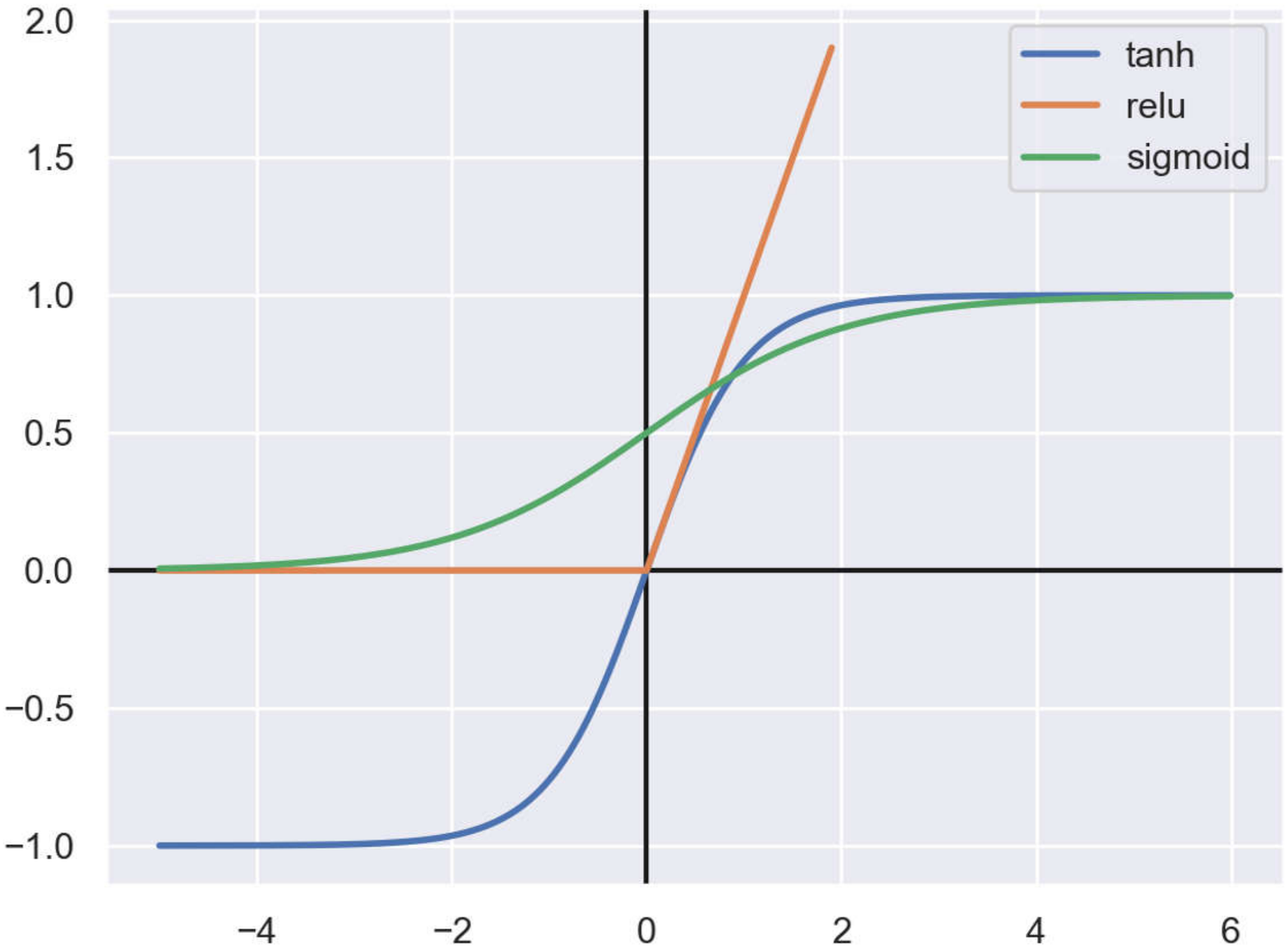}
\vspace{-3  mm}
  \caption{ \small Three different activation functions (e.g., referred to in Fig. \ref{fig:comb} as $\sigma$ functions  in various DL architectures).}
  \label{fig:activation}
  \vspace{- 5 mm}
\end{figure}
 
{\bf Feedforward Neural Networks}
Artificial NNs are powerful structures designed to imitate the human brain. An NN consists of one or more layers, each of which has a number of parallel neurons (nodes), see Fig. \ref{fig:comb}-(a). The neuron computes a weighted combination of the input features and then passes it through a (usually non-linear) transformation, a.k.a. "activation function". Fig. \ref{fig:activation} shows examples of such nonlinear activation functions. Thus a neuron can be viewed as a generalization of logistic regression (see above), or a perceptron (uses threshold based activation function).  

When NNs have no feedback loops they are referred to as Feedforward NN. The simplest architectures are sometimes referred to as fully-connected multi-layer perceptrons (MLP), where each node (perceptron) in a layer is connected to all other nodes in the following layer, see Fig. \ref{fig:comb}-(a). One reason for their popularity can be attributed to the universal approximation theorem, which states that with at least one hidden layer and non-linear activation functions (such as sigmoid), NNs can approximate any function with arbitrary accuracy\cite{hornik1991approximation}. However, to achieve such results with single hidden layer, it could be necessary to have an very large number of neurons \cite{goodfellow2016deep}. Increasing the depth of the network can alleviate this problem, as the number of possible network states grow exponentially with the depth of the network \cite{goodfellow2016deep}.

Even to train a single neuron, similar to the logistic regression example, we need an optimization algorithm. For the training of multiple neurons and deep architectures, efficiency of such training is critically important. NN training commonly uses backpropagation, which consists of forward pass of the training input values $\boldsymbol{x}^i, i\in \{1,.., n\}$. Then, utilizing the chain rule, we back-propagate the gradient of the loss function over the layers; subsequently algorithms such as gradient descent can be used to update the parameters. Since many DL problems use large datasets, the parameters update is usually done over mini-batches (compared to using all the data at once to calculate the gradient in each iteration as a single batch); this is usually referred to as Stochastic Gradient Descent (SGD). There are various SGD based optimizers; methods can use different functions of the gradients, e.g., the weighted average of the gradients. The Adam (Adaptive Moment Estimation) optimizer is one of the most widely used algorithms \cite{goodfellow2016deep}.

{\bf Convolutional NN (CNNs):} are efficient Feedforward NNs architectures, see Fig. \ref{fig:comb}-(d), that have shown outstanding performance in CV \cite{krizhevsky2017imagenet}. Compared to a fully connected NN, CNNs use parameter sharing, which allows building deeper networks with a much smaller number of parameters. Key components in CNNs are the "filters" (sometimes referred to as kernels or features map) and the convolution operation. The filters have tunable weights that are multiplied with the input data during the convolution process. However, different from the fully connected NN, the size of the filter could be much smaller than the size of the input data. To demonstrate that, let us consider an example from CV: For an image of size $1024 \times 768 \times 3$ (or generally $n \times m \times k$), corresponding to height, width and color dimensions of an image,\footnote{Note here we deviated from our initial definition of an example $\boldsymbol{x}\in\mathbb{R}^d$ to $\boldsymbol{x}\in \mathbb{R}^{n\times m \times k}$, as each data-point (i.e., example) represents a tensor.} we could apply a filter of size $2 \times 3 \times 3$ (or generally $a \times b \times c$), where $a<<n, b<<m$. Note that the third dimension is sometimes referred to as "channels". The filter is then "convolved" with the input data, where the filter can start from, for example, the left upper corner of the image, then the filter's weights and the input data values in the over lapping region are correlated (element-wise multiplied followed by summation), the filter is then shifted by $s$ pixels to the right and then correlated with the overlapping pixels. The value of $s$ is usually referred to as the {\em stride}. The process continues to produce an output of size $ \floor{\frac{n-a}{s}+1} \times \floor{\frac{m-b}{s}+1}\times 1$, the last dimension (number of channels) is equal to the number of filters used at this layer. 
 
We should here point out two interesting observations from the  example above: (1) the parameter sharing, where all the pixels share the same set of weights. (2) Connection sparsity, where each output value is produced by $a\times b \times c$ elements. These properties contribute to interesting features of the CNNs: translation invariance and efficiency. The translation invariance refers to the robustness of the CNNs to translation, and thus produces similar responses regardless of the shift of the input data. The efficacy allows to build deeper architectures, and thus reap the advantages of the network depth as discussed above. Also the reduction of the number of trainable parameter reduces the chances of over-fitting, making the network trainable with a smaller number of data examples. For example, \verb|AlexNet|, which is one of the most popular image classification models, is built with 5 convolutional layers with about 60 million parameters; it has achieved remarkable performance in classifying images in \verb|ImageNet| dataset into 1000 classes.  We finally note that many problems in localization can be re-framed as image classification, e.g., RSSI values are placed into a 2D grid and the output is the corresponding fingerprint; as will also be discussed in later sections. 

{\bf Recurrent NN} 
Recurrent Neural Networks (RNNs) are a type of NN widely used in modelling sequential data. They have shown great promise in time series data processing and NLP. RNNs  capable of performing tasks in which the current output is dependent on both the previous outputs and the current input. RNNs can be thought of having the ability to “remember” what has been calculated so far.
RNNs have a hidden state $\boldsymbol{h}(t)$, which can store information about the past computations. The hidden state is a function of the previous hidden state and the current input and is calculated by 
$$\boldsymbol{h}(t) = \sigma_h(V_x \boldsymbol{x}(t) + V_h \boldsymbol{h}(t-1)+ \boldsymbol{b}_h),$$
where $\sigma_h()$ is the activation function, $V_h$ and $V_x$ and $b_h$ are trainable parameters. The output at time $t$, $y(t)$, can then be calculated using $\boldsymbol{h}(t)$ as:
$$	y(t) = \sigma_y(W_h \boldsymbol{h}(t) + \boldsymbol{b}_y),$$ where $\sigma_y$ is the activation function.
The weights in RNNs are shared across time. For example, $V_h$ and $W_h$ ($\boldsymbol{b}_h$ and $\boldsymbol{b}_y$) are shared among all connections in a layer. At each point in time, the back propagation through time (BPTT) algorithm is used for training and adjusting the weights in each iteration.
 However, RNNs usually suffer from gradient vanishing or explosion, where the gradient values decay (or grow) as we propagate the gradient through time. An even bigger problem is the decaying influence over time i.e., the model cannot remember too far in the past. Hence vanilla RNNs are not widely used in practical applications. 
In practice, decaying influence or degradation over time is addressed by using “gates” in the RNN cells. Long Short Term Memory (LSTMs) and Gated Recurrent Units (GRUs) are the two most popular gated RNN models, see Fig. \ref{fig:comb}-(c) for an example of an LSTM architecture, where the network uses three gates to control what to remember in the memory cell, when to use it, and when to forget it. There are many variations and designs of RNNs, such as bi-directional RNNs. Examples of successful deep architectures based on RNNs are sequence-to-sequence networks \cite{sutskever2014sequence}, Conv-LSTM \cite{shi2015convolutional}. In localization, RNNs are of obvious importance for tracking the location of targets over time. 

{\bf Auto-Encoders}
Auto-encoders (AEs) are NN architectures where the output of the of the network is a "copy" of the input data, see Fig. \ref{fig:comb}-(b). Thus, AEs are usually considered as  unsupervised learning architectures. We can distinguish three main components of the AEs, (i) the encoder $g(.)$, (ii) the latent variable $\boldsymbol{h}$, and (iii) the decoder $f(.)$. The encoder maps the input $\boldsymbol{x}$ data to the hidden state $\boldsymbol{h}$, the decoder maps $\boldsymbol{h}$ back to $\boldsymbol{x}$. In practice, we are interested in $\tilde{\tilde{\boldsymbol{x}}} = f(g(\tilde{\boldsymbol{x}}))$, where the relations between $\boldsymbol{x}$, $\tilde{\boldsymbol{x}}$ and $\tilde{\tilde{\boldsymbol{x}}}$ depend on the application at hand. For {\em denoising} AEs, we want the AE to denoise the input signal, such that $\tilde{\tilde{\boldsymbol{x}}} \approx \boldsymbol{x}$ given that the data used at the input is noisy version of $\boldsymbol{x}$, i.e., $\tilde{\boldsymbol{x}} = \boldsymbol{x} + \boldsymbol{n}$, where $\boldsymbol{n}$ represents the noise vector. In this case, the loss function defined in eq. (\ref{eq:cost}) has $y = \boldsymbol{x}$, i.e., the "label" here is the noise-free data. The training procedure follows the standard backpropagation algorithm that we described earlier.

In another important application, we might be interested in the latent representation $\boldsymbol{h}$, which is sometimes referred to as "code". This can be seen as dimensionality reduction when the size of $\boldsymbol{h}$ is much smaller than that of $\boldsymbol{x}$. In this case our goal is to train the AE such that $\tilde{\tilde{\boldsymbol{x}}} \approx \boldsymbol{x}$, where here $ \tilde{\boldsymbol{x}}= \boldsymbol{x}$. For these applications, we hope that $\boldsymbol{h}$ provides a robust concise representation or a projection on the underlying manifold. Note that when the neurons of the AEs use linear activation functions, AEs can be shown to be equivalent to PCA. With nonlinear activation functions, AEs can provide powerful non-linear dimensionality reduction.
   
There are several other kinds and architectures of AEs, e.g., penalizing the hidden representation $\boldsymbol{h}$ leads to a sparse AE, i.e., by adding an $L_1$ regularization to the cost function $r(\boldsymbol{h}) = |\boldsymbol{h}|_1$, compare to (\ref{eq:cost}). This can be used to learn important features and be used as method for pre-training other NN based architectures \cite{goodfellow2016deep}. Another example is a generative AE network named variational AE (VAE), which is derived based on the variational inference for generative models. VAEs take samples from a random distribution $p(\boldsymbol{h}|\boldsymbol{x})$ (i.e., the encoder) to generate examples of $\boldsymbol{x}$. The VAE utilizes what is known to be the {"reparameterization trick"}, that allows to separate the stochastic sampling process from the parameters of $q(\boldsymbol{h}|\boldsymbol{x};\Phi)$, so that we backpropagate through the parameters and train the model as we do for Feedforward NNs.

{\bf Restricted Boltzman Machine and Greedy Pretraining}
Restricted Boltzman machines (RBM) are simple undirected graphical models that are used to learn the distribution of the data $p(\boldsymbol{x})$. They consist of one layer of observable data and one layer of hidden data, with no connections within the same layer. The probability relation between the vertices are captured with an energy based model 
$$ p(\boldsymbol{x},\boldsymbol{z}) = \frac{1}{Z} \exp^{-E(\boldsymbol{x},\boldsymbol{z})} $$
with $E(\boldsymbol{x},\boldsymbol{z})$ represent the "energy" between the variables which have trainable parameters, and $Z$ is a, usually intractable, partitioning function that ensures that $p(\boldsymbol{x},\boldsymbol{z})$ is a proper probability. The bipartite nature of the graphs makes RBM relatively easy to be trained by approximating the gradients of Maximum Likelihood using "Contrastive Divergence" (iterative procedure uses Gibbs sampling to sample the model) \cite{goodfellow2016deep}. 

Greedy layer-wise pre-training was one of the early methods to allow training deep networks. The network is built layer-wise, where initially, the first hidden layer represents the hidden layer RBM (or the hidden layer is of that for one layer AE). Once trained, the hidden layer will be the input layer of a new hidden layer that will be added on the top of the current network. This process continues until the network structure is constructed. This provides good initial weights and acts as regularization \cite{goodfellow2016deep}. However, the modern training and regularization methods have over-shadowed these techniques. 


{\bf Techniques to Enhance DL Solutions}
 The hierarchical and deep structure of DL solutions exacerbates some of the existing ML challenges and brings many new ones. For instance, in DL, training networks with tens of million of parameters is not unheard of (AlexNet has over 60 million parameters), thus aggravating the over-fitting problem. In the last decade many novel techniques were proposed to alleviate some of these challenges. Here we {\em briefly list} some of them:
\begin{itemize}
    \item {\em Dropout} is one of the novel techniques to combat over-fitting, where some of the neurons are shut off at random during training. This increases the network robustness as it is supposed to provide a solution even if some of its internal connections are not available, or highly noisy (think of product noise). One interesting view of dropout is that it can be viewed as bagging with exponentially many NNs\cite{goodfellow2016deep}. 
    \item {\em Noise injection:} here we can corrupt the data with several realizations of noise. While artificial addition of noise seems counter-intuitive - in particular as many systems try to reduce the noise during preprocessing step earlier - it 
    increases the number of available data and increases the network's robustness to small variations. 
    \item {\em Data augmentation.} Since overfitting can occur due to the limited number of available examples, one solution is to increase the size of the dataset. However, collecting new data is usually costly. In data augmentation we aim at generating new data examples from the existing ones. Noise injection above is one such technique. Another example is shifting, rotating or cropping images in CV. We here emphasize that data augmentation techniques that are valid in one domain may not be applicable for other problems.
    \item {\em Early stoppage.} One intuitive and efficient solution to overfitting is to stop the iterative training algorithm before it starts fitting the parameters over the noise and the outliers. Finding the right stopping time can be assessed by evaluating the performance over a validation set. 
    \item {\em Utilizing related data through pre-training.} Typically data from a different domain, or unlabeled data could be available in abundance. Pre-training over such data could be very helpful. Briefly, some of these benefits include:
    \begin{itemize}
        \item Provide good weight initialization, since during the training process we run an iterative optimization.  
        \item Utilize low level features that could be shared over different domains. This is also related to the concept of Transfer Learning.
        \item Allow the network to estimate the underlying density of the data and possible correlation. This is usually viewed to be in the domain of unsupervised learning.  
    \end{itemize}
\end{itemize}

\subsubsection{Available DL Platforms}
One reason behind the accelerated research and development of DL based solutions can be attributed to accessibility of suitable open-source platforms and libraries in common programming languages, some of which were initially developed by the industry. For DL, \verb|Python| is the most widely used programming language, as per a \verb|GitHub| survey. 
This can be attributed to the fact that \verb|Python| syntax is easy to learn and apply,
so that more tools and frameworks for DL are available in \verb|Python| compared to other languages. 
\verb|C++| is the second-most popular language for ML, and in particular applied 
where efficiency is key and speed is needed. 
\verb|C++| can fully exploit the power of GPUs at the very low level of programming. Python does not provide such flexibility. \verb|Java| and other languages, including \verb|R| and \verb|MATLAB|, are also used at times. 

The most widely used libraries are \verb|Numpy|, \verb|SciPy| and \verb|Pandas|; all of which are \verb|Python| libraries. These are normally used for intense vector math and are very efficient. 

\verb|TensorFlow| and \verb|PyTorch| are two of the most widely used DL libraries. They support a wide array of operations and provide high flexibility for doing almost anything in DL. Another famous framework is \verb|Keras|, which can be used with a backend of \verb|TensorFlow| or \verb|Theano|. Below are some comparisons of these three:

\begin{itemize}

   \item While \verb|TensorFlow| and \verb|PyTorch| are DL libraries, \verb|Keras| is a framework made just for DL. This means \verb|TensorFlow| and \verb|PyTorch| can be used for vector math, but Keras is made to make DL easier.
   \item \verb|TensorFlow| works on a backend of \verb|Theano|, whereas \verb|PyTorch| works on a backend of \verb|Torch|, both of which can be used as DL libraries themselves. \verb|Keras| can be used with a backend of \verb|Theano| as well as \verb|TensorFlow|.
   \item Prior to \verb|TensorFlow 2.0|, \verb|TensorFlow| did not support eager execution. This is because \verb|TensorFlow| used static computational graphs, which cannot be changed once created. So execution of operations would require a \verb|tf.Session()| or an Interactive Session. \verb|PyTorch| and \verb|Keras| both don’t need such a Session (because \verb|PyTorch| graphs are dynamic).
   \item  \verb|PyTorch| is generally used in research projects and \verb|TensorFlow| is used in production systems. 
   \item  The three frameworks are all open-source.
\end{itemize}
Note that we focused here on the three frameworks above that generally have the largest number of users along with a very strong online community support. However, there are other frameworks such as \verb|Caffe|, \verb|MxNet| and \verb|Theano|.


 \subsection{Wireless Channels}\label{sec:prelWireless}
 RF-based localization systems rely on the properties of the wireless propagation channel such as the attenuation or the flight time of a signal traveling from the transmitter (TX) to the receiver (RXs). 
 In free space, where a single electromagnetic wave carries the signal without interaction with other objects, these properties can be easily related to the location of TX and RX. 
 However, real-life propagation channels are much more complicated, a fact that creates the main challenges for precision localization. In this subsection we thus review the most important propagation characteristics that have an impact on localization systems. Furthermore, since many localization systems are integrated into communications systems, we also present a brief overview of the most salient features of those systems. 
 
 A wireless RF signal emitted from the TX antenna(s) interacts with the environment through reflection, scattering, and diffraction at various objects in the environment before it arrives at the RX side. These processes give rise to, and determine the properties (such as amplitude, phase, delay, and angle) of, the multi-path components (MPCs) \cite{ProfMolischText}. The different MPCs, which are usually modeled as plane waves, have different amplitude $ |\alpha |$, phase shift $ \phi$, delay $\tau$, angle-of-departure (AoD) $\Omega$ from the TX, and angle-of-arrival (AoA) $\Psi$ at the RX. The {\em propagation channel}, more precisely, the double-directional channel impulse response, can thus be written as a sum of $N(t)$ plane waves (MPCs) \cite{steinbauer2001double}
 \begin{equation}
   h(t,\tau, \Omega, \Psi)  = \sum_{l=1}^{N(t)} \alpha_l \delta(\tau-\tau_l)\delta(\Omega-\Omega_l)\delta(\Psi-\Psi_l),
   \label{eq:DD-IR}
 \end{equation}
 where $ |\alpha |,\tau,\Omega,\Psi$ are constant within a small area (typically a few meter) called ''stationarity region". When a TX or RX moves over larger distances, also the $ |\alpha |$ (due to shadowing and changing pathloss), and the $\tau,\Omega,\Psi$ (due to changes in the geometry) change. Furthermore, when the considered frequency changes by a large amount (typically $10 \%$ of the carrier frequency) called "stationarity bandwidth", the $ |\alpha |$ become a function of frequency, or - equivalently - the $\delta(\tau-\tau_l)$ is replaced by a function describing the delay dispersion of a single MPC $\xi(\tau-\tau_l)$; this situation is relevant for ultrawideband (UWB) systems \cite{molisch2009ultra}. Note that any impulse response can be expanded into a sum of plane waves as given above, yet in some cases it is more informative (and sometimes better related to the underlying physics) to describe a large number of weak components by a continuous "diffuse multipath component" DMC \cite{richter2005estimation}. Finally, we note that a further generalization can be achieved by incorporating the polarization of the MPCs, which converts (\ref{eq:DD-IR}) into a matrix equation \cite{shafi2006polarized}. We also note that a Fourier Transform (FT) with respect to $\tau$ provides the channel frequency response (CFR)  $h(t,\tau, \Omega, \Psi) \xrightarrow{\mathcal{F_{\tau}}} H(t,f, \Omega, \Psi)$.
A FT with respect to $t$ provides a transformation to the Doppler shift $\nu$, resulting in the Doppler-variant impulse response, also known as ''spreading function" $h(t,\tau, \Omega, \Psi) \xrightarrow{\mathcal{F}_{t}} H(\nu,\tau, \Omega, \Psi)$. 
\subsubsection{ Impact of the antenna and system}
 The above description describes the propagation channel only, without incorporating the effect of the antennas or the RF components. The signals that are actually observed at the antenna connectors of an array are related to the double-directional impulse response as 
 \begin{equation}
     h_{m,n}(t,\tau)=\int d\Psi G_{{\rm RX},n} (\Psi) \int d\Omega  G_{{\rm TX},m} (\Omega) h(t,\tau, \Omega, \Psi)
 \end{equation}
 where $G_{{\rm TX},m} (\Omega)$ and $G_{{\rm RX},n}(\Psi)$ are the complex antenna pattern of the m-th TX and n-th RX antenna element, respectively. For the case of single-antenna elements, we simply set $m=n=1$.   
 We can see from this that the measured impulse response is not only a function of the propagation channel, but also of the antennas, and that measurements with different antennas thus result in different impulse responses even for the  same propagation channel, i.e., the same location. 
 
 Furthermore, the necessarily finite bandwidth of the transmit waveform and receive filter leads to a superposition of the MPCs that arrive at approximately (within one inverse bandwidth) at the same time; this superposition can be constructive or destructive depending on the different phase shifts. Thus, the actually measured "channel impulse response" (CIR)  impulse response is 
 \begin{equation}
 h_{{\rm meas},m.n}(t,\tau)=h_{\rm sys}(\tau) \ast h_{m,n}(t,\tau) ~.     
 \end{equation}
where $"*"$ is the convolution operator, and $h_{\rm sys}$ is the effective system impulse response. Obviously, the smaller the bandwidth, the more MPCs are superimposed by this convolution operation, and the temporal resolution (ability to find the delay of the MPCs decreases). 
A similar observation can be made for directional channel characteristics: a suitable FT can move the observations from different antenna elements (location space) into the ''beamspace", providing the same results as if we would have a set of co-located directional antennas that are pointing into different directions \cite{sayeed2002deconstructing}. Each of those virtual directional antennas superposes the MPCs that are in its beamwidth. Thus, the smaller the antenna aperture (and thus, the larger the beamwidth), the more MPCs are superposed, and the worse the directional resolution. 

\subsubsection{{Small Scale vs Large Scale Fading}  }
The observables, $h_{{\rm meas},m.n}(t,\tau)$, experience what is typically called small scale and large scale fading. Small scale fading is caused by the fast variations of the phase (due to movement of the TX, RX and/or objects in the environment) and resulting change in constructive/destructive summation. The amplitude (power) variations of the signal due to small scale fading can be 30 dB or more. 
A movement of the TX or RX on the order of a wavelength is usually sufficient to change a constructive to a destructive interference or vice versa. This may be advantageous, e.g., for a fingerprinting-based localization system because it provides a very sensitive measure of location. Yet, small-scale fading also has major drawbacks. For example, 
even when TX and RX are completely static, the movement of interacting objects (e.g., moving pedestrians, cars, etc.) can significantly change the small scale fading state. Thus, for a fingerprinting system, the observables are sensitive to small environmental changes beyond the control of the system. In addition to fading, multipath propagation also leads to delay dispersion or, equivalently, frequency selectivity. The former means that even when we send a a very short pulse from the TX, the received signal extends over a considerable delay range; the latter means that at different frequencies, we observe different values of the small-scale fading. Multipath propagation has been generally considered an obstacle in deterministic localization procedures, though recent work has established ways of turning it into an advantage, see \cite{witrisal2016high} and references therein.

Besides the small scale fading, there are also larger-scale variations that are caused by the fact that the other MPC parameters (power, angle, delay) change as the TX and/or the RX move over larger distances. The large-scale characteristics, such as the angular power spectrum (power as a function of the incident directions, averaged over the small scale fading), are thus indicative of the general area in which a device is located. On an even larger scale occurs the pathloss, which is a power loss due to the "thinning out" of the area power density as waves propagate further and further away from the TX. In the case of a pure ''free-space" scenario, this power loss is proportional to the square of  the TX-RX distance $d$; more generally it can be modeled as $d^{n_{\rm PL}}$, where the pathloss coefficient $n_{\rm PL}$ depends on the environment and whether a LOS connection exists between TX and RX.

\subsubsection{Channel Models} There are a number of models for wireless propagation channels that can be used to test localization algorithms. Most important among these are "statistical channel models", which prescribe the small-scale and large-scale statistics of the propagation channel in closed form (or tabular form), based on which different realizations of the channel can be created. The most widely used of those models are the 3GPP channel models, originally designed for 3G cellular communications systems \cite{calcev2007wideband,3GPP38901}, and repeatedly extended over the years, so that they are now used for 4G and 5G cellular systems as well. An efficient and widely adopted implementation is provided by the Quadriga website quadriga-channel-model.de, see also \cite{jaeckel2014quadriga}. The model generates double-directional channels for a ''drop" of a UE (user equipment, mobile station), i.e., placing of a UE in a particular location within a cell. The "baseline" direction is the LOS between the base station (BS) and the UE, and all angles are defined relative to this direction. The double-directional impulse response consists of a number of \emph{paths}, each of which has a  particular delay, which is chosen at random, according to a given (parameterized) probability density function. Power is assigned according to the path delay, with the average power decreasing with increasing delay. Furthermore, each paths consists of 20 sub--paths, which all have the same delay, but slightly different angles, such that each path has an angular spread of, e.g., 5 degrees. Each of the sub--paths has the same amplitude, and random phases; their superposition thus provides not only an angular spread, but also small--scale fading when either different values of the random phases are chosen or the UE moves. In the latter case, the geometric relationship between the antenna array orientation, direction of the sub-paths at the UE, and the UE movement vector allow to compute the temporal changes of the impulse response. 

While, of course, the propagation channel is independent of which applications are run over it, the admissible simplifications of a channel {\em model} do depend on the application. 
Thus, it must be kept in mind that the 3GPP model was designed to test communications systems, and not localization systems. For example, the average power in the first path is assumed to be the largest, even though extensive measurements have shown situations with a ''soft onset" (e.g., \cite{karedal2004uwb}), and other situations where there is no discernible energy at the LOS arrival time, and the first component is only several ns later \cite{kristem2014experimental}. Furthermore, changes of the channel when the UE moves over large distances are not modeled accurately (the recently introduced ''spatial consistency" simulation approach solves the problem only partially), and the purely stochastic modeling of shadowing based on a simple (exponential correlation) shadowing model also fails to model interesting higher-order relations between channels at widely separated locations. Last but not least, the channel model has a finite number of MPCs that are all fairly strong, and no DMC. As a consequence of all these simplifications, the channels generated by the 3GPP model might not show the same complexity as real-world (measured) channels, which might have significant impact on localization in general, and ML-based localization in particular. 

A channel model that is somewhat more realistic for localization systems is the IEEE 802.15.4a model \cite{molisch2004ieee}, which was designed explicitly for the testing of both communications and localization. It is also a statistical channel model, but considers the "soft onset" of MPCs. However, it does not include directional characteristics, and is furthermore not intended for outdoor usage. 

\section{Types of Localization}\label{sec:LocGaol}
The ML-based localization solutions that have been proposed in the literature differ depending on the considered localization level, as well as the target engagement. In the following subsections, we discuss those two aspects; Table \ref{tab:LocType} lists papers based on these categorizations. We defer the discussion of the conventional localization systems vs. the ML-based ones to Sec. \ref{sec:ConvVsMLSystems}, i.e., after we introduce the basic RF features in next section.

\subsection{Localization Level}
We here differentiate between three types: (i) Region classification, (ii) Fingerprint classification, (iii) and coordinates estimation.
Region classification means that the system aims to identify a sub-region of the study area in which the target is located, such as the building, floor, or room. This can be interpreted as finding the location with a certain, more or less rough, quantization, or, in other words, a classification problem. This result can be either part of a multi-step localization, i.e., forming the starting point for a more accurate coordinate-level localization, or used by itself. As a matter of fact, in many situations this result will be sufficient. For example, in buildings with many small rooms (e.g., hotels) identifying the target at room level will be sufficient.
The main advantages of this approach are (a) low misclassification error, and (b) possibly reduced computational effort. 

Alternatively, in fingerprint localization, the solution could be to match the target to the closest fingerprint in the database, also known as Reference Point (RP). Also this solution approach can be viewed as classification problem with $N$ distinguished classes, where $N$ is the number of RPs in the database. The localization error (in meter) that is engendered by such an approach depends on many factors such as the separation distance between RPs in the database, the environment and the matching method. Note especially that a small error in the fingerprint could lead to a large error in the localization. 
To provide a better accuracy, several work have considered the fusion of the closest $K$ locations of target locations. This reduces the probability of associating with a far-away fingerprint, and furthermore reduces the quantization error, since a location interpolation between the set of nearby RPs can be performed. 

The most natural localization solution is to provide the explicit coordinates of the target; in the context of ML this can be seen as a regression problem. The approach can either use as input the signals at/from all the anchors, and deduce the target location directly, or it can learn the relation between the input features and the distance in considered dimensions, e.g., the mapping between the observed time of signal arrival and the distance, which then can be fed to a classical trilateration algorithms.

There have been several other solutions that use ML to aide the localization process, such as LOS vs Non-LOS (NLOS) discrimination \cite{van2015machine_1,huang2020machine}, or scheduling of localization signals \cite{peng2019decentralized_124}; however, these applications are not at the core of this survey.

\subsection{Target Engagement}
Depending on the application, localization systems could differ in the level of target engagement in the process. We here classify the localization to be 
\begin{itemize}
    \item {\em Active localization}: here the target is equipped with a wireless transceiver, and participates in the localization process. Due to the operating principle, different targets can be easily differentiated. While normally only anchors and targets exchange information, improved accuracy can be achieved when multiple targets cooperate with each other ({\em cooperative} localization). 
    For future reference, we furthermore define
    \begin{itemize}
       \item {\em One-way localization:} in this approach, the target either sends out a localization (training signal) that is received by the anchor node(s) of the system, or conversely it receives localization signals from the anchor nodes. 
       \item {\em Two-way localization:} in this approach, the target both transmits to and receives from the anchors signals that can serve to localize. 
    \end{itemize}
    \item {\em Passive localization}: here, the target does not have a wireless transceiver, but rather only reflects a signal; this is sometimes referred to as device-free localization, and more often as radar. Different targets are not necessarily easy to distinguish from their signal characteristics - for example, cars of similar type have similar radar signatures.\footnote{ML for radar is outside the scope of the current paper, as it focuses on different aspects such as detection and identification of objects with possible presence of jamming systems, some of which are related to remote sensing and image processing. A comprehensive survey is provided, e.g., in \cite{lang2020comprehensive}. The coupling of the two could be interesting research direction.}  
    \item{\em Semi-passive localization}: a borderline case are passive RFID tags. While they are passive in the sense that they do not contain transceivers and only reflect signals, they do so in a way that is unique for a particular tag, and thus allow distinction of the signals from different targets. 
\end{itemize}
Table \ref{tab:LocType} shows a list of relevant papers.
 
\begin{table}
\begin{tabularx}{0.5\textwidth} { 
   >{\centering\arraybackslash\hsize=1.05\hsize}X 
   >{\centering\arraybackslash\hsize=0.95\hsize}X 
   >{\centering\arraybackslash}X 
   >{\centering\arraybackslash}X }
 \hline
  & \footnotesize Classification Region & \footnotesize Classification RP & \footnotesize Coordinates \\
 \hline
  \vspace{1.5mm}  Passive & ~\cite{vlfANN ,  hybloc} 
 & ~\cite{InterpretingCNN ,  devicefree ,  gbrbm ,  8673800 ,  HumanDetection ,  DCNN ,  wu2018accurate  ,  sanam2018improved ,  zhao2019accurate} 
 & ~\cite{FineGrainedSubcarrier ,  hybloc  ,  yang2018clustering ,  chen2019dpr}\\
 \hline

 \vspace{7mm} Active & ~\cite{laperls ,  8651736 ,  6805641 ,  8362712 ,  hapi ,  8008794 ,  yang2012locating ,  kim2018scalable ,  kim2018hybrid ,  secckin2019hierarchical ,  oussar2011indoor} 
 & ~\cite{schmidt2019sdr,nguyen2005kernel,FPfusing ,  NLOS_UWB ,  OnDevice ,  lorawan ,  IndustrialWSN ,  uav ,  location_aware ,  probabilistic_localization ,  8780770 ,  DNN/KNN ,  Discriminative ,  DataAugmentation ,  8362712 ,  Timotheatos2017FeatureEA ,  FeatureSelection ,  RelationLearning ,  FullBandGSM ,  hiwl ,  7064997 ,  JUIndoorLoc ,  8661625 ,  Wang2017WiFiFB ,  Chunjing2017WLAN ,  8422182 ,  cottone2016machine ,  khatab2017fingerprint ,  rezgui2017efficient ,  liu2019hybrid ,  ahmadi2017exploiting ,  gu2015online ,  yan2017hybrid, salamah2016enhanced ,  zhang2019improving ,  belmonte2019swiblux ,  li2019smartloc} 
 & ~\cite{schmidt2019sdr,laperls  ,  MassiveMIMO ,  7743685 ,  MonoDCell ,  7938617}  ~\cite{OnDevice ,  telco ,  sdr ,  6805641 ,  wsn_svr ,  TargetTracking ,  tdoa ,  RadioMaps ,  tlfcma ,  LowOverhead ,  magnetic_inertial ,  patent ,  dabil ,  8240410 ,  8108573 ,  10.1145/3321408.3321584 ,  rfid ,  kmgpr ,  csiRNN ,  8765368 ,  Eloc ,  grof ,  hapi  ,  WeightedAmbientWiFi ,  eee112 ,  ZhangIndoor ,  8008794 ,  8584443 ,  8712551 ,  8931646 ,  LiUnsupervised ,  AntonioLocalization ,  RamaLocalization ,  yang2012locating ,  8115921 ,  Qian2019Supervised ,  Zhang2019Wireless ,  wu2019mobile ,  feng2019received ,  jaafar2018neural} \cite{felix2016fingerprinting} \cite{  kim2018scalable ,  yan2018noise ,  wang2019robust ,  comiter2017data ,  nguyen2017performance ,  musa2019decision ,  comiter2017structured ,  zou2015fast ,  figuera2012advanced ,  guo2018accurate ,  bi2018adaptive ,  elbes2019indoor ,  comiter2018localization ,  zhang2018heterogeneous ,  adege2018indoor ,  adege2019mobility ,  yan2017hybrid ,  malik2019indoor ,  bae2019large ,  yi2019neural ,  zhang2018enhancement ,  tewes2019ensemble ,  kim2018hybrid ,  xiao2012large ,  widmaier2019towards ,  homayounvala2019novel ,  bhatti2018machine ,  zhang2019improving ,  secckin2019hierarchical ,  belmonte2019swiblux  ,  yang2012locating ,  xiao2018learning ,  mahfouz2015kernel ,  qiu2018walk ,  patel2020millimeter}\\
 \hline
  \vspace{0.25mm}  Cooperative &  & ~\cite{alteredAP} & ~\cite{7504333 ,  mmwave ,  OpticalCamera}\\
 \hline
\end{tabularx}
\caption{\small Example of papers with different types of localization and localization levels. Some papers present different levels of localization.}
\label{tab:LocType}
\end{table}

 
Active localization is the method most often used in ML-based localization due to its relatively simple setup and wide range of applications. The capabilities of the target device play a role in shaping the localization problem. 
\begin{itemize}
\item In cellular systems, BSs are typically the anchors, and UEs the agents (targets) of the localization. A variety of localization methods, some of them explicitly supported in the standards, have been developed, see Sec. IV. Most of these methods operate in the downlink, i.e., the BSs send out localization signals, and the UE determines its location from the received signals - this is in particular important for privacy reasons. The accuracy can be improved by 
acquiring side information using the UEs' built-in sensors. 

\item Wireless LANs (WiFi) are fairly similar to cellular localization, with Access Points (APs) taking the role of BSs, and stations (STAs) taking the role of UEs. The similarity is most pronounced in enterprise networks, where multiple APs are under the control of the network and can contribute to the localization in a coordinated way. In home settings, often only a single AP is available that can coordinate with the STA, while the other APs are not under the control of the same network/user; signals from those act as interference for communications though they might still be useful for localization purposes, e.g., for fingerprinting. 

\item Wireless sensor networks (WSNs) have gained more attention over the past few years due to the proliferation of the Internet of Things (IoT). Sensor nodes have limited processing capabilities. Therefore, in WSNs, active localization based on uplink transmission is preferred, so that the computations necessary for the localization can be done at the BSs, the gateway or other elements of the infrastructure. 

\item In contrast to traditional WSNs, which tend to have low mobility, 
vehicles are usually in highly dynamic environments. Localization there can be done with Road Side Units (RSU). Modern vehicles are usually equipped with many sensors and cameras that can be used as side information. For autonomous vehicles, it is envisioned to achieve reliable sub-meter localization accuracy, which could be difficult with GNSS-only systems. 

\item In passive localization, static transceivers could be set such that they sense the changes in the environment. The ML solution can be used to interpret such changes into a localization. This type of localization can be used for surveillance, occupancy detection, gesture detection, and object counting, see, e.g., \cite{liu2017wicount}. The difficulty of such problems lies in differentiation between target and non-target objects. 

\item Targets cooperation should usually improve the localization accuracy \cite{wymeersch2009cooperative}. It is a good approach particularly in infrastructure-less or highly dynamic environments. Overall, cooperative localization has received the least attention in ML-based solution.
\end{itemize}

\section{Features}\label{sec:Feat}

Although localization has a wide range of applications and has drawn huge commercial interest, standalone localization systems - with the notable exception of GNSS - are rare. Rather, existing wireless communications systems are adopted to provide also localization information, as a supplementary service. This minimizes the setup cost and increases the deployment speed, but on the other hand creates certain limitations on accessible features. In this section, we first introduce the basic wireless features that are used in localization. We describe how these features are used in ''classical" localization systems, and also survey which papers use them for 
ML-based localization. We then discuss their availability in different wireless standards and technologies that have been used thus far. We also summarize some of the used side information along with the wireless signals, and some of the methods for data transformation. A corresponding classification of papers is provided in tables \ref{tab:Standard} and \ref{tab:Technology}.
\subsection{Feature Type}
As introduced earlier, an ML-based localization solution is a mapping from the observed features $\boldsymbol{x}$ to the location $y$
\begin{align}
    y=  f(\boldsymbol{x}) ~~~.
\end{align}
Here, $y$ could be any system defined location output, such as class (e.g., room or RP number) or coordinates (e.g., 3-dimensional coordinates). The input to the ML solution, i.e., the features $\boldsymbol{x}$, are a design choice that largely depends on the accessibility of such data, usually constrained by the system, and the admissible complexity of the solution. Similarly, the choice of the features used for classical localization algorithms is limited by what is available in particular systems. 

\subsubsection{Signal Strength}

The simplest and most fundamental feature obtained by a receiver is the receive power, which we define here as the total power received over the bandwidth of the system, $P_{\rm RX}(t)=P_{\rm TX}(t)\int |H(t,f)|^2 df$.\footnote{Needless to say, the integration becomes summation in real system as discrete samples are available.}
The receive power is also often referred to as Received Signal Strength (RSS). Related to it is the Received Signal Strength Indicator (RSSI), which is a -- usually vendor specific -- quantization of the RSS value. In many cases, only this RSSI is available to localization system, in particular (i) if the information is measured at the UE in a cellular system, and needs to be fed back to the BS for localization, or (ii) the localization system does not have access to the actually received signals, but only the system parameters available via the APIs; this situation occurs when the localization system manufacturer and the chip manufacturer are different. However, in the remainder of this paper, we do not distinguish between receive power, pathloss, and RSSI anymore. 

If the transmit power is known, this allows to deduce the corresponding (wideband) pathloss. This requires that either the TX does not employ power control, or that the power control settings are known to the RX. For instance, in LTE, the Reference Signal Received Power (RSRP) is the time-averaged received signal of all {\em reference signals} from the serving BS. Reference Signal Received Quality (RSRQ), as the name suggests, indicates the quality of the received signal from the serving BS; some of these measurements are reported to the BS, which could be used for localization as used in \cite{ray2016localization_94,zhang2020transfer_49,zhu2016city_50r,butt2020rf_57,zhang2019deeploc_108}.

The RSS information can be used in one of the following two ways: (i) when the channel model is known, RSS can be mapped to distance. Thus localization could be achieved with trilateration with at least three anchors. In pure LOS scenarios, the received power maps to the distance by Friis' law
\begin{equation}
    P_{\rm RX} = P_{\rm TX} |G_{\rm TX}|^2 |G_{\rm RX}|^2 \left ( \frac{\lambda_w}{4 \pi d}  \right )^2
\end{equation}
where $\lambda_w$ is the wavelength. 
However, in most practical situations the applicable channel model is unknown, making the distance inference prone to errors (note that famous statistical channel models for the received power, such as the Okomura-Hata model, do not constitute a good basis for distance inference because they are averaged over measurements in a large number of environments); even if it were known, the fading variations make a mapping from RSS to distance almost impossible. (ii) RSS can be used as the basis of fingerprinting, in particular when the RSS from several BSs to the target device provides unique fingerprint points. The importance of RSS lies in the fact that it is widely available in most systems and easy to acquire, while the more detailed channel state information might not be always available, see below. 


Due to multipath, shadowing, as well as hardware impairments, and variations between different devices, the value of the RSSI could vary significantly between different measurements. Thus many works suggest recording the signal over a time window. To overcome the signal variation, many of solutions take several time measurements in the training and testing data, measurements with different devices, e.g., \cite{shokry2018deeploc_50,goswami2011wigem_75,choi2019unsupervised_123, pandey2019handling_127}, posture or directions e.g., \cite{bi2018adaptive,pandey2019handling_127, akram2018censloc_18}. Other solutions apply pre-processing techniques to denoise and stabilize the values. 
 

\subsubsection{Time of Arrival}
Time of arrival (ToA), is one of the basic quantities that have been used in standard localization systems (e.g., GNSS systems). In pure LOS (free-space) channel, the ToA can be converted to the distance between TX and RX. In a multipath channel, the ToA usually refers to the time of the first detectable MPC. The ToA estimation process usually relies on the TX sending out a training signal at a pre-determined time, from which the RX determines the impulse response. The largest/only (in the case of free-space propagation) or first (in the case of multipath) peak of the determined impulse response is used to determine the pseudo-range, i.e., the runlength (in seconds) between the TX and RX.

While determination of the ToA sounds simple in principle, there are a number of problems that arise both from hardware imperfections and the characteristics of the wireless propagation channel. First of all, the above description requires TX and RX to use clocks that are accurately synchronized to each other, which might be difficult to achieve in practice. The requirement can be circumvented by not measuring the runtime from TX to RX, but rather using a two-way signal exchange to determine the {\em round-trip} time from TX to RX and back, thus eliminating constant offsets in the clock times. Possible clock drift can be eliminated by multi-packet message exchange \cite{chui2009time}. An alternative approach is the use of Time Difference of Arrival (TDoA), where only the clocks of the different BSs are synchronized, and the localization determines the difference of the pseudoranges from UE to the different BSs, so that the timing offset of the UE clock cancels out; the same principle is also used in GNSSs.

Even with perfect clock synchronization, errors do occur due to noise. Assuming Additive White Gaussian Noise (AWGN), the amount of ranging error can be bounded by the Cramer Rao Lower Bound (CRLB)
\begin{equation}
var(\widehat{d})\geq \frac{c_{0}^{2}}{8\pi \gamma \beta ^{2}}
\end{equation}%
where $c_{0}$ is the speed of light, $\gamma $ is the Signal to Noise Ratio (SNR), and $\beta $ is the effective
bandwidth%
\begin{equation}
\beta =\left[ \frac{\int f^{2}|S(f)|^{2}df}{\int |S(f)|^{2}df}\right] ^{1/2}
\end{equation}%
with $S(f)$ the amplitude spectrum of the signal.
While this bound mainly depends on SNR and carrier frequency $f_c$, it is worth noting that it only holds at very high SNR; at intermediate SNRs, the Barankin bound, which is
larger by a factor $12(f_{c}/B)^{2}$ than the CRLB, and thus is proportional to the {\em bandwidth}, provides a better approximation. This is intuitive, as a finite bandwidth leads to a "smearing out" of the impulse response, and thus greater difficulty in determining where the true peak is.

ToA based methods obviously suffer from problems when the LOS to/from one or more of the anchors is blocked. In a pure LOS situation, this means that no signal (or a signal with insufficient SNR) is received; this situation often occurs for GNSS systems, e.g., in street canyons. Consequently no measurement result is available, which may (or may not) prevent localization by trilateration as discussed below. In multi-path channels, a blocked LOS will often still result in a pseudorange estimate, since the time of arrival of a reflected MPC might be interpreted as a pseudorange, leading to a positive bias of the range estimate. While in some situations having such an estimate might be useful, in other cases it can actually be worse than having no estimate at all, because erroneous information is entered into the trilateration. It is thus important to identify whether a blocked-LOS situation occurs or not, and take this information into account in the trilateration algorithm. A variety of methods for identification of blocked-LOS have been established in the literature, see \cite{aditya2018survey} for a survey. Besides "classical" methods for identification of blocked LOS, ML-based methods have been suggested, e.g., \cite{marano2010nlos,huang2020machine}. 

The amount of range error that occurs in a blocked-LOS situation depends both on the environment and the particular algorithm used for ToA determination. Firstly, there are some environments in which the delay between the LOS pseudorange and the pseudorange of the first identifiable component is very large, e.g., when a building blocks the connection between TX and RX \cite{kristem2014experimental}. There are also propagation channels with a ''soft onset", meaning that the first path is not the strongest, e.g. \cite{karedal2004uwb}. In such a situation, the algorithm that determines the range, and the criterion for identification of the ToA become important. For example, detection of the strongest peak in the impulse response can give much larger errors than identification of the first path. Various methods for finding the first path have been suggested, such as ''search forward" from the nominal ''zero delay" of the arriving signal, or ''search backwards" from the time of arrival from the strongest path \cite{sahinoglu2008ultra}. Finally, note that the pseudorange is also increased when a LOS has to propagate through a dielectric material, since the group velocity in such a material is slower than in air. 

The accuracy of the ToA estimation is also reduced due to finite bandwidth of the system. Firstly, the bandwidth reduces the achievable accuracy in pure LOS scenarios due to the ''smearing out" of the received location signal, which makes finding the peak more difficult. Secondly, in a multipath environment, several MPCs may fall into a resolvable bandwidth and thus superpose constructively or destructively. Thus, the location of the maximum in the impulse response may change when the UE or scatterers move - even when the movement is so small that the delays of the MPCs themselves do not change. 

Several ML based solutions try to identify the LOS scenarios, e.g., \cite{NLOS_UWB, tdoa, musa2019decision}, other have ML solutions that estimate the offset and provide mitigation \cite{wu2019neural_116,IndustrialWSN,van2015machine_1}. To relax the synchronisation constraint, TDoA metrics have been used in the literature \cite{tdoa,mmwave,xue2019deeptal_36,li2017robust_60,de2019enhanced_129}.

\subsubsection{Angle of Arrival}
 Another classical way of determining the location of an object is triangulation, based on the AoA at (at least) two anchors. Thus, the AoA can be an important feature for localization algorithms. Similar to ToA, also AoA suffers from practical difficulties.
 
 First and foremost, determination of the AoA requires real or virtual antenna arrays. The former case implies a physical array with multiple antenna elements that receive the signal (quasi-)simultaneously. These arrays need to fulfill certain conditions in terms of antenna spacing (has to be smaller than half a wavelength to provide unambiguous angles) as well as calibration (the relationship of the {\em complex} antenna patterns at the different antenna elements needs to be accurately known). In the virtual array case, a single antenna is sufficient, which is either moved to different locations, or pointed into different directions, at different times - synthetic aperture radar is a famous example of such an approach. However, in this case, the movement of the device creating the virtual array needs to be precisely known. 
 
 The antenna pattern can usually be determined through calibration of the device. However, the presence of conducting or dielectric objects near the array leads to pattern distortion. Thus, a device being held by a user can have a very different pattern than the calibration result obtained without that user. As a consequence AoA tends to be used mainly at the BSs.
 
 A further challenge is the limited resolution. If an analysis is done by Bartlett beamforming (Fourier techniques), then the resolution is limited by the aperture of the antenna array (similar to the bandwidth limiting the ToA resolution). If High Resolution Parameter Estimation (HRPE) techniques such as MUSIC, ESPRIT, or SAGE, are applied, higher resolution can be achieved, though these algorithms are sensitive to errors in the calibration, and also require high computational complexity \cite{ProfMolischText}. 
 
 Just like for ToA, the biggest challenges arise when the LOS is blocked. In this case, the system might determine the direction of the strongest AoA, but this might not be identical to the LOS direction. And in contrast to the ToA case, where the ''earliest significant" MPC is often associated with an attenuated LOS, no such identification is possible for the AoA case. For these reasons, it is rare to use {\em only} AoA for determination of location; rather, it is used to augment estimations based on other features. This is particularly true for classical localization approaches. 
 
 To overcome the high demands on signal processing and calibration, many ML papers use coarse estimates of AoA. For example, \cite{wang2018deep_23} noticed that the {\em phase} difference between antenna pairs is correlated with the true AoA, Ref. \cite{zhang2019efficient_122} use the recorded {\em directional} RSSI value, with simplified angle calculation method in \cite{passafiume2015music}, to estimate the AoA. Note that with a large number of antennas, finer estimates of AoA become easier to obtain.

\subsubsection{Channel State Information}

Channel State Information (CSI) is often defined as the complex value of the received signal, i.e., amplitude and phase values, for all subcarriers, \footnote{We assume here and henceforth an OFDM system, since it is by far the most popular implementation of wideband transmission. Equivalently, the complex samples of the impulse response, spaced at most at the Nyquist sampling rate, can be used.} at all antenna elements. This CSI is clearly the most comprehensive "raw" information obtained by the system; the other features mentioned above, such as RSSI, ToA, and AoA are compressed, derived versions of it. CSI can thus serve as the basis of compressed information, possibly expanding the number of derived features. Alternatively, CSI can be used directly in fingerprinting. The operating principle of CSI-based fingerprinting is the same as for RSSI-based fingerprinting; yet much more accurate fingerprinting can be established because of the richer information contained in the CSI. For example, it is possible to obtain localization from the CSI of the link between target and a single anchor, as was done in \cite{wu2018accurate,schmidt2019sdr,widmaier2019towards,8673800,HumanDetection,wang2019fast_83,sun2018single_90}. Note that such an approach is not practical for RSSI-based fingerprinting because the same RSSI can occur at multiple locations around an anchor; after all RSSI is only a single scalar. Conversely, it is very unlikely that two locations have the same CFR, since this would require that the complex channel response is the same at all the subcarriers, i.e., agreement of all entries of a complex {\em vector}. The main challenges of using CSI are (i) complexity of using such rich information, and (ii) availability of the information at the APIs used by the localization system.

\subsection{Wireless Technology}\label{sec:Technology}

Advanced wireless technologies, in particular multi-antenna systems and ultra-wideband systems, are especially helpful in localization. While they can be interpreted as simply providing particular features (direction, ToA) with increased accuracy, we briefly survey their background and their application in ML-based localization. 

\subsubsection{Multi-Antenna Systems}
When the TX and/or RX are equipped with multiple antennas, the system can collect several copies of the wireless signal. For localization, it can be used to: (i) Improve the estimate of the features by enhancing the signal to interference and noise ratio (SINR) through beamforming and interference nulling \cite{ProfMolischText}. (ii) Improve the resolution of AoA estimation. (iii) Enhance the separability of the fingerprints, by providing a larger dimensionality of the fingerprint. Multi-antenna systems are often called multi-input multi-output (MIMO) systems - in an abuse of notation, this term is even used when multiple antenna elements are only at one link end.
 
In the ML-based localization literature, different methods were proposed to utilize multiple antennas. Several papers use the difference of values between antennas pairs to create stable features, e.g., phase difference \cite{janssen2019comparing_21,ghourchian2017real_45}, as it eliminates the constants that are common to the antennas, making the features less sensitive to specific device models. Furthermore, large arrays may allow for sparse representation of the channel in the angle domain.
The advent of 5G has triggered interest in large antenna systems. Thus, many recent works proposed localization for massive MIMO \cite{wu2019learning_78,prasad2018machine_62,sun2019fingerprint_79,vieira2017deep_80,decurninge2018csi_81,arnold2018deep_82,wang2019fast_83,prasad2017numerical_84,pirzadeh2019machine_87,sun2018single_90,flordelis2019massive}.  

\subsubsection{Ultra Wide Band (UWB) Systems}
Time resolution increases with the bandwidth of the wireless system, which can allow for finer ToA estimates. It has further the advantage of combating the small scale fading, as fewer MPCs might fall into the same delay bin. UWB systems are systems with at least 500 MHz bandwidth; they are permitted (in the US) to operate in the frequency range $3.1-10.6$ GHz subject to constraints of the power spectral density \cite{di2006uwb}. Systems with similarly large, or even larger, bandwidths can also exist at higher carrier frequencies, such as the 60 GHz unlicensed band. The large bandwidth leads to resolvable delay bins that are fractions of nanoseconds, thus providing a localization accuracy in the order of centimeters \cite{sahinoglu2008ultra,gezici2005localization}. 

These advantages have also motivated the use of UWB in ML-based localization, e.g., \cite{kram2019uwb_68,xue2019deeptal_36,krishnan2018improving_71,wymeersch2012machine_72}, It is possible to acquire and use directly the CIR as a feature, e.g., see \cite{bregar2018improving_77}. It is also possible to estimate with improved accuracy a number of compressed channel parameters, such as Power Delay Profile (PDP), delay spread, and other quantities that could capture the structure of the environment. For instance, \cite{wymeersch2012machine_72} uses delay spread, rise time, mean excess delay (among others) for ranging error mitigation in UWB systems. Since mmWave systems are expected to use multi-antennas (to compensate for the increased path loss at higher frequencies), recent works explore suitable features, e.g., \cite{patel2020millimeter} uses the set of beamformed signals and PDP as input to a DL algorithm. In \cite{mmwave}, the authors use hybrid delay and angle measurements in a cooperative mmWave system.

\subsubsection{Multi-Carrier and Multi-Band Systems}
As we have seen in Sec. II.B, impulse response and CFR of a channel are equivalent, and related through a simple FT. Yet, which of those to use can still make a difference in implementations of ML algorithms. Most practical cellular and WiFi systems are multi-carrier signals, so that they naturally measure the CFR, or more precisely, the samples of the CFR at discrete frequencies, the subcarrier frequencies used in the multicarrier signaling.
When CSI at each sub-carrier is available, CFR is usually used as feature \cite{wu2019learning_78,berruet2018delfin_35,wang2019fast_83}. Note that in classical localization algorithms, FT and use of the impulse response for localization is more common, even when the system is based on multicarrier communications.

On a larger scale, wireless systems might use different frequency bands to utilize the inherent propagation advantages over different bands or because they might simply coexist, e.g., 2.4GHz and 5GHz in WiFi or centimeter-wave (cmWave) and mmWave bands in 5G networks. Features over these bands can be independent or complement each other, e.g., wide coverage in the cmWave band, and better delay resolution in the mmWave band. Some recent works started to investigate such multi-band systems, such as $2.4$GHz and $5$GHz in \cite{akram2018censloc_18,own2019signal_125}. Ref \cite{own2019signal_125} derived features from RSSI at $5$GHz for LOS identification, then uses RSSI in $5$GHz and $2.4$GHz for LOS and NLOS localization, respectively.

\begin{table}  
\begin{scriptsize}
\begin{tabularx}{0.48\textwidth} { 
   >{\centering\arraybackslash\hsize=.75\hsize}X 
   >{\centering\arraybackslash\hsize=1.25\hsize}X }
 \hline

  & Papers \\
 \hline
   \vspace{2mm} OFDM & ~\cite{uav,wang2016csi_5,wu2019csi_7,li2018channel_8,wang2017cifi_9,li2019convolutional_12,hsieh2019deep_28,wang2015deepfi_33,berruet2018delfin_35, wu2017passive_44,ghourchian2017real_45,wang2017resloc_56,butt2020rf_57,fukushima2019evaluating_73,wu2019learning_78,sun2019fingerprint_79,vieira2017deep_80,decurninge2018csi_81,arnold2018deep_82,sun2018single_90,sobehy2019csi_117,lei2019siamese_130,sanam2018improved}\\
 \hline
  \vspace{2mm}  MIMO & ~\cite{niitsoo2019deep_2} \cite{wang2016csi_5,li2018channel_8,wang2017cifi_9,li2019convolutional_12,wang2018deep_23,hsieh2019deep_28,wang2015deepfi_33,berruet2018delfin_35, wu2017passive_44,ghourchian2017real_45,wang2017resloc_56,butt2020rf_57,fukushima2019evaluating_73,arnold2018deep_82,chen2018multipath_111,sobehy2019csi_117,zhang2019efficient_122,lei2019siamese_130,jing2019learning_131,sanam2018improved}\\
 \hline
  \vspace{1mm}  Massive MIMO & ~\cite{MassiveMIMO,widmaier2019towards,prasad2018machine_62,wu2019learning_78,sun2019fingerprint_79,vieira2017deep_80,decurninge2018csi_81,wang2019fast_83,prasad2017numerical_84,sun2018single_90,de2019csi_121}\\
 \hline
 Distributed MIMO & ~\cite{MassiveMIMO,prasad2018machine_62,prasad2017numerical_84,pirzadeh2019machine_87}\\
 \hline
 UWB & ~\cite{NLOS_UWB,IndustrialWSN,8712551,7313038,musa2019decision,van2015machine_1,xue2019deeptal_36,kram2019uwb_68,krishnan2018improving_71}\\
 \hline
\end{tabularx}
\caption{\small Technologies used in a sample of papers.}

\label{tab:Technology}
\end{scriptsize}
\end{table}

\subsection{Standards Type} 

Localization in wireless systems can be either based on signals and protocols that are explicitly designed for the purpose of localization, or they can make use of "incidental" signals of systems designed for communications. 
When the localization system is built based on a wireless communication standard, the available features are restricted by the specifications of the underlying standard, e.g., the bandwidth, the central carrier frequency and the maximum supported number of antennas. Furthermore, the communication protocols could also restrict the access to certain features, e.g.,, does the target have to be in "connected mode" to observe the feature? This is important as it limits the number of usable anchors. In this subsection we provide a brief review of some of the standards commonly used in ML-based solution; details can be found in number of other dedicated paper \cite{zafari2019survey,del2017survey,lin2017positioning,faragher2015location}. 

The most famous example of an explicit localization protocol is GPS - actually GNSS systems have as their only purpose localization, so that clearly their protocols and signals are designed for localization. The basic principle of GNSS is TDOA; the particulars of the localization signals and synchronization procedures can be found, e.g., in \cite{zekavat2013handbook}. 

Another important explicit ranging protocol is used in LTE. There, different BSs send out ranging signals (either on demand, or at regular intervals) that can be used by the UE to determine its location by trilateration. Those ranging signals are similar to the ''reference signals" (pilot signals) used for channel estimation. The difference is that during the time that one BS sends out a localization signal, the surrounding BSs are silent, thus drastically improving the SINR and consequently the precision of the ranging. However, ranging in LTE can also use other, non-standardized, procedures, such as CSI in either uplink or downlink (which is acquired as a matter of course in the normal operation of the system). Localization can be based either on the directly measured CSI, or (When that is not available at the APIs), by using compressed CSI such as RSRP and RSRQ of the serving BS and possibly one or more of the neighbor BSs, as conveyed in the measurements reports (MRs) that are sent by the UEs either periodically or triggered by certain events or procedures. A number of proposed solutions use MRs for ML-based localization, such as radio map construction or channel model based localization. LTE cellular networks have been used in number of ML solutions, such as \cite{margolies2017can,zhang2019deeploc_108,zhu2016city_50r,ray2016localization_94,lee2018dnn_95}. 

Another standard that has an explicit protocol for ranging is IEEE 802.15.4a, which is based on UWB signalling. The ranging signal there is a pseudorandom sequence with special correlation properties \cite{zhang2009uwb}. It has been widely used for deterministic localization experiments, but is not widely used in ML solutions, possibly due to the lack of extensive deployments. 

Besides cellular systems, WiFi (802.11) is the most widely used standard for data transmission. However, until recently there had not been an explicit localization protocol. A key challenge is that neither the location of the APs is precisely known, nor are the APs generally synchronized to each other. Furthermore, in home networks, only a single AP is under the control of a particular network, so that data packets to/from other APs, which are encrypted, cannot be used to help with the localization. The only contribution that APs of other users can provide are the beacon signals, which can be detected by a UE without being authorized to log in and decrypt. Note that since the deployment is usually indoor, and the LOS is blocked, determination of the distance based on the received signal power is not easily possible. Yet the beacon signals provide RSSI that can be used for fingerprinting and/or ML solutions. 

WiFi based ML localization tends to use RSSI measurements to the serving AP, as well as RSSI to other APs as input features due to the easy availability. However, since many WiFi releases, such as IEEE802.11n, use OFDM and MIMO technologies, a number of recent work started to use CSI (with the serving AP) over different sub-carriers and antennas. Enhancements to timing measurements were introduced in 802.11-2016 (IEEE 802.11 REVmc) through Fine Time Measurements (FTM) protocol, this allow to use better ToA estimates as features for localization which is used, e.g., in \cite{choi2019unsupervised_123}.

With the growing commercial interest in proximity based Services, localization using Bluetooth Low Energy (BLE) has attracted considerable attention \cite{faragher2015location}, especially after major companies such as Apple and Google released their BLE protocols, iBeacon and Eddystone, respectively. Although it is possible to locate the target via BLE RSSI readings (as in WiFi), due to practical reasons, the focus was on ranging, as it was also encouraged by iBeacon protocol, where the goal is to identify the proximity of the target. 

In the literature, other standards have been used as well for localization, such as Zigbee \cite{rfid,hiwl,LiUnsupervised,RamaLocalization} and FM signals \cite{8108573}. However, these systems are not as popular as the ones above, possibly due to their smaller installed userbase. Note that there are many other emerging standards and protocol that could be potentially used for localization, such as LoRaWAN and SigFox for IoT (see \cite{lin2017positioning, zafari2019survey} for good surveys), and next generation wireless standards, $5$G and beyond.

\begin{table*} 
\centering
\begin{scriptsize}
\begin{tabularx}{1\linewidth} { 
   >{\raggedright\arraybackslash\hsize=.75\hsize}X 
   >{\centering\arraybackslash\hsize=1.65\hsize}X 
   >{\raggedleft\arraybackslash\hsize=0.95\hsize}X 
   >{\raggedleft\arraybackslash\hsize=0.75\hsize}X 
   >{\centering\arraybackslash\hsize=0.95\hsize}X 
   >{\centering\arraybackslash\hsize=0.95\hsize}X }
 \hline
  & \small WiFi &  \small Cellular & \small BLE ~~ &  \small WSN$^{\mathsection}$ & \small Other \\
 \hline
 \small \vspace{9mm} RSSI & ~\cite{alteredAP , FPfusing , laperls , 7743685 , OnDevice , 6805641 , LowOverhead , magnetic_inertial , patent , OpticalCamera , probabilistic_localization , kmgpr , DNN/KNN , 8765368 ,Discriminative , 8362712 , FeatureSelection , hapi , hybloc , 7064997 ,WeightedAmbientWiFi , ZhangIndoor , 8008794 , 8584443 , JUIndoorLoc ,8931646 , yang2012locating , 8115921 , Qian2019Supervised , 8661625 ,Zhang2019Wireless , Chunjing2017WLAN , 8422182 , wu2019mobile ,feng2019received , zhang2017efficient , ge2019hybrid ,felix2016fingerprinting , sun2008adaptive  , khatab2017fingerprint ,kim2018scalable , yan2018noise , wang2019robust , mei2018novel ,nguyen2017performance , zou2015fast , rezgui2017efficient ,figuera2012advanced , guo2018accurate , bi2018adaptive , elbes2019indoor ,liu2019hybrid , adege2018indoor , adege2019mobility , gu2015online ,yan2017hybrid , malik2019indoor , zhang2018enhancement , tewes2019ensemble,sung2019neural , kim2018hybrid , xiao2012large , salamah2016enhanced ,homayounvala2019novel , zhang2019improving , secckin2019hierarchical ,belmonte2019swiblux , li2019smartloc , dashti2016extracting ,yang2012locating , xiao2018learning , qian2019convolutional_10 ,ibrahim2018cnn_13  , akram2018censloc_18 , turgut2019deep_27 ,hsieh2019deep_28 , gan2017deep_32 , zou2018deep_37 , zhang2016deep_38 ,zhou2017robust_46 , sorour2014joint_47 , ouyang2011indoor_51 ,liu2009low_52, gu2015semi_54 , pulkkinen2011semi , zhou2017semi_55 ,wang2020robust_59, aikawa2018wlan_61 , abbas2019wideep_63 ,huang2020widet_64, 8733822 } ~\cite{ ciftler2020federated_70 , zhou2017grassma_76 , zhou2017indoor_86 , chen2018deep_88 , jung2015unsupervised_89 , wang2012no_92 , wu2012will_93 , rizk2019device_96 , jang2018indoor_99}  
 
 &   ~\cite{MassiveMIMO , MonoDCell , 8765368 , DataAugmentation ,  FullBandGSM , wu2019mobile ,  oussar2011indoor , chen2019dpr ,  qiu2018walk ,  rizk2018cellindeep_15 ,  zhang2020transfer_49 , shokry2018deeploc_50 ,  ray2016localization_94 ,  lee2018dnn_95 ,  rizk2019solocell_98 , zhang2019deeploc_108} 
 
 & ~\cite{location_aware ,  dabil ,  8240410 ,  xu2019efficient ,  ahmadi2017exploiting ,  belmonte2019swiblux ,  li2019deep_31 ,  gan2017deep_32 ,  ebuchi2019vehicle_126, peng2016iterative}
 
 & ~\cite{wsn_svr ,  TargetTracking ,  LiUnsupervised ,  RamaLocalization ,  cottone2016machine ,  yang2018clustering ,  li2017measurement ,  bhatti2018machine ,  mahfouz2015kernel ,  chen2011semi_53 ,  ahmadi2015rssi_58 ,  belmannoubi2019stacked_110 , nguyen2005kernel} 
 
 & ~\cite{7504333 ,  lorawan ,  uav ,  8108573 ,  gbrbm ,  8780770 ,  rfid ,  Timotheatos2017FeatureEA ,  grof  ,  hiwl ,  eee112 ,  DCNN ,  7313038 ,  xu2019efficient ,  zhang2018heterogeneous ,  zhao2019accurate ,  aikawa2019cnn_16 ,  zheng2016cold_19 ,    njima2019deep_22 ,  wang2016device_39 ,  alhajri2019indoor_74}\\
 \hline
 \small \vspace{3mm} CSI & ~\cite{InterpretingCNN,devicefree ,  Eloc ,  8673800 ,  HumanDetection ,  RelationLearning ,  Wang2017WiFiFB ,  wu2018accurate ,  schmidt2019sdr ,  tewes2019ensemble ,  sanam2018improved ,  wang2016csi_4 ,  wang2016csi_5 ,  li2018channel_8 ,  wang2017cifi_9 ,  li2019convolutional_12 ,  wang2018deep_23 ,  hsieh2019deep_28 ,  wang2015deepfi_33 ,  yazdanian2018deeppos_34 ,  berruet2018delfin_35 ,  zhou2018device_40, wu2017passive_44 ,  ghourchian2017real_45 ,  wang2017resloc_56 ,  fukushima2019evaluating_73 ,  sen2012you_91 ,  zhang2019efficient_122} &  \cite{sdr ,  liu2017wicount ,  decurninge2018csi_81} &  & ~\cite{FineGrainedSubcarrier} & ~\cite{csiRNN ,  widmaier2019towards ,  van2015machine_1 ,  niitsoo2019deep_2 ,  niitsoo2018convolutional_11 ,  sobehy2019csi_117}\\

 \hline
 \small  \vspace{1mm} ToA &  ~\cite{ge2019hybrid ,  choi2019unsupervised_69} &  &  &  & ~\cite{NLOS_UWB ,  8712551 ,  7313038 ,  zhang2018heterogeneous ,  van2015machine_1 ,  li2017robust_60}\\
 
 \hline
 \small  \vspace{2mm} Other & ~\cite{roshanaei2009dynamic,FPfusing ,  tdoa ,  tlfcma ,  AntonioLocalization ,  comiter2017data ,  comiter2017structured ,  wu2019csi_7}
 & ~\cite{8651736 ,  vlfANN ,  7938617 ,  telco ,  tdoa ,  FullBandGSM ,  butt2020rf_57} &  & ~\cite{IndustrialWSN} 
 & ~\cite{    RadioMaps ,  mmwave ,  8780770 ,  musa2019decision ,  comiter2018localization ,  yi2019neural ,  patel2020millimeter ,  xue2019deeptal_36 ,  bregar2018improving_77}\\
 \hline
\end{tabularx}
\end{scriptsize}
\caption{\small Technology and basic features. $^{\mathsection}$Note that for WSN a number of works use unsupervised learning methods based on a presumed knowledge of the distance between the nodes, which can typically be estimated using ToA or RSSI, we here limit the discussion of these solution due to these implicit assumptions about the RF signal and presence of dedicated survey paper \cite{saeed2019state} (see Sec. \ref{sub:unsuper} for more details).}
\label{tab:Standard}
\end{table*}

\subsection{Side Information}\label{sec:SideInfo}
One advantage of using ML is its ability to integrate efficiently different, and seemingly non-homogeneous, sets of features in the solutions. For many localization problems, it is possible to acquire additional information besides the RF signal. For instance, modern cellphones are equipped with light sensor, acoustic sensors, accelerometers, gyroscopes, and magnetometers. Some of these sensors can be integrated into what is referred to as the Inertial measurement unit (IMU). Using these measurements, it is possible to detect the motion, identify the orientation, or environment change, which can help in the localization and tracking process.

Maps and environment structures, such as floor plans, have been used as side information. It can help imposing physical constraints on the solution, such as walls and entrances. This results in realistic and smooth trajectories for tracking applications. Maps are usually integrated using map matching algorithms, e.g., using HMMs, where the transition between the hidden states are governed by the map constraints, see for instance \cite{newson2009hidden}.
Images can be used for localization, e.g., using photos of local "landmarks" as RPs. Note that each of the above can be used as the only feature for localization. However, in this survey, we highlight solutions that use them along with RF signals; a list of such ML-localization papers is presented in table \ref{tab:SideInfo}.

\begin{table}
\begin{scriptsize}
\begin{tabularx}{.45\textwidth} { 
   >{\centering\arraybackslash\hsize=0.70\hsize}X 
   >{\centering\arraybackslash\hsize=1.1\hsize}X 
   >{\centering\arraybackslash\hsize=1.1\hsize}X 
   >{\centering\arraybackslash\hsize=1.1\hsize}X 
   >{\centering\arraybackslash\hsize=1.1\hsize}X }
 \hline
\small  Maps & \small  Magnetic field & \small  Accelerometer \\
 \hline
  ~\cite{RadioMaps,sorour2014joint_47,zhou2017indoor_86,jung2015unsupervised_89,wu2012will_93,yang2012locating}  & ~\cite{magnetic_inertial,7064997,OpticalCamera,Discriminative,bae2019large,belmonte2019swiblux,qiu2018walk,gan2017deep_32,wang2012no_92,zhang2017deeppositioning_106,shao2018indoor_120} & ~\cite{6805641,TargetTracking,magnetic_inertial,OpticalCamera,Discriminative,belmonte2019swiblux,yang2012locating,qiu2018walk,wang2012no_92,wu2012will_93}\\
\hline
\end{tabularx}   

\end{scriptsize}
\caption{\small Side information}
\label{tab:SideInfo}
\end{table}

\subsection{Feature Representation}
As we will discuss below, in some systems, it is possible to collect several values of the basic features, e.g., RSSI values from nearby BSs/APs. Several ML-based localization solutions use statistical quantities, such as the mean, variance, kurtosis, and skewness as compact feature representation. These quantities are usually more stable, but they reduce the granularity of the observations. Thus they can be used for coarse localization or used jointly with other features.

Dimensionality reduction techniques have been widely used, including linear methods such as PCA and Linear Discriminant Analysis (LDA) (a supervised dimensionality reduction technique) and several non-linear techniques that we discuss as part of the later sections. One goal of these techniques is to provide compact and robust representations of the features without the loss of useful data, thus making learning easier and more stable.

A number of solutions use signal processing techniques to transform the features to other domains. The goal is to utilize some advantageous properties in the new domain. For instance, the FT is used to transfer the directional CFR acquired in an OFDM-MIMO system from the frequency-antenna domain to the delay-angle domain in the form of CIR, which has been observed to improve localization performance. One easy explanation is that it is straightforward to locate the target when the delay and the AoA of the LOS component are identified. These transformations have been used in several works, e.g., \cite{wu2019learning_78,sun2019fingerprint_79,wang2019fast_83}. Wavelet transform is another method that has been used to denoise the received signal and to provide higher dimensional representation, e.g., from 1D time series vector to 2D time-frequency "images", as in \cite{soro2019joint_118, huang2020widet_64} where they use the wavelet to transform the RSSI readings in WiFi system.

Another group of papers aims to reduce the correlation between the acquired features, mainly RSSI, through AP selection, e.g., \cite{kushki2007kernel,7504333,probabilistic_localization,chen2006power}. Examples of other transformations include using visibility graphs, \cite{lacasa2008time}, to transform a series of CSI values on different sub-carriers into a network representation that could
reveal the internal relationship of data and improve classification results as used in \cite{wu2018accurate}.

\subsection{Features Utilization in Conventional and ML-based Localization Systems}\label{sec:ConvVsMLSystems}

Before going into the details of ML-based solutions, it is useful to review how conventional localization algorithms use the above features, and where the ML methods were introduced in a number of ML-based localization. 
The most common principles are trilateration/triangulation and fingerprinting (proximity and direct localization methods can be viewed as special cases of the two). Trilateration uses estimates of the distance between anchors and target to determine the location of the target. In the case of active localization with ToA estimation, each range estimate corresponds to a circle; for TDoA to a hyperbola, and for passive estimation with ToA to an ellipse. The intersection of those curves then provides the location estimate. The accuracy of the location is determined by the accuracy of the range estimate as well as the Geometric Dilution of Precision (GDOP) due to ''grazing" intersection of the curves corresponding to the different anchors. When AoA is available at a particular anchor, it provides additional information, namely that the target must lie on the line corresponding to this angle. Taking into account the uncertainty of the range estimate due to noise, various linear and nonlinear techniques have been developed for the computation of the most likely intersection point of the circles \cite{zekavat2013handbook}. 

Naive application of this approach can lead to significant errors, in particular in the presence of LOS blockage and/or multi-path. The (positive) bias of the range estimates from NLOS links leads to circles that are larger than the true distance between anchor and agents. Consequently, the circles corresponding to the different anchors might not intersect in a single point at all. A variety of methods have been developed to perform localization under such circumstances; for example POCS \cite{gholami2011robust}, and approaches based on consistency of the estimated solutions \cite{aditya2017localization}. A further improvement can be achieved by employing ''soft information" \cite{conti2019soft}. While most trilateration approaches find the mean and variance of the range estimate, and combine those, soft information uses the probability density function or log-likelihood of the range estimate as the input, and from that determines the most likely location. In the presence of non-Gaussian, asymmetrical ranging errors, such as occur in the presence of blockage and multi-path, this can lead to a considerable improvement of the accuracy.

Many recent localization systems utilize ML to improve trilateration/triangulation systems, e.g., to predict the type of environment (e.g., indoor vs outdoor \cite{ray2016localization_94,qiu2018walk}, LOS vs NLOS \cite{musa2019decision,tdoa,NLOS_UWB,van2015machine_1,krishnan2018improving_71,huang2020machine}), or to correct the bias in the range estimates \cite{wymeersch2012machine_72,bregar2018improving_77,wu2019neural_116}, or for a direct range estimation \cite{7313038,savic2014kernel}. 

The fingerprinting can be summarized as follows. The system operates in two phases: an offline and an online phases. During the offline phase, the environment is surveyed at predefined locations (RPs) and certain features of the RF signal are recorded; location of the RPs and the recorded features at those points are then stored in a database (radio-map). The majority of papers are based on RSSI signals from multiple APs, however, as discussed above, recent methods start to employ CSI as well. During the online phase, when the location of a target is requested, the system matches the observed features to the ones in the database. Matching is based on deterministic or probabilistic methods. In the latter the system finds the most probable RPs for the given observations. The output of the fingerprinting system (i.e., target's location), is usually a function of the location of one or more of the matched RPs. The research related to fingerprinting systems revolves around the best methods to construct the radio map, the matching techniques, and final location estimate (i.e., observed feature to location mapping). For instance, one of the main challenges in the practical implementation is the requirement for a fast search for the fingerprint in the database that best matches the current observation. Straightforward linear searches are too computationally intensive in particular when large databases are available; rather a hierarchical search is preferable. 

Localization usually refers to the determination of the location from observations (features) at a single time instant. However, localization systems often have measurements at multiple time instances available which - when associated with a moving target - allows to {\em track} the target trajectory. Such tracking, or in general utilization of previous observations (historical data),  is usually done by means of {\em Kalman filters} or particle filters \cite{zekavat2013handbook}. They predict the change of location in a timestep based on the previous movement, and subsequently correct the estimate from measurements of the new location. Both the prediction and the correction are probabilitistic, and their combination provides an estimate that filters out some of the noise in the measurements.   

ML has been used in fingerprinting solutions since its infancy, where standard ML methods have been utilized as matching mechanism, e.g., \cite{bahl2000radar,liu2007survey}. Since then it has been utilized in other aspects as well, for instance for feature extraction and radio-map construction \cite{wang2016csi_4,wang2017cifi_9,khatab2017fingerprint,wang2019robust,yang2012locating,Timotheatos2017FeatureEA,FullBandGSM,eee112,8931646,yang2012locating}, radio-map updating \cite{alteredAP,kmgpr,RadioMaps,location_aware}, hierarchical solutions \cite{mittal2018adapting_109,kram2019uwb_68,jaafar2018neural,ibrahim2018cnn_13,secckin2019hierarchical}, and robust matching \cite{FPfusing,DNN/KNN,probabilistic_localization}. This is expected because fingerprinting systems, similar to ML, are data-driven and both {\em  may} \footnote{Note that some methods have been proposed to construct the radio-map online, and some ML solution may operate fully online ("online learning").} operate in a training (offline) phase, and an online phase. Nevertheless, ML-based solutions span a wider scope as, for instance, ML could be used as an end-to-end (i.e., direct) localization solution, without the need for an explicit radio-map construction (compare the "instance learning" to other methods in Sec. \ref{sec:MLPrelm}). Tracking problem may utilize ML solutions to assist the traditional tracking methods (e.g., improving Kalman filter based tracking as in \cite{TargetTracking,7743685,magnetic_inertial,hapi,8008794,wang2015floor}). Furthermore, many methods to track the targets make assumptions about the location evolution process; the advances of recurrent DL methods can potentially eliminate the need for such restrictive assumptions.

\section{Learning Solutions}\label{sec:LearnStrcut}
As discussed in Sec. \ref{sec:MLPrelm}, we can distinguish a number of ML classes:
\begin{itemize}
    \item Supervised Learning: features and labels are present during the training phase.
    \item Semi-Supervised Learning: labels are available for a subset of the features.
    \item Unsupervised Learning: no labels are available.
    \item Other Learning Approaches: labels may be available but training is done differently, for example, in sequential or distributed fashions.
    
\end{itemize}
We can also distinguish between the standard and the DL approaches. We here refer to approaches that use hierarchical learning structures as DL; with this definition, NNs with a number of hidden layers can be classified as DL. Note that it is not easy to draw a line between the two, thus it is possible to classify some of the reported papers below differently. Furthermore, as discussed in the previous section, probabilistic methods are used for fingerprint matching. Based on the discussion in Sec. \ref{sec:MLPrelm}, one might view, for instance, all Bayesian methods, mixture models, and HMM as part of the standard ML methods. However, in this survey, we do not focus on those methods as they lie in the gray area between model-based analytical solutions and ML (though we do cover a number of DL probabilistic methods).

In this section, we present a representative set of different structures for each of the aforementioned ML approaches, and summarize a number of works that adopted them for localization. In each subsection we order the reviewed papers based on the feature complexity, which is usually related to the technology and the standards.  

\subsection{Supervised learning}\label{subsec:supr}
In this section, we start by reviewing works that use standard ML algorithms, followed by those using DL algorithms. As highlighted above, the organization is done based on the perspective of the used algorithm; for DL this could provide a glimpse at the considered problem as it influences the DL architecture. For standard ML this is generally not the case; however, we follow the same structure, as it is motivated by the goal of this survey (emphasis on recent ML solutions which tend to be DL based), and to maintain the organizational consistency. 
\begin{table*} 
\centering
\begin{scriptsize}
\begin{tabularx}{1\textwidth} { 
   >{\centering\arraybackslash\hsize=.85\hsize}X 
   >{\centering\arraybackslash\hsize=1.6\hsize}X 
   >{\centering\arraybackslash\hsize=.85\hsize}X 
   >{\centering\arraybackslash\hsize=.85\hsize}X 
   >{\centering\arraybackslash\hsize=.85\hsize}X }
 \hline
  &   Supervised &   Semi-Supervised &   Unsupervised  &  Other \\
 \hline
  \vspace{10mm}
 Standard ML & ~\cite{alteredAP,FPfusing,InterpretingCNN,7504333,7938617,NLOS_UWB,OnDevice,telco,sdr,6805641,TargetTracking,tdoa,RadioMaps,magnetic_inertial,IndustrialWSN,uav,patent,mmwave,8240410,8108573,OpticalCamera,probabilistic_localization,gbrbm,8780770,8673800,rfid,kmgpr,DNN/KNN,Discriminative,8362712,FineGrainedSubcarrier,Timotheatos2017FeatureEA,FeatureSelection,FullBandGSM,grof,hapi,hybloc,7064997,WeightedAmbientWiFi,eee112,8008794,8584443,JUIndoorLoc,7313038,8931646,yang2012locating,zhang2017efficient,cottone2016machine} \cite{mei2018novel,wu2018accurate,nguyen2017performance,musa2019decision,rezgui2017efficient,figuera2012advanced,guo2018accurate,bi2018adaptive,zhang2018heterogeneous,ahmadi2017exploiting,adege2018indoor,yan2017hybrid,li2017measurement,sanam2018improved,salamah2016enhanced,homayounvala2019novel,bhatti2018machine,zhang2019improving,secckin2019hierarchical,li2019smartloc,dashti2016extracting,yang2012locating,oussar2011indoor,mahfouz2015kernel,roshanaei2009dynamic,ma2008cluster,roshanaei2009dynamic,torteeka2014indoor,peng2016iterative} & ~\cite{tsui2009unsupervised,laperls,8651736,MassiveMIMO,7743685,wsn_svr,ZhangIndoor,8115921,LiUnsupervised}  ~\cite{Chunjing2017WLAN,chen2019dpr} & ~\cite{location_aware,dabil,gbrbm,hiwl,li2019smartloc} & ~\cite{tlfcma,LowOverhead,8661625,feng2019received} \cite{sun2008adaptive,guo2018accurate,yang2018clustering,xiao2012large,qiu2018walk}\\
 \hline
 \vspace{1mm}
Convolutional based & ~\cite{Eloc,DCNN,Qian2019Supervised,Zhang2019Wireless,comiter2018localization,liu2019hybrid,schmidt2019sdr,zhang2018enhancement,tewes2019ensemble,zhao2019accurate,patel2020millimeter} &  &  & \\
 \hline

Recurrent & ~\cite{ML_thesis, 8733822, MonoDCell, vlfANN, telco, Qian2019Supervised, uav, csiRNN, wu2019mobile, xu2019efficient, yan2018noise, elbes2019indoor, adege2019mobility, bae2019large, patel2020millimeter} &  &  & \\
\hline
 \vspace{3mm}
Other & ~\cite{sdr,DNN/KNN,HumanDetection,Timotheatos2017FeatureEA,RelationLearning,hapi,8584443,RamaLocalization,Wang2017WiFiFB,8422182,jaafar2018neural,ge2019hybrid} \cite{felix2016fingerprinting,khatab2017fingerprint,kim2018scalable,wang2019robust,mei2018novel,comiter2017data,comiter2017structured,guo2018accurate,liu2017wicount,adege2018indoor,schmidt2019sdr,malik2019indoor,yi2019neural,sung2019neural,kim2018hybrid,widmaier2019towards,belmonte2019swiblux,xiao2018learning} & ~\cite{8651736,lorawan,10.1145/3321408.3321584,DataAugmentation,Qian2019Supervised} & & ~\cite{lorawan,khatab2017fingerprint,zou2015fast,gu2015online}\\
 \hline
\end{tabularx}
\end{scriptsize}
\caption{\small Type of ML model used}
\label{tab:MLType}
\end{table*}
\subsubsection{Standard Machine Learning Solutions}
In this section we review different localization solutions based on standard ML techniques. For brevity we review representative works that capture different aspects of the research in this area. Note that many research works present several ML algorithms when evaluating their solution's performance (i.e., use some of them as benchmarks). We here report the main solutions of those papers.

{\bf K-Nearest Neighbors (KNN):}
In KNN, the coordinate of the target is approximated by the average location of the $K$ closest RPs in the fingerprint database. When the goal is to identify the region, e.g., the room, the predicted region is the one where the majority of the $K$ neighbors are located, see Sec. III for more details about these two cases. In Weighted KNN (WKNN) the locations of the $K$ closest neighbors are given different weights before averaging them, i.e., KNN is used as a fusing technique. Due to this intuitive structure and relative simplicity, KNN has been considered in many localization solutions, especially fingerprint based localization. When using KNN, research has been dedicated to fingerprint database construction methods, adaptively selecting $K$, and deriving meaningful similarity metrics. Here we review a few works to highlight these aspects. Other work can be found in various surveys for fingerprints, e.g., \cite{he2015wi,liu2007survey}.

A typical early example for this approach is \cite{bahl2000radar}, where the collected RSSI values are stored to construct the radio map. During the online phase, i.e., the localization phase, the newly observed RSSI value from three indoor APs are compared against the stored RSSI dataset. The comparison is usually done using Euclidean distance,

$$\mathcal{L}(\boldsymbol{x}_i, \boldsymbol{x}_0) = \sqrt{\sum_{k=1}^N |x_{i,k}-x_{0,k}|^2}$$
where $N$ is the number of APs (three in \cite{bahl2000radar}) and  ${x}_{i,k}$ and ${x}_{0,k}$ are, respectively, the RSSI value from the $k^{\rm th}$ AP at the $i^{\rm th}$ RP and the target.\footnote{We stress that $x$ is the RSSI, and not a location coordinate; this is to stay consistent with Sec. \ref{sec:MLPrelm} that generally denotes features as $x$.} The standard KNN uses fixed $K$ value; however, due to RSS variations or the placement of the RPs  (e.g., enforced by the environment structure), using fixed $K$ might result in large localization estimation errors. To alleviate that, a number of works considered modifications of KNN, e.g.,  \cite{bi2018adaptive,ma2008cluster,shin2012enhanced,peng2016iterative,torteeka2014indoor}. The methods range from sub-selecting some of the $K$ nearest neighbors as in \cite{ma2008cluster,torteeka2014indoor,shin2012enhanced,bi2018adaptive,altintas2011improving}, to modifying the similarity (i.e., the distance) metric \cite{peng2016iterative}. For instance, the solution in \cite{ma2008cluster} groups the $K$ nearest RP neighbors in different clusters (according to their locations), then the final location is the average of the RPs in the "delegate" cluster. The work in \cite{peng2016iterative} suggests using both the Euclidean distance and the cosine similarity  between the RSSI values as a similarity metric, which they also use to construct the weights for a WKNN. 

KNN has been also used with different features. In \cite{alhajri2019indoor_74} the authors  built a cascaded localization system based on KNN, where a KNN is used to first identify the environment, then for localization they apply KNN with different features (RSSI, the CFR and its auto-correlation function). They note the importance of considering hybrid features in different environments, and that they outperform the RSSI-only approach.  

The AoA can be used as a feature in MIMO system; for instance \cite{roshanaei2009dynamic} uses AoA to sub-select a group of the $K$ RPs that satisfy an AoA constraint. Ref. \cite{sun2018single_90} proposes a fingerprint-based single-site localization method for massive-MIMO OFDM systems. However, storing the raw channel observations and searching through them requires large storage and incurs high computational complexity. The paper proposes an efficient database construction by utilizing the sparse representation of the channel in the angle-delay domain, which can be efficiently compressed. To search over different fingerprints, it uses two levels of fingerprint classification and clustering. The similarity between the observed channel and fingerprints is captured by a proposed joint angle and delay similarity metric that depends on the level of overlap of the scatterers. Finally, localization is performed by applying a WKNN to the K nearest fingerprints, where the weights are functions of the proposed metric that correspond to the AoA and ToA. Along the same lines, \cite{wang2019fast_83} proposes a method to construct the database of the angle-delay domain representation by compressing the database and applying a fast method to retrieve the candidate RPs using hashing algorithms; localization is obtained from a WKNN based on the minimum Euclidean distance of the features.

In Ref. \cite{OpticalCamera}, a crowd-sourcing indoor localization algorithm via an optical camera and orientation sensor on a smartphone is introduced. In this solution fingerprint based localization uses KNN over RSS from WiFi, which is used as a coarse estimate of the location to speed up the search space of images based localization; this can be further constrained with information from orientation sensors.   

In summary, since WKNN is an intuitive technique, many advanced ML solutions use it as a final step to predict the location, where different methods are proposed to define the "distance" between the fingerprints and the weights. Methods to calculate the distances and weights include Euclidean distance and Kernel functions, respectively. Additional details can be found in later sections.


{\bf Kernel Based Methods:}
A kernel function $\mathcal{K}(.,.)$ captures the similarity between vectors, and thus can be integrated in instant based learning techniques, see Sec. \ref{Sec:Prem}. One natural way to capture the similarly between two vectors $\boldsymbol{x}_i$ and $\boldsymbol{x}_j$ is based on the distance between them. The Radial Basis Function (RBF) is a kernel that is a function of the distance. An example of that is the Gaussian Kernel
 \begin{align}
     \mathcal{K}(\boldsymbol{x}^i,\boldsymbol{x}^j) = \exp^{-\frac{||\boldsymbol{x}^i-\boldsymbol{x}^j||^2}{2\sigma^2}},
 \end{align}
where $\sigma$ controls the speed of similarity decay between $\boldsymbol{x}^i$ and $\boldsymbol{x}^j$. Other kernels include linear, polynomial, and  sigmoid kernels.  

 One of the powerful algorithms that uses kernels is the Support Vector Machine (SVM). It aims at finding the best hyperplane to separate different classes. In particular, the hyperplane is chosen such that it maximizes the margin between the different classes. In SVM the predictions usually appear in the form of inner products between training instances and observed feature; this allows SVM to use the kernels to find efficiently this product in higher (possibly infinite) dimensional space, which can increase the separability of the features, see Sec. \ref{sec:MLPrelm}.

  SVM, along with its regression counterpart SV-Regression (SVR), and other kernel methods have been used extensively in localization. In one of the early works that uses kernel based localization,  \cite{nguyen2005kernel} uses SVM for localization in a WSN that has a few sensors with known locations. Then the sensors measure the RSSI between one another, in order to obtain an initial region classification using an SVM with Gaussian kernel. Finally, finer coordinates are obtained by averaging the centers of the regions  each sensor belongs to. Ref. \cite{oussar2011indoor} uses linear and Gaussian kernels for room level localization using cellular RSSI.  Ref. \cite{zhang2017efficient} first reduces the dimensionality of the RSSI features through PCA, then localizes the target using SVM with an RBF kernel. In \cite{rezgui2017efficient} the authors propose a method to address the diversity of the devices by first sorting the RSSIs from the most reliable APs, then use SVM to classify the target location. Ref. \cite{kushki2007kernel} proposes real time AP selection to minimize the number of correlated APs used, and then uses a kernel as a metric measure to construct a weight based on the similarity between the RSSI observation and the RPs, which is then used to weight the coordinates of the RPs to estimate the location.
  
  SVM and other kernel methods have been used with other features as well. In \cite{zhang2018heterogeneous} the authors used ToA and RSSI in Gaussian kernel Ridge regression for localization. Ref. \cite{li2017robust_60} uses TOA as fingerprints, and proposes a Gaussian Kernel based solution; the proposed approach is insensitive to random synchronization and measurement timing errors. Ref. \cite{kram2019uwb_68} proposes a localization solution in UWB systems, by extracting different features from the channel impulse response, such as the energy decay time. The idea is to compress the impulse response and capture different aspects of the environment. The features are then fed to an SVM algorithm for hierarchical region classification. Using CSI of 30 sub-carriers, \cite{wu2018accurate} constructed a visibility graph that captures the frequency correlations between adjacent sub-carriers, which are used for localization with SVM. CSI based activity recognition and localization using SVM and kernel regression is proposed in  \cite{wu2019csi_7}, where the SVM is first used to classify the target in one of the possible activity classes, then the localization is done with the associated localization regression model. In a device-free localization system, \cite{sanam2018improved} used the CSI in an OFDM-MIMO system with different detecting points to sense the channel and send it to a central sever. Localization with RSSI and accelerometer readings was used in WSN target tracking in \cite{TargetTracking}; the RSSI based kernel method provides a coarse location estimate, which is then utilized along with the accelerometer readings to get the instantaneous localization using a Kalman filter.
 
 Kernel methods have been used in other aspects of localization, such as an SVM for ranging error estimation based on CSI in an UWB system in \cite{wymeersch2012machine_72} or SVM for pose recognition in \cite{zhang2019improving}, which is then used to match for the appropriate RPs.  Indoor vs outdoor classification was done in \cite{ray2016localization_94} before location estimation using particle filters. To improve the robustness of the localization process against noise, \cite{7504333} proposes a solution for AP selection and classification, then uses SVR to reconstruct the RSSI values of the non-selected APs.

The choice of the kernel has an important impact on the solution. Ref. \cite{pan2005accurate} suggests using kernel canonical correlation analysis to better capture the correlation in the signal (RSSI) space and the coordinates before the localization, where Gaussian and Matern kernels are used for signal space and physical space, respectively. In \cite{wu2007location} the authors propose using a sum-of-exponentials kernel that is tolerant to missing values and sensitive to feature differences in RSSI based cellular positioning. Ref. \cite{figuera2012advanced} uses a kernel that incorporates the spatial structure of the training set; it also proposes a kernel that utilizes the possible correlation between the dimensions in the coordinates. In \cite{yan2017hybrid} a hybrid kernel consists of a local kernel (an RBF kernel) and a global kernel (a polynomial kernel); jointly they take into account the impact of nearby and distant RPs. In \cite{NLOS_UWB} an Import Vector Machine (IVM) was used for NLOS classification for ToA  based ranging in UWB systems; the authors point out that ISM has less complexity and represents better classification probability.

 


{\bf Gaussian Process Based Methods:}
There are several other ML-solutions based on Gaussian Process (GP)\cite{prasad2017numerical_84,bekkali2011gaussian,homayounvala2019novel,MassiveMIMO,kmgpr,FullBandGSM,7313038}. GP can be viewed as Bayesian alternative to the kernel methods\cite{murphy2012machine,rasmussen2003gaussian}, where the goal is to infer the posterior distribution of the labeling function for given observations. In the training phase, we seek good values for the mean and the covariance of the GP. The latter is usually captured with kernel functions, such as RBF or Matern.

In \cite{bekkali2011gaussian} the authors propose to use GP regression on the training data to build a continuous distribution of the RSSI for each AP. For localization, for given RSS observations, they apply a Maximum Likelihood Estimator algorithm to infer the coordinates of the target. To capture the correlation between different locations they studied three different kernel based correlation functions:  Gaussian, Matern, and Quadratic kernels. Ref. \cite{kumar2016gaussian} uses a GP to model the probability distribution of the RSSI values, then for a given RSSI observation, and using Bayes rule, the location is estimated by a weighted combination of the RPs' locations. Ref. \cite{atia2012dynamic} describes a method for dynamically estimating and calibrating the RSSI radio map using GP; for this the standard deviation of the trained GP model is used to measure the accuracy of the estimated position; the final location is estimated using WKNN. We here point out that there are number of other works that use GP (and other ML techniques) to capture the RSSI correlation and distribution \cite{bui2017survey}, which can be used for different purposes in wireless systems. 

For a distributed massive-MIMO system,  \cite{MassiveMIMO} introduce numerical approximation GP methods that result in a test Root Mean Square Error (RMSE) very close to the Cramér–Rao bound, which is verified with simulated data. For a UWB system, \cite{7313038} uses a number of channel parameters that are extracted from the PDP, such as TOA, RSS, and RMS delay spread for ranging. The first step consists of using a kernel PCA, in which the selected channel parameters are projected onto a nonlinear orthogonal high-dimensional space; a subset of these projections is then used as an input for GPR to provide a ranging estimate.

{\bf Trees and Ensemble Methods}
  Decision trees derive their classification rules by splitting the observation-labels space. Decision trees have been used in a number of localization problems, e.g., for LOS identification \cite{musa2019decision}, coordinate prediction \cite{ahmadi2015rssi_58}, and localization after AP selection and clustering  \cite{chen2006power}. However, many of the papers use them for simple classification problems, or end up using many decision trees in their solutions. 

Ensemble methods use multiple learning solutions to obtain the final decisions; they have been shown to provide excellent performance even when built as an ensemble of basic ML solutions, e.g., decision trees.
Random Forest (RaF) is one such ensemble learning method; it incorporates multiple decision trees. It has been used frequently in localization. For example, \cite{de2019enhanced_129} applies Volume Cross-Correlation on the CSI to acquire the TDOA, then uses a RaF for classification-based localization using multi-path information. The proposed solution uses a combination of both ray-tracing and measurements to enhance the localization performance. Ref. \cite{hybloc} uses the RSSI in WiFi networks to provide the location in terms of both coordinates and room-level prediction. In the offline phase, fingerprint data are pre-processed, ensemble classifiers (based on GMM) are trained for room prediction, and a RaF regressor is trained for location prediction. In the online phase, data is pre-processed, soft cluster membership is determined, and room and location are predicted. Ref. \cite{FPfusing} utilizes a multi-antenna system to build fingerprints with different features, such as RSSI, power spectral density, and other statistical features. For each feature a RaF is trained as classifier. To increases the robustness of the location estimates, the solution uses multiple samples and classifiers, where an entropy metric is used to choose a robust classifier and a stable time instant. Then the location is the mode of the location predictions constrained to be within the union of the predicted locations from the selected classifier and at the selected time instant. 

Other ensemble learning methods have been used as well, such as gradient boosting regression forest (GBRF) \cite{8931646}, and AdaBoost \cite{7064997, FeatureSelection,8673800}. Due to fluctuation of the RSSI signals, the RSSI distance might not reflect the true location distance; to address this \cite{8931646} proposes a ﬁngerprinting method by transforming raw RSSI into features with a learned non-linear mapping function using GBRF. The idea is to pair the RPs in positive and negative pairs (based on a predefined threshold), then create a loss function that ensures that the similarity is preserved. The mapping function (here GBRF) is then trained. Finally, the localization is performed using WKNN with the new mapping function. Ref. \cite{8673800} employs Adaboost for passive localization, where the phase information in CSI is used to construct the ﬁngerprint map. Through continuous iteration with the Adaboost algorithm, the sample weights of the training sets are continuously adjusted to prepare for classiﬁcation. The final location estimate is weighted average of the locations of the top four predicted RPs.        



Finally, we point out that a number of the above-discussed papers compare the performance of their solution against other ML algorithms. Furthermore, a number of works studies the performance of different ML solutions, such as Ref. \cite{FullBandGSM} for indoor localization using cellular signals. Ref. \cite{fukushima2019evaluating_73} compares KNN, RaF and SVM for device-free localization using CSI. Ref. \cite{8108573} suggests localization using frequency modulation and digital video broadcasting terrestrial signals, comparing KNN and, SVM and SVM with Ensemble Learning.


\subsubsection{Deep Learning Solution }
Nowadays, NNs are among the most popular ML architectures. This can be partly attributed to the fact that many of the successful DL architectures are based on them. Thus we start this section with a review of the localization solutions based on Feedforward NNs.\\
{\bf Feed-Forward NNs:}
Due to a number of attractive properties of NNs (see Sec. \ref{sec:MLPrelm}),  many NN based localization solutions have been proposed; in the following we highlight representative samples.

Ref. \cite{lee2018dnn_95} uses the RSRP of the three strongest BSs along with their locations in an LTE network as input to localize the user. Ref. \cite{shokry2018deeploc_50} uses an NN, with three hidden layers, that takes RSSI from a number of BSs. Data augmentation techniques, e.g., masking some of the BS values, improve the robustness of the solution. In  \cite{ge2019hybrid }, two networks are trained for two different features, namely TOA and RSSI from a WiFi systems, and the location is a weighted combination of the outputs of the two networks. Ref. \cite{jaafar2018neural} uses a hierarchical localization solution, where for each indoor location, an NN is trained. During the online phase, the real-time RSSI measurements (from WiFi or/and cellular) are first used to identify the environment, and then they are passed as input to the corresponding trained NN.

Using directional antennas in a WiFi system, \cite{zhang2019efficient_122} builds fingerprints using RSSI (or CSI amplitude) and uses an NN with two hidden layers as a classifier to predict the closest fingerprint to the observed values. Using a rough estimate of the AoA provides an improved localization accuracy.  For an $8\times 2$ MIMO system, \cite{sobehy2019csi_117}  uses CSI amplitude at 16 antennas and 924 sub-carriers as input to several NNs with different hyperparameters (i.e., an ensemble learning solution); the final location is a combination of the output from all networks. This paper also studied different combining techniques; a method that uses both the median and weighted average of the location estimates provided the best result. 

The above architectures usually require the backpropagation algorithm for training, which is generally time-consuming, Extreme Learning Machine (ELM) is a simplified feed-forward NN architecture with one hidden layer. The weights of the first layer are set to random values, while the weights for the hidden layer are usually calculated with the least square fit. A number of works try to utilize the simplicity of such architecture in localization, \cite{zou2015fast, decurninge2018csi_81,feng2019received,xiao2012large}. In \cite{zou2015fast}, during the offline phase, a WiFi RSSI fingerprint database is created, and an ELM is trained on the data. During the online phase, fingerprints are still collected at some known locations, and the solution is updated according to the new fingerprints too. Then the latest ELM is used to predict the coordinates. In Ref. \cite{xiao2012large}, during the offline phase, the data is first clustered using k-means clustering, an ELM is used to classify the data to one of the clusters, and a dedicated ELM for each cluster is trained. During the online phase, the RSSI values are first classified and then the ELM of the associated cluster is used.

A framework that utilizes features from different technologies for target tracking was introduced in \cite{belmonte2019swiblux}. The inputs to the NN are: the RSSI measurements from WiFi, Bluetooth and XBee (a wireless connectivity module used for IoT) from different nodes, plus yaw readings when available. The yaw readings can be extracted by filtering measurements of some wearable sensors (accelerometer, magnetometer and gyroscope). The probabilities of fingerprints are passed through a Gaussian Outliers Filtering process, and the output can then be weighted and combined to provide the location estimate, which in turn is fed to a particle filter for target tracking.
  
For MIMO-OFDM systems, \cite{wang2016csi_4} proposes a fingerprint-based localization scheme (which is called DeepFi) that utilizes the magnitude of the CSI at 90 sub-carriers from three antennas. It uses the {\em weights} of NNs with four hidden layers to represent the fingerprints. Training the NNs is done by a greedy learning algorithm using a stack of RBMs (see sec. \ref{sec:MLPrelm}). After the pre-training and supervised fine tuning, the output of the NN is a reconstruction of the input data.\footnote{Note that this can be as well viewed as an Autoencoder structure, which we discuss below. } During the online phase, using a number of RSSI realizations (packets), a Gaussian RBF kernel is used to represent the likelihood probability of the observed data when $i^{\rm th}$ location (RP) is true, from which Bayes rule can be used to calculate the posterior probability of location $i$. The final location is the weighted average of all RP locations. Ref. \cite{wang2015phasefi_43} proposes a scheme (called PhaseFi) that utilizes the phase value to construct a fingerprint database, where a linear transformation of the phase value as a calibration step, improves the stability of the phase values. The weights of a three-hidden-layer-NNs serve as fingerprints. 

NNs in passive localization have been considered in \cite{HumanDetection}, where amplitude and the calibrated phase of CSI are used as a hybrid complex input feature to the NN to detect the presence of a human. In mmWave communication systems, \cite{mmwave} integrates NNs in a cooperative WLS estimator. NNs were used for ranging error mitigation in \cite{wu2019neural_116}, where the authors use RSSI values as input to an NN, which then predicts the ranging error; a localization algorithm such as least squares can then use the adjusted range values.
 


{\bf Convolution NN (CNN):}
Recently, several works proposed CNN based localization schemes, since those have shown good performance in computer vision, as discussed in Sec. \ref{sec:MLPrelm}. 

Since the number of observed RSSI values in urban areas is large, several works reshape the observed RSSI array into 2D or 3D images and use it as input to a CNN network. Ref. \cite{jang2018indoor_99} uses a CNN, applied to the RSSI image, to classify the location in one of several buildings and floor levels. Modifications of the RSSI images were proposed in \cite{mittal2018adapting_109,liu2019hybrid}: Ref. \cite{mittal2018adapting_109} augments the features with correlation values (between the RSSI from the AP at all RPs and their locations). The augmented image is used as input to a hierarchical localization structure, where initially a CNN predicts the floor number, then a CNN associated with the chosen floor predicts the corridor number, and finally a CNN associated with that corridor is used to estimate the coordinates. Ref. \cite{liu2019hybrid} proposes hybrid RSSI features that contain the ratio of the contribution of the RSSI value to the fingerprint. Viewing the RSSI values as time series, Ref. \cite{soro2019joint_118} uses CWT (see Sec. \ref{sec:Feat}) to produce a 2D time-frequency image; the CNN predicts the closest reference points to the target, and KNN is then used to infer the coordinates of the target. In another hierarchical localization model, \cite{ibrahim2018cnn_13} uses RSSI values from all APs over different time instances to form RSSI-time 2D images to predict the building, floor and coordinates. BLE RSSI observations over time are used in \cite{ebuchi2019vehicle_126} for vehicle and pedestrian detection/localization in a smart parking system. In \cite{own2019signal_125}, the RSSI values from several APs are used to predict the location with two capsule networks, one for the RSSI values in $2.4$ GHz and one for RSSI values in $5$GHz. Capsule networks are CNN architectures with added capsule modules to track the hierarchy of the objects in the images \cite{sabour2017dynamic}. 

\begin{figure}
\centering
\vspace{-20 mm}
 \includegraphics[width=1\linewidth]{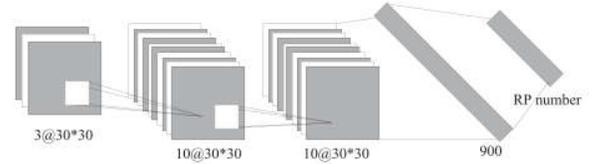}
\vspace{-20  mm}
\caption{ \small The structure used in \cite{chen2017confi}, with three CNN layers and two fully connected layers. The input features are the amplitude of the CSI at $30$ sub-carriers and $30$ time instant, these values are captured from three antennas to constitute an image with three channels. They use $10$ kernels at the first layer. Since the input image is relatively small and to use deep structure, the authors propose to maintain the size of the image for the two consequent CNN layers (through padding and stride equal to one), the output is probability of the RP.}
\label{fig:confistructure}
\vspace{-5 mm}
\end{figure}


With the improved efficiency of NNs through CNNs, it is possible to utilize high dimensional features.  The "ConFi" scheme \cite{chen2017confi} is a multi-layer CNN network that uses amplitudes of the CFR at the available sub-carriers (in a WiFi system) at a number of time instances to form an image; realizations at different antennas can be used at different CNN channels (see Sec. \ref{sec:MLPrelm}), and the output is the probability that a target is located at a given RP, see Fig. \ref{fig:confistructure}. The final location is estimated by averaging the location of the RPs with their predicted probabilities. The so-called "CiFi" scheme \cite{wang2018deep_23} uses coarse but stable estimates of the AoA, based on the pair-wise phase difference between antennas. The provided example uses the $30$ available sub-carriers and $960$ time instances to form $16$ $60 \times 60$ images for training. During the online phase it predicts the location by fusing the location of the most probable reference points.  Ref. \cite{jing2019learning_131} uses the learned spatio-temporal features through dual stream 3D CNNs to approximate the posterior distribution of the mobile device’s location with a GMM. It creates two 3D images of the calibrated phase and amplitude of the CSI, where the height, the width, and the depth (number of channels) are, respectively, the amplitude/phase from different packets, different sub-carriers (30 available) and Tx-Rx antenna combinations. In a three-antenna system, \cite{li2019convolutional_12} creates three-channel images that have the amplitudes of all sub-carrier $\times$ packet samples in one channel, and pair-wise phase differences between the antenna signals in the two other channels. They use ShuffleNet, a computationally efficient deep CNN architecture, to predict the most probable RPs.

For massive MIMO systems, \cite{wu2019learning_78,vieira2017deep_80} uses images in the angle-delay domain. For a Uniform Planar Array, \cite{wu2019learning_78} proposes 3D images that have power values in horizontal, vertical, and delay domains. The images are fed to a CNN network that uses the inception module and different kernel sizes (due to different sparsity at different domains). The inception module has been used in famous Deep NNs architectures such as AlexNet, where the output of different kernels are concatenated, which allows better feature extraction. 

CNNs have been used for device-free localization in \cite{huang2020widet_64,hsieh2019deep_28, InterpretingCNN, DCNN}. For instance, \cite{huang2020widet_64} creates images using the time sequence of RSSI values and their CWT; the solution then detects the presence of users indoor. In \cite{hsieh2019deep_28}, the authors use 1D CNN networks with either RSSI values from all antennas over several packets or CSI amplitudes from all sub-carries and antennas. They notice the superior performance of CSI based solutions to detect the region of the target. For ranging in UWB systems, Ref. \cite{bregar2018improving_77} uses a CNN network for ranging error mitigation, proposing two approaches,: one for NLOS detection and another one to estimate the ranging error. In an industrial environment, where ranging based on ToA is difficult, due to the increased multi-path propagation, Ref. \cite{niitsoo2019deep_2} proposes to use time-calibrated complex CIRs to create images. The authors modified several of popular deep networks such as AlexNet and GoogLeNet. They found that GoogLeNet gives a good complexity-to-performance trade-off. They also propose a distributed framework to reduce the overhead. 

\begin{figure}
\centering
\vspace{0 mm}
 \includegraphics[width=\columnwidth, trim={6.5cm 2cm 5cm 2cm},clip]{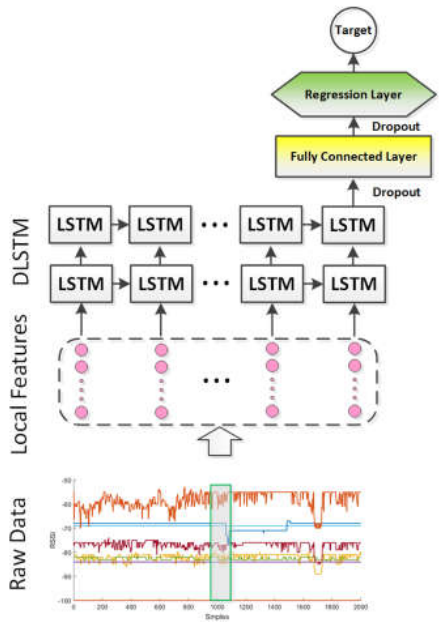}
\vspace{-5 mm}
\caption{  \small The structure used in \cite{8733822}: Several consecutive RSSI samples from each AP are collected ($n$ samples and $r$ APs), for a given time window (of size $w$) and for every AP, $5$ statistical features are computed: (minimum and maximum RSSI values, and the first, the second and the third quartiles), this results in $5 r$ local features and a sequence of length $n/w$. The resultant sequence is fed to two LSTM layers to extract high level features, which are then fed to fully-connected layers and regression. Dropout layers combat over-fitting.}
\label{fig:DLSTM}
\vspace{-5 mm}
\end{figure}

{\bf Recurrent NN (RNN) Based Architectures:}
Several recent works utilized sequential RSSI values for localization. Ref. \cite{turabieh2019cascaded_14} uses a sequence of RSSI values as input to cascaded RNNs to estimate both the building and the floor numbers. In a system with multiple BLE anchors, \cite{xu2019efficient} proposes efficient construction of the fingerprint database for real-time systems with RSSI samples, where LSTM is used for localization. In a cellular system, Refs. \cite{MonoDCell,rizk2019solocell_98} use the RSSI history from only the associated cell tower to track users' location with an LSTM, where the inputs are (cell tower ID, RSS) sequential pairs and the output is the 2D coordinates. To train the LSTM, Ref. \cite{rizk2019solocell_98} proposes using KNN and HMM to generate synthetic measurements based on the previously observed ones. In \cite{8733822}, the authors apply a sliding window to the sequence of RSSI samples; in each window they calculate five features (minimum, maximum, and three quartile  values) from each AP. They then feed them as a sequence of vectors to an LSTM, see Fig. \ref{fig:DLSTM} for more details.

In cellular and WiFi systems, \cite{wu2019mobile} uses RSSI values from cellular and WiFi systems, the solution takes the RSSI fingerprint from multiple APs/BSs at multiple timesteps and passes them to a single-layer RNN which then gives the coordinates. In systems with Unmanned Aerial Vehicle (UAV)-BSs,  \cite{adege2019mobility} proposes using RSSI from three UAV-BSs and nearby WiFi APs to predict the location. The RSSI features are first reduced to 30 features using PCA, and then combined with the rest of the features. The new feature vector is then passed as input to the GRU Network for location estimation. In a related setup, Ref. \cite{tarekegn2019applying_107} uses RSSI values from the reachable WiFi APs and three temporarily deployed UAV-BSs to build a radio-map, and uses LDA to select the subset of APs to reduce computation time. It then feeds the RSS sequences to an LSTM to predict the location. In an MIMO-OFDM system, \cite{csiRNN} utilizes the CSI amplitudes at different antennas and sub-carriers for localization. After prepossessing, shifting and polynomial regression (for smoothing) of the CSI, the correlation matrix, CSI amplitude and SNR are used as candidate features for several ML solutions. The authors point out that LSTM outperforms other solutions when feeding the data sequentially data (i.e., utilizing user's trajectory).

Composites of CNNs and RNNs have been considered in \cite{zhang2019deeploc_108, qian2019convolutional_10}. Ref. \cite{zhang2019deeploc_108} uses the consecutive MR samples in a cellular system to generate a smooth trajectory
consisting of predicted locations. It first divides the entire region into cells, then creates images with height and width similar to the grid structure and fills it with features from the MRs. Next, it feeds these images to a CNN to estimate a score for every potential location based on the extracted spatial features. Then it uses the windowed scores as input to a multi-layer LSTM and finally with regression to generate the trajectory. Note the role of the CNN here is to learn local spatial features from each individual MR. Ref.
\cite{qian2019convolutional_10} uses sequential RSSI values from nearby APs for localization. The data are fed to a 1D-CNN that captures the features. Then a GRU is used to capture the time dependency. The output of the RNN is fed to a Mixture Density Network (MDN) that learns the conditional probability distributions of the locations (see sec. \ref{sec:MLPrelm} for mixture models).

A somewhat different approach is taken by 
 ''DeepTAL" \cite{xue2019deeptal_36}, which handles the TDOA measurement error or missing data in an asynchronous system, where the system predicts the target state (moving /static) and outputs the TDOA. The TDOA and the difference of the TDOA are used as the training input. The location is then solved with a quantum-behaved particle swarm optimization algorithm.

\begin{figure}
\centering
\vspace{-15 mm}
 \includegraphics[width=1\linewidth]{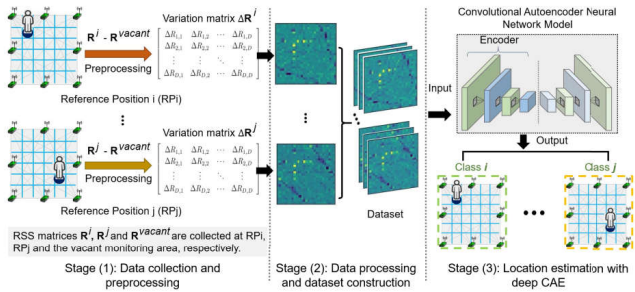}
\vspace{-15 mm}
\caption{ \small The system proposed in \cite{zhao2019accurate} formulates the device-free localization as a classification problem to predict the location of the target in one of $L$ RPs, where the area is divided into $L$ grids. The proposed network is based on a Convolutional network that is pre-trained to extract useful features as an encoder of a Convolutional AE architecture. The encoder is then attached to fully connected layers and softmax. The input image is constructed using the RSSI measurements from $D$ APs that act as a TX and an RX in turn. They collect the RSSI between all the APs when (i) no target is present in the area, (ii) when a target is present. The RSSI images are the difference between the two cases.}
\label{fig:CAE}
\vspace{-5 mm}
\end{figure}

{\bf Auto Encoders:} 
In the localization literature, AEs have been used heavily. Although AEs are usually unsupervised learning techniques, in the localization literature, they have been used in conjunction with many of the earlier supervised solutions. One popular approach is to use the AE to extract robust feature representations, where the output of the {\em encoder} will be the input of a localization module. Ref. \cite{khatab2017fingerprint} integrates AE into ELM based classification localization, where the AE is used to extract high level features, and the output of the AE's encoder is used to replace the random projection of an ELM. As discussed in later sections, it is relatively easy to integrate different regularization approaches when training the ELM, which in this case depend on the encoder output. A hierarchical tuning mechanism is used to train the solution, first tuning the classifier parameter, and then adjusting the encoder parameters. Ref. \cite{song2019novel_119} uses the output of the Stacked AE (SAE) with a one-dimensional CNN to provide the location based on the floor/building and the target coordinates. Similarly, \cite{kim2018hybrid} proposes a multi-output network that uses the encoded features to predict the building, the floor, and the coordinate. An SAE followed by two FC layers and then an argmax to predict the multi-label classification is used in \cite{belmannoubi2019stacked_110}. Ref. \cite{zhang2017deeppositioning_106} uses the AE to provide a low dimensional representation of the RSSI and magnetic field signals. AEs can also be applied to device-free localization \cite{zhao2019accurate,wang2016device_39,gbrbm}. Convolutional AE (CAE) uses the output of the encoder as input to a classifier that predicts the RP where the target may be \cite{zhao2019accurate}. The input to the CAE is an image constructed with the difference between the (target free) RSSI values and the online RSSI values; Fig. \ref{fig:CAE} elaborates on the method. Ref. \cite{wang2016device_39} estimates the target’s location, activity, and gesture, based on the RSS measured signals. The approach first denoises the signal with a four-level wavelet decomposition, then uses Sparse-AE to project the high dimensional data to low-dimensional features and finally feeds the learned features into the softmax classification and regression model. The AE could also be used for data transformation: the device heterogeneity problem, can be tackled, e.g., by an AE that maps the features observed by a test device to features that corresponds to the device used for acquiring the database \cite{rizk2019device_96}.

In an alternative approach, the AE could be used in fingerprint-based localization to assess the similarity between the observed features and the collected fingerprints. In "BiLoc," a fingerprint localization solution uses AoA estimates and the amplitude averaged over two antennas as input to AEs, and then employs the weights of the trained AEs as fingerprints \cite{wang2015phasefi_43}. The online phase uses the AEs to reconstruct the observed features; the level of similarities (calculated with RBF) will define the probability of the target being in a specific location. The final target location is the average of the weighted location of the fingerprints. Ref. \cite{yazdanian2018deeppos_34} uses the location along with the latent variable of the AE to reconstruct the observed signal. The location is based on the most similar RP. Ref. \cite{abbas2019wideep_63} builds SAEs corresponding to each fingerprint. In the online phase, it tries to reconstruct the observed features and compare the similarity using RBF to provide a probabilistic location estimate. 

In another approach, the AEs are used to provide a pre-training method before complete training to fine-tune the parameter. As highlighted in Sec. \ref{sec:MLPrelm}, the pre-training technique is not limited to AEs; in fact, one of the early works uses RBMs \cite{hinton2006reducing}. For pre-training, not all the data need to be labeled. These methods have been used in \cite{zou2018deep_37, zhang2016deep_38}, while \cite{zhang2016deep_38} uses a DNN to provide coarse location estimate and an HMM to refine results. In \cite{zou2018deep_37} the solution is composed of a NN, CNN, and followed by a probabilistic
technique for fusing the candidate location estimate. 



\subsection{Semi-Supervised learning}\label{sub:semisuper}

As introduced in Sec. \ref{sec:MLPrelm}, in semi-supervised learning, the solution uses a combination of limited (but important) labeled data and a large subset of unlabeled data. Similar to supervised learning, assumptions such as smoothness are required to allow generalization. With high dimensional data it could be difficult to assess the similarity between points \cite{chapelle2009semi}, and the manifold assumption could offer a remedy for this. In fact many of the semi-supervised learning solutions utilize this assumption. The taxonomy of this section is different from Sec. \ref{subsec:supr} as the approaches are usually different. 

\subsubsection{Manifold Learning }
Manifold alignment is a popular ML framework. It allows a transfer of information between datasets under the condition that there is an underlying common manifold \cite{ma2011manifold}. There are many applications for this, such as dimensionality reduction, and data visualization, see Sec. \ref{sec:MLPrelm}. Graphs are sometimes used to approximate the point relations on the manifolds \cite{chapelle2009semi}. Different manifold alignment algorithms are usually different in how neighborhood graphs and the corresponding distance between nodes are established. Manifold alignment has been used in the localization literature. In semi-supervised learning, the knowledge about the "distance" between the observed data can be used to build the radio map. Works including \cite{zhou2017semi_55,zhou2017grassma_76} use the Laplacian Eigenmap manifold alignment approach. In this technique, a weighted graph between the data points (including labeled and unlabeled data) is constructed to preserve the local geometry of the data. Then the mapping uses the eigenvectors of the graph Laplacian. For example, labeled RSSI data from nearby APs as well as unlabelled timestamped traces can be used to construct the graphs (which incorporates the physical relations of labeled fingerprints), using the labeled data as hubs to construct the graphs \cite{zhou2017grassma_76}. Ref. \cite{zhou2017robust_46} also uses the RSS and RSS traces (based on crowdsourcing), which can integrate the spatial correlation property into the graph. In Ref. \cite{gu2015semi_54} the authors utilize ELM for deep feature extraction, which they use along with the graph Laplacian to define the objective of the semi-supervised classification (localization) problem; the weights on the graph are calculated with Gaussian kernels between the RSSI vectors. Ref. \cite{pourahmadi2012indoor} proposes modification of the objective functions to preserve wireless propagation model based estimated distances. 

Ref. \cite{yoo2019time_132} uses manifold regularization \cite{belkin2006manifold} to develop a semi-supervised RSSI localization solution, which first produces pseudo-labels by optimizing a time-series graph Laplacian SVM; the pseudo-labels are then integrated into a learning framework. It combines the manifold regularization into a transductive SVM \footnote{ Transductive SVM belongs to the set of transductive learning, where the goal is not to learn the general mapping function, but rather it is restricted to the given test data \cite{chapelle2000transductive}.} to balance the contribution of labels and the pseudo-labels \cite{chapelle2000transductive}. In \cite{liu2011selm_133} the authors use ELM for localization. The solution for ELM parameters takes into account the manifold regularization based on Laplacian graphs in WiFi. Ref. \cite{jiang2018fselm_134} adds the RSSI measurements from BLE; since WiFi and BLE signals exhibit different propagation conditions (different transceivers capabilities and transmission powers) two graph Laplacians are constructed to capture the smoothness in WiFi and BLE.

\begin{figure}
\centering
\vspace{-8 mm}
 \includegraphics[width=1\linewidth]{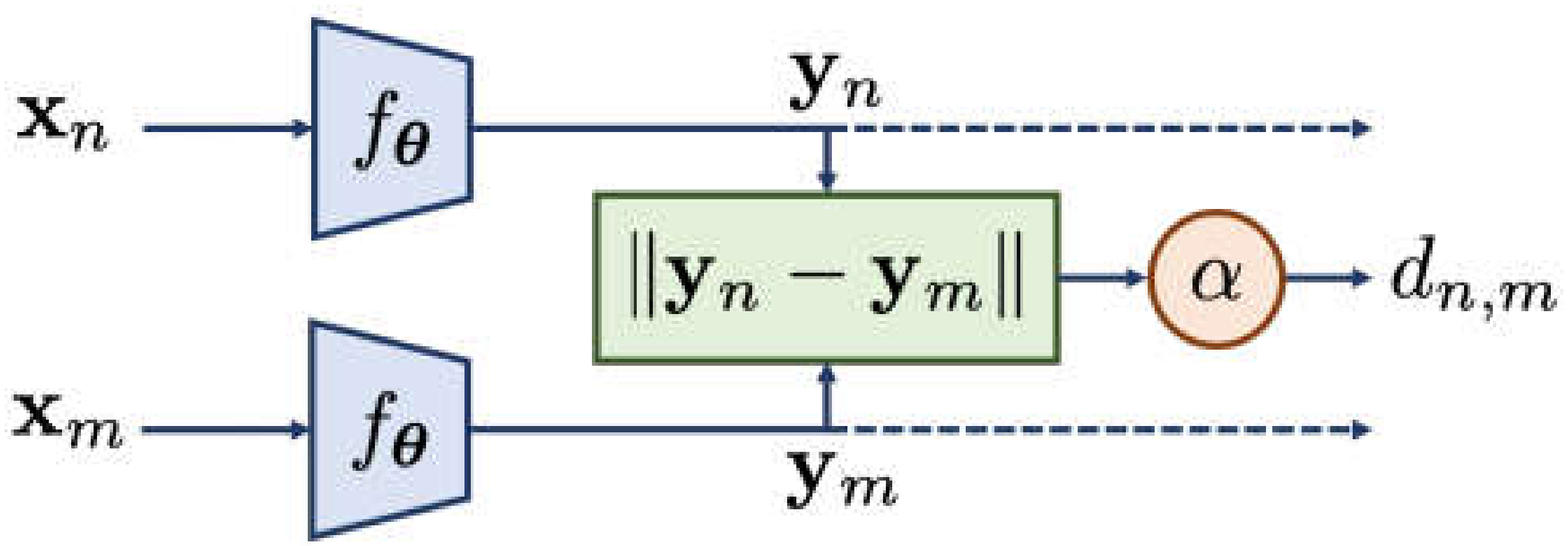}
\vspace{-25 mm}
\caption{\small The proposed Siamese network in \cite{lei2019siamese_130}. The network uses two {\em identical} NNs $f_\theta$ to map high dimensional input features (derived from CSI) $\boldsymbol{x}_n$ and $\boldsymbol{x}_m$ to lower dimensional representations $\boldsymbol{y}_n$ and $\boldsymbol{y}_m$. The goal of the network is to preserve the distance between $\boldsymbol{x}_n$ and $\boldsymbol{x}_m$, and $\boldsymbol{y}_n$ and $\boldsymbol{y}_m$ ($\boldsymbol{d}_{m,n}$). The parameter $\alpha$ is a scaling value used in a semi-supervised learning scenario for distance scale matching\cite{lei2019siamese_130}. }
\label{fig:Siamese}
\vspace{-5 mm}
\end{figure}

A Siamese network architecture uses two identical NNs to compare two inputs; it was used in \cite{lei2019siamese_130} as supervised or semi-supervised CSI-based localization solutions , see Fig. \ref{fig:Siamese}. Stemming from the fact that location influences the values of large scale parameters, the idea is to use two identical feedforward NNs to map the input features to lower dimensional representations (e.g., location). When labeled data is available the loss function should include the ground truth CSI to coordinate mapping while preserving the feature distances.

\subsubsection{Generative and Statistical Models}
Ref. \cite{ouyang2011indoor_51} proposes a hybrid generative/discriminative classification and regression learning algorithm, employing a naive Bayes and EM algorithm to construct the generative model. It uses the naive Bayes method to learn the initial probabilistic model parameters from a limited number of labeled samples. The EM algorithm is then employed to gradually improve the parameters using the unlabeled samples. The least-square SVM (LS-SVM) is then trained on the labeled data to perform discriminative learning.

Motivated by the success of semi-supervised learning with deep generative models \cite{kingma2014semi}, Refs. \cite{chidlovskii2019semi_100,Qian2019Supervised} use Variational AE based semi-supervised localization that can utilize both labeled and unlabeled RSSI observations. It consists of two components: a latent-feature discriminative model M1 and a generative semi-supervised model M2 \cite{kingma2014semi}. Both rely on variational inference, which aims at replacing a complex unknown distribution $p$ with a simpler and more tractable distribution $q$ (usually Gaussian). The similarity of $p$ and $q$ (captured by the KL-divergence) is maximized by optimizing the "variational lower bound". In M1 the encoder learns the latent representation of the observation (RSSI in \cite{chidlovskii2019semi_100}), in M2 it uses both the observation and the labels (if present) to learn the mapping. Combining M1 and M2 provides a powerful deep generative model that uses all the available data for location inference \cite{chidlovskii2019semi_100}.

\subsubsection{Other Methods}

In practical setups, the environment changes over time, which makes the collected data and the trained model outdated after some time. The goal of \cite{ghourchian2017real_45} is to keep the trained classification solution of a device-free localization up to date. It proposes a method to label unlabeled data and then use them when drift occurs to re-train the model. Thus the solution automatically re-trains when the uncertainty level rises significantly. The uncertainty is based on KL divergence, which is used as a distance metric to track substantial changes in the distribution of the features. More solutions for the same problem are presented in Sec. \ref{subsec:TL}.

Ref. \cite{liu2009low_52} proposes a label propagation (LP) algorithm in classification-based localization. By representing labeled and unlabeled data as vertices in a connected graph, the algorithm iteratively propagates labels to unlabeled vertices through weighted edges; the weights of the edges depend on the distance of the observed signal strength values. It can then infer the labels of unlabeled data after this propagation process converges. Throughout the iterative process, the true labels of the collected training data are maintained. A similar LP is carried out for the new observations. 

 In a different research direction, Ref. \cite{7743685} uses semi-supervised learning for trajectory learning in a map-less setup. The overall localization solution is a combination of GP for location likelihood learning (probability of RSSI given the location) and a particle filter. The learned trajectory is used to identify the prior of a particle filter. The semi-supervised part of the solution is a weighted version of unsupervised and supervised dimensionality reduction techniques, PCA and LDA respectively, which is employed to learn the landmarks (e.g., a room) for the given RSSI signals. Once identified, the landmarks are used to identify the trajectories' starting and end points.


\subsection{Unsupervised learning}\label{sub:unsuper}
As discussed in the previous subsection, unlabeled data can still be useful as they contain the structure and distribution of the features. For this reason, many solutions pre-train the model with unlabeled data. This step can be used in combination with any other ML solution. In \ref{subsec:supr}, we discuss several algorithms that use SAE to pre-train deep networks. Other architectures have been used as well, such as RBM and Deep Belief Networks (DBN). For instance, \cite{le2018unsupervised_101} uses DBN for deep feature learning from RSS measurements (after noise reduction); the features are then fed to another ML solution (classification/regression) for location estimation. Unsupervised learning was used to enhance localization solutions, e.g., for AP selection (e.g., \cite{chen2006power} using k-means clustering), data filtering (e.g., clustering erroneous crowdsourcing data for outlier detection in \cite{park2010growing}), and training-device to testing-device mapping (e.g., by introducing an online step to train linear mapping between the two devices based on coarse labels \cite{tsui2009unsupervised}). However, in this section we focus on solutions that use mainly unlabeled data for localization.

 In addition to pre-training, unlabeled data could be used for radio-map construction. The "WILL" scheme \cite{wu2012will_93} for room number prediction creates a logical floor plan using RSSI traces and accelerometer readings (to identify mobility) from any user within the service area. The plan construction is based on clustering techniques on RSSI stacking difference (RSSI reading difference from one AP to all other APs); an example of one of the used clustering is k-means clustering. Each virtual room on the logical map is assigned a representative fingerprint for the localization phase. The paper also provides a matching algorithm to map the logical plan to the ground-truth floor plan that is usually available to real-estate administrators. Ref. \cite{wang2012no_92} proposes a framework for unsupervised localization that uses readings from accelerometer, compass, gyroscope, as well as WiFi readings the users report while moving naturally inside a building. It first tries to identify some “seed landmarks”, which are certain structures in the building, such as stairs, elevators, entrances, that force the users to behave in predictable ways. It then dead-reckons the devices starting from a known reference location. Since the sensors might show some consistent measurements in the indoor areas, e.g., dead-spots or sudden increase of reflections, the algorithm employs K-Means clustering to extract unique sensor signatures that can increase the localization accuracy. Note that the framework relies on one-time global truth information, e.g., the location of a door, or staircase, or elevator. Ref. \cite{jung2015unsupervised_89} aims to fit the RSSI traces into the structure of the environment without use of labels. Localization is obtained from use of HMM and global-local optimization that considers the solutions that do not violate the signal propagation to restrict the search space. 
 
 In a different set of approaches, \cite{choi2019unsupervised_69,choi2019unsupervised_123} devise customized loss functions that can be used to train the localization model without the labels. The proposed solution in \cite{choi2019unsupervised_123} can run in a semi-supervised or in an unsupervised way. It assumes that the user can estimate the distance to the APs with RSSI measurement (through path-loss model, multinomial fit or NN), or using an FTM protocol. The paper proposes several cost functions that can be used to assess the ranging accuracy of the APs, e.g., the distance between the predicted location and ranging distance (for unsupervised) or difference between the predicted location and true location for the few labeled points (for semi-supervised). Finally, the gradient of these cost functions can be used to train the models. Ref. \cite{choi2019unsupervised_69} uses the unlabeled data with the customized cost function to predict the location of a subset of APs with known locations.
 
Refs. \cite{goswami2011wigem_75,location_aware} propose graphical model based solution (see Sec. \ref{sec:MLPrelm}). The location (points on a grid) (in both works) along with the transmitted discrete power levels (in \cite{goswami2011wigem_75}) are modeled as latent variables. Assuming the RSSI values to be approximately normally distributed and to be independent from one APs to another, the solution uses GMM to model the probability of the received RSSI. The latent parameters are then estimated using EM algorithm. To solve the identifiability problem (to match the predicted grid pint to the true one), in \cite{goswami2011wigem_75} the knowledge of the APs locations along with simple pathloss model is used to initialize the EM algorithm.

Since a number of solutions use fusion to predict the final location of the target, \cite{guo2018accurate} proposes an unsupervised learning method to improve the fusion, using an extended candidate location set (concatenation of the top classifiers' outputs), and designing a joint estimate of weights and location under an unsupervised optimization framework. Reliable predictions are assigned higher weights than the unreliable ones, so the true location of the user should be close to reliable predictions.

Finally, Multidimensional Scaling (MDS), a dimensionality reduction technique, has been used extensively for WSN localization \cite{saeed2019state,shang2003localization}. Similar to LLE (see Sec. \ref{sec:MLPrelm}), MDS performs projection to a lower dimensional space while preserving the known inter-wireless nodes distances; the projection results in the node configuration in space. The distance can be estimated using wireless signals, e.g., RSSI or ToA readings; this has been recently utilized as a cooperative localization technique for RFID (and in general IoT) nodes \cite{gao2017indoor}, and cognitive radio \cite{saeed2014robust}. We here limit the discussion of these methods as wireless signals are usually used to only calculate the inter-node distances; we refer the interested reader to other recent survey in this subject \cite{saeed2014robust}.

\subsection{Transfer Learning}\label{subsec:TL}
Transfer learning (TL) is used in ML to exploit the knowledge acquired in the source domain for the learning task in the target domain \cite{pan2009survey}. There might be several reasons to use TL: When the acquisition of the data that matches the specific application domain could be difficult, there might be abundant data in one domain, but limited data in the target domain. In localization, the system might be configured or set up in one environment and deployed in a different environment, e.g., in dynamic systems. Alternatively, the data could be available in both domains, but retraining the model is difficult. TL could be used to arrive at a good solution in the target domain without retraining the solution from scratch.

Some of the early works to apply TL in localization using RSSI are summarized in \cite{pan2008transfer_48}. TL over space (e.g., different parts of the building), devices and time, is considered, where in each of these problems the distribution of data could be different over time, space or target devices. For TL over space, TL is formulated as two optimization problems, where in the first the underlying semantic manifold of the signal is extracted, which can be used as constraints for the second one, where the unlabeled data in the target domain are labeled. For TL across devices, the problem can be formulated as multi-task learning. For a TL over time, \cite{pan2007adaptive} proposes a solution based on modified manifold regularization; this is motivated by the fact the distributions of the data (RSSI), while different over different time instance, are expected to be similar in low-dimensional space (the physical space). The solution uses labeled fingerprints and a few unlabeled data in the offline phase, which can be used to learn the localization function. During the online phase, it collects a few measurements on some of the fingerprints and additional unlabeled data points. With the modified manifold regularization it learns the joint mapping function for localization to update the mapping function learned during the offline phase.

Manifold alignment has been used for TL. Ref. \cite{sorour2014joint_47} uses a simulated (e.g., by ray-tracing) radio propagation map as source domain, then use it along with a few calibration fingerprint in the target indoor environment. This study constrains the two domains to have similar spatial correlation of the RSS values, and then used LLE (see Sec. \ref{sec:MLPrelm}) to find a low dimensional representation, and find the nearest location to the observed features. Manifold alignment for TL was used in \cite{sun2008adaptive} as well, where the source domain has fingerprint data measured at different times and using different devices. In \cite{8661625}, with the source and the target domains, respectively, containing labeled RSSI fingerprints and online RSSI readings, the transfer learning is performed by mapping the source and target domains into a latent feature subspace and maintaining both global and local structural consistency. In the latent subspace, conventional matching or ML algorithm can be used to yield the location estimates.

In another approach, \cite{zou2017adaptive} uses the labeled fingerprints (collected offline) with additional online APs readings as source domain, and the online target reading as the target domain. Using the data from both domains a transfer learning kernel is used to learn a domain-invariant kernel, which can be used with SVM for localization. In \cite{LowOverhead}, in order to reduce the offline training overhead in the new environment, the TL-based framework consists of two parts: metric learning and metric transfer. The metric learning part learns the distance metrics from source domains by maximizing the statistical dependence between the signal features and the corresponding labels. The metric transfer part identiﬁes the most suitable metric for the target domain by minimizing the data discrepancy between target and source domains. Sometimes the TL solution might not account for the entire environment structure: Ref. \cite{tlfcma} uses fuzzy C-means clustering aimed at minimizing the effect of environmental changes on the surrounding area; based on this the radio map can be reconstructed.

TL with DL solutions have been proposed recently. Ref. \cite{de2019csi_121} uses TL between different antenna setups in a massive MIMO system, where the input to a deep CNN network is time-domain CSI (raw values, amplitude, and phase). Once the network is trained, the lower layers of the network could be retrained with limited data points for the new antenna setup. TL between different environments is considered in \cite{xiao2018learning}. The authors use NNs to solve the RSSI based localization problem. It is done through pre-training the model in one environment and then performing TL to another one, where the source and target domains are separate propagation environments. The solution is applied to two different floors of one building, which have a similar structure and AP placement. The TL only uses $30\%$ of the available data in the target floor (domain).

\subsection{Other learning Structure}
\subsubsection{Reinforcement Learning}
Reinforcement Learning (RL) frameworks usually have at least one agent that has to take sequential actions in a given environment such that the cumulative reward is maximized. RL is one of the hot research areas in ML. It has also gained considerable attention in wireless communications for dynamic systems, such as resource allocation. Recently a few papers started to explore RL for localization. Ref. \cite{li2019deep_31} proposes a system in BLE that takes RSS values and previous locations information, and considers the reward to be whether the solution reaches known RSS values or an RP. Ref. \cite{dou2018top_112} views the localization in a WiFi system as Markov Decision Process (MDP), then used a model-free algorithm (Deep Q-Learning) to find the policy that progressively localizes the target with "right" mapping from observable states to actions. In the solution, the states include the RSSI values, action history, and the coordinate of the current center. The action space contains five motion directions to move the expected location window with different radius parameters. The reward is a function of the intersection between the window and the ground truth. Ref. \cite{peng2019decentralized_124} uses RL to schedule the exchange of signals in cooperative localization. The solution views the links as agents, the measurement decisions as actions. They use the distance, covariance values and the number of nodes that did not achieve a localization quality threshold to be the observations.
 
\subsubsection{Federated Learning}\label{subsec:Fed}
In federated learning (FL), the training is carried out at the target or distributed computing edges. This collaborative framework has many advantages, such as reduced training load and, more importantly, maintaining users' privacy, as the user does not need to share private data with the network. FL usually assumes the presence of a centralized entity that updates the model by combining the locally (at the user side) trained models and broadcasts it back. There have been limited research papers that use FL for localization. However, this is expected to change as location information is one of the basic private pieces of information. One recent work \cite{ciftler2020federated_70} proposes an FL scheme that improves the reliability and the robustness of RSS fingerprint-based localization, while preserving the privacy of the participants, using an NN that predicts the coordinate of the users. After the model is trained at the users' sides, the central node weighs the NN weights by their number of used samples and broadcasts the updated model back to the users.

\section{Datasets}\label{sec:Data}

The success of ML is predicated on the availability of suitable datasets. In this section we first summarize the types of datasets that were used in the literature for localization, and then we list the datasets that are currently publicly available. 
\subsection{Used Datasets}
We can distinguish three main types of datasets that were used in the localization literature:
\begin{itemize}
    \item {\em Measurements}. Here the authors carry out experiments using real devices; the collected data are used to train and test the proposed models. One of the following two data collection methods is used
    \begin{itemize}
        \item {\em Off the shelf hardware}. Commercially available devices have been the most prevalent choice, especially when only the RSSI signals are used. For instance, mobile phones can be used to collect the RSSI signal in WiFi or cellular systems, e.g., \cite{DataAugmentation, wu2012will_93,rizk2019device_96,zhao2019accurate}. Sometimes the development of special software is needed to extract the desired features. For example, in \cite{wu2012will_93} the authors developed an application to collect WiFi signal and sensor data. Mobile service providers could use systems logs available in their networks as in \cite{ray2016localization_94,zhang2020transfer_49}. A number of datasets with annotated RSSI signals are publicly available; Ujiindoorloc is one of the early datasets \cite{torres2014ujiindoorloc}, where annotated RSSI measurements have been recorded in a number of buildings and for a large number of APs.
        
        Collecting more advanced features might require more specialized hardware or software. For instance, to collect the raw CSI signal, several papers rely on Intel WiFi link $5300$ NIC chipset \cite{IntelNIC5300}. They use modified chipset firmware \cite{wang2015phasefi_43,sen2012you_91, li2019convolutional_12,wang2017cifi_9}, which provides access to the CSI at three antennas. To use the FTM protocol, \cite{choi2019unsupervised_123} uses the Intel AC8260 WiFi chipset that supports FTM functionality \cite{IntelAC8260}. For UWB, the DecaWave DW1000 transceivers, \cite{Decawave}, are popular \cite{xue2019deeptal_36,krishnan2018improving_71,goswami2011wigem_75,bregar2018improving_77}. 
        A number of papers introduced modifications to existing localization systems. For instance, the authors in \cite{pseudolite} test the Pseudolite system , where they have to deploy the system indoor. Switched-beam antennas with a USRP board can acquire directional RSSI in WiFi systems \cite{zhang2019efficient_122}. A testbed with a CC2530 chipset for device-free localization-based Zigbee has been used in \cite{wang2016device_39}.
        
        \item {\em Channel sounders} are specialized measurement equipment constructed for the precision measurement of wireless propagation channel characteristics \cite{ProfMolischText}. They have two main advantages: (i) higher accuracy and reproducability of the measurement results, since they are carefully calibrated and use designs that reduce impact of noise, interference, etc.; (ii) since they are custom devices, they can be constructed for scenarios that fall outside the operation range of commercially available systems, such as 
        carrier frequency, bandwidth, antennas structure, etc. 
        However, their use is usually limited due to the cost, effort, and the needed expertise to develop them. A few ML-based localization systems used channel sounders, e.g., \cite{arnold2018deep_82, sobehy2019csi_117,de2019csi_121,wymeersch2012machine_72, flordelis2019massive}.
    \end{itemize}
    \item {\em Ray-tracing software.} Ray tracing, or more generally deterministic channel modeling, uses 3D models of the environment and the electromagnetic characteristics of the objects in them to provide accurate simulations of the double-directional impulse response. The limitations are mainly due to (i) limitations in the representation of the propagation effects, such as diffuse scattering, and (ii) inaccuracies and limited resolution of the environmental database. Ray tracing is widely used by cellular network operators for network planning, and the accuracy of RSSI prediction is quite high. However, other aspects such as angular spread, are modeled with less accuracy, in particular at higher frequencies. One widely used ray tracing tool for academic investigations is Wireless InSite \cite{wiweb}, where it was used for localization in, e.g., \cite{ebuchi2019vehicle_126}.
    
    \item {\em Statistical Simulations.} A number of authors use statistical channel models to generate data to evaluate their solutions. Such statistical models and their limitations are discussed in Sec. \ref{sec:prelWireless}.
    System evaluations have been based on the 3GPP model \cite{prasad2018machine_62},  \cite{chen2018multipath_111},\cite{wu2019learning_78}, the COST 2100 channel model under the 300 MHz parameterization for MIMO channels \cite{vieira2017deep_80}, a geometric stochastic channel with parameters based on LTE-Advanced \cite{sun2018single_90,sun2019fingerprint_79}, the ITU indoor office channel model \cite{xiong2019decorrelation_115}, or LOS and NLOS models based on the Berlin UMa scenario \cite{lei2019siamese_130}. A number of these papers have used the model implementations of QuaDRiGa \cite{jaeckel2014quadriga}, a public-domain channel simulator that implements a number of the above-cited as well as other channel models; it allows spatial consistency of the simulated channel, and permits imposition of geometric structure in the considered environment. 
\end{itemize}
A number of works use more than one dataset to train and test their model, for instance, \cite{arnold2018deep_82} uses ray-tracing data to pre-train the solution before it is fine-tuned and tested on measurement data. Ref. \cite{sorour2014joint_47} uses a statistical channel simulator as a source domain in a TL approach. Ref. \cite{de2019enhanced_129} uses both measurements and ray-tracing to enhance TDoA based localization.

\begin{table*}
\centering
\begin{scriptsize}
\begin{tabularx}{1\textwidth} { 
   >{\centering\arraybackslash\hsize=1.6\hsize}X 
   >{\centering\arraybackslash\hsize=0.85\hsize}X 
   >{\centering\arraybackslash\hsize=0.85\hsize}X 
   >{\centering\arraybackslash\hsize=0.70\hsize}X 
   >{\centering\arraybackslash\hsize=0.85\hsize}X }
 \hline
  \scriptsize Measurement & \scriptsize Simulation & \scriptsize Ray tracing & \scriptsize Open Dataset \\
 \hline
 ~\cite{alteredAP,8651736,7743685,MonoDCell,vlfANN,7938617,OnDevice,lorawan,telco,6805641,wsn_svr,RadioMaps,tlfcma,LowOverhead,magnetic_inertial,IndustrialWSN,uav,location_aware,8240410,8108573,OpticalCamera,devicefree,10.1145/3321408.3321584,8780770,8673800,rfid,kmgpr,8765368,Discriminative,DataAugmentation,8362712,Eloc,HumanDetection,RelationLearning,FullBandGSM,grof,hapi,hiwl,7064997,WeightedAmbientWiFi,eee112,8008794,8584443,JUIndoorLoc,7313038,8931646,LiUnsupervised,yang2012locating,8115921,8661625,Wang2017WiFiFB,Chunjing2017WLAN,8422182,wu2019mobile,feng2019received,zhang2017efficient,xu2019efficient,cottone2016machine, sun2008adaptive,yan2018noise,wang2019robust,mei2018novel,wu2018accurate,nguyen2017performance,musa2019decision,zou2015fast,rezgui2017efficient,figuera2012advanced,guo2018accurate,bi2018adaptive,elbes2019indoor,comiter2018localization,liu2019hybrid,liu2017wicount,ahmadi2017exploiting,adege2018indoor,adege2019mobility,schmidt2019sdr,gu2015online,yang2018clustering,yan2017hybrid,malik2019indoor,bae2019large,tewes2019ensemble,sung2019neural,li2017measurement,sanam2018improved,xiao2012large,zhao2019accurate,salamah2016enhanced,widmaier2019towards,homayounvala2019novel,zhang2019improving,belmonte2019swiblux,dashti2016extracting,yang2012locating,oussar2011indoor,chen2019dpr,mahfouz2015kernel,qiu2018walk,van2015machine_1,niitsoo2019deep_2,wang2016csi_4,wang2016csi_5,wu2019csi_7,li2018channel_8,wang2017cifi_9,niitsoo2018convolutional_11,li2019convolutional_12,rizk2018cellindeep_15,aikawa2019cnn_16,wang2018deep_23,hsieh2019deep_28,li2019deep_31,gan2017deep_32,wang2015deepfi_33,yazdanian2018deeppos_34,berruet2018delfin_35,xue2019deeptal_36,zhang2016deep_38,wang2017biloc_42,wu2017passive_44,ghourchian2017real_45} ~\cite{zhou2017robust_46,sorour2014joint_47,zhang2020transfer_49,shokry2018deeploc_50,ouyang2011indoor_51,liu2009low_52,gu2015semi_54,pulkkinen2011semi,zhou2017semi_55,wang2017resloc_56,aikawa2018wlan_61,abbas2019wideep_63,huang2020widet_64 ,8733822, dayekh2010cooperative_66,kram2019uwb_68,choi2019unsupervised_69,krishnan2018improving_71,alhajri2019indoor_74,goswami2011wigem_75,zhou2017grassma_76,decurninge2018csi_81,arnold2018deep_82,zhou2017indoor_86,chen2018deep_88,jung2015unsupervised_89,sen2012you_91,wang2012no_92,wu2012will_93,ray2016localization_94,lee2018dnn_95,rizk2019device_96,rizk2019solocell_98} ~\cite{wang2018deepmap_103,liu2018autloc_104,zhang2017deeppositioning_106,tarekegn2019applying_107,zhang2019deeploc_108,mittal2018adapting_109,adege2018applying_113,sobehy2019csi_117,song2019novel_119,shao2018indoor_120,de2019csi_121,choi2019unsupervised_123,own2019signal_125,ebuchi2019vehicle_126,pandey2019handling_127,de2019enhanced_129}
 
 & ~\cite{khatab2017fingerprint,FPfusing,InterpretingCNN,MassiveMIMO,7504333,NLOS_UWB,sdr,TargetTracking,tdoa,patent,FineGrainedSubcarrier,ZhangIndoor,8712551,AntonioLocalization,jaafar2018neural,ge2019hybrid} ~\cite{comiter2017data,comiter2017structured,zou2015fast,comiter2018localization,zhang2018heterogeneous,yi2019neural,bhatti2018machine,xiao2018learning,mahfouz2015kernel,njima2019deep_22,chen2011semi_53,li2017robust_60,prasad2018machine_62,vieira2017deep_80,prasad2017numerical_84,pirzadeh2019machine_87,sun2018single_90,chen2018multipath_111,xiong2019decorrelation_115,wu2019neural_116,own2019signal_125,lei2019siamese_130} 
 
 & ~\cite{felix2016fingerprinting,mmwave,AntonioLocalization,gan2017deep_32,butt2020rf_57,wang2019fast_83,de2019enhanced_129,bekkali2011gaussian} 
 
 & ~\cite{mmwave,gbrbm,kmgpr,Timotheatos2017FeatureEA,FeatureSelection,hybloc,Qian2019Supervised,wang2019robust,nguyen2017performance,zhang2018enhancement,tewes2019ensemble,kim2018hybrid,xiao2018learning,patel2020millimeter,qian2019convolutional_10,ibrahim2018cnn_13,turgut2019deep_27,wang2020robust_59,ciftler2020federated_70,jang2018indoor_99,nowicki2017low,belmannoubi2019stacked_110,dou2018top_112,song2019novel_119} ~\cite{csiRNN,kim2018scalable,secckin2019hierarchical,li2019smartloc}\\
 \hline
\end{tabularx}
\end{scriptsize}
\caption{\small Type of datasets used.}
\label{tab:Dataset}
\end{table*}

\subsection{Publicly Available localization Datasets}
Open datasets help researchers to develop solutions without the need for the exhaustive data collection process, and also to measure the effectiveness of the proposed approach against other approaches. One of the best examples of open datasets in DL is the MNIST dataset \cite{mnist}, which is a collection of $60,000$ images of handwritten numbers. This dataset is often used by researchers in CV to determine the effectiveness of their algorithms and also by many budding researchers in the field to try out and learn more about CV. 

For localization, there are number of publicly available datasets. UJIIndoorLoc is one of the most frequently used datasets in Indoor Localization. It consists of RSSI readings from WiFi, collected using about $25$ different mobile phones at three multi-floor buildings from the Jaume I University, from $520$ Wireless Access Points. The dataset has $21,049$ data points split across training and validation sets. Another version, the UJIIndoorLoc-Mag was also released, which consisted of sensor readings such as accelerometer, magnetometer and rotation sensor. It consists of $40,159$ measurements.

Many open datasets in indoor localization use WiFi RSSI as one of the key features for localization. Generally they are fused with some other attributes such as cellular RSSIs. An examples of WiFi plus Cellular RSSI readings is the {\em PerfLoc} dataset which consists of $900$ data points. This dataset also consists of sensor readings from a large variety of sensors available on an android phone. It was collected in 4 multi-floor buildings.

Recently a number of BLE based datasets have been made public, {\em BLE RSS Measurements Dataset} for Research on Accurate Indoor Positioning. It has over $4700$ fingerprints of BLE RSSI data collected from two zones in the Jaume I University. Android smartphones were used for collecting data from off the shelf Bluetooth beacon systems.

\begin{table*}
\begin{scriptsize}
    \centering
    \begin{tabular}{p{3cm}|p{1cm}|p{3cm}|p{2cm}|p{2cm}|p{2cm}|p{2cm}}
        \hline 
        Dataset name & Released & Features used & Wireless technology used & Location points & Online availability & Environment\\
         \hline \hline 
        Long-Term WiFi Fingerprinting Dataset ~\cite{ref1} &  2017 & RSSI, location coordinates & WiFi & 212 reference points with 63,504 measurements & Yes ~\cite{mendoza_silva_german_martin_2020_3748719} & Indoor (Multi-floor) \\
        \hline
        WiFi Crowdsourced Fingerprinting Dataset ~\cite{ref2} &  2017 & RSSI, local location coordinates & WiFi & 4648 fingerprints & Yes ~\cite{lohan_elena_simona_2017_889798} & Indoor (Multi-floor) \\
        \hline
        WLAN Indoor Ranging Dataset ~\cite{ref3} & 2018 & WLAN Ranging Signals & WLAN (IEEE 802.11g/n) & Approx. 320 files each with a 300x30000 MATLAB Matrix & Yes ~\cite{wlan_indoor_ranging_dataset} & Indoor \\
        \hline
        Rural Sigfox data-set ~\cite{ref5} &  2018 & RSSI, Location coordinates & Low Power WAN (Sigfox) & 25,638 rows & Yes ~\cite{aernouts_michiel_2018_1193563} & Outdoor \\
        \hline
        Urban Sigfox data-set ~\cite{ref5} & 2018 & RSSI, Location coordinates & Low Power WAN (LoRaWAN) & 14,378 rows & Yes ~\cite{aernouts_michiel_2018_1193563} & Outdoor \\
        \hline
        Urban LoRaWAN data-set ~\cite{ref5} & 2018 & RSSI, Location coordinates, Low Range Spreading Factor (SF), Horizontal Dilution of Precision (HDOP) & Low Power WAN (LoRaWAN) & 123,529 rows & Yes ~\cite{aernouts_michiel_2018_1193563} & Outdoor \\
        \hline
        Residential wearable RSSI and accelerometer measurements dataset ~\cite{ref6} & 2018 &  RSSI, accelerometer readings, battery level of wearable devices, tagged video of user & BLE (Bluetooth 4.0) & 42 MB of csv files & Yes ~\cite{Byrne2018} & Indoor (4 different residences) \\
        \hline
        UJIIndoorLoc-Mag ~\cite{ref7} &  2015 & Magentometer readings, Accelerometer readings and Rotation Sensor readings, Location coordinates & None (Sensor readings only) & 40,159 measurements & Yes ~\cite{Dua:2019, UJIIndoorLoc_Mag} & Indoor \\
        \hline
        UJIIndoorLoc ~\cite{ref8} &  2014 & RSSI from 520 APs, Location co-ordinates & WiFi & 933 different locations, 21049 different readings & Yes~\cite{Dua:2019, UJIIndoorLoc} & Multi-building, Multi-floor\\
        \hline
        AmbiLoc ~\cite{ref9} &  2017 & RSSI from FM, TV and GSM signals, coordinates & FM Radio, TV, GSM & 2697 samples & Yes ~\cite{AmbiLoc} & Multi-building, Multi-floor \\
        \hline
        PerfLoc ~\cite{ref10} &  2016 & Motion sensors, RSSI from WiFi and Cellular, GPS & WiFi, Cellular & 900 points & Yes & Multi-building, Multi-floor \\
        \hline
        KIOS Dataset ~\cite{ref11} & 2013 & RSSI from all available APs & WiFi & 105 different reference points, 2100 fingerprints & Yes ~\cite{KIOS} & Indoor \\
        \hline
        BLE RSS Measurements Dataset ~\cite{ref12} &  2018 & RSS from BLE Beacons, coordinates & BLE (Bluetooth 4.0) & 4752 different samples & Yes ~\cite{mendoza_silva_german_martin_2018_1618692} & Indoor (2 university zones) \\
        \hline
        JUIndoorLoc ~\cite{ref14} &  2019 & RSSI Data & WiFi & 25,364 samples & Yes ~\cite{JUIndoorLoc} & Multi-floor \\
        \hline
    \end{tabular}
\end{scriptsize}
\caption{\small Examples of publicly available dataset.}
\label{tab:PublicDataset}
\end{table*}

In outdoor localization, examples are the {\em Sigfox} and {\em LoRaWAN} datasets which consists of three datasets: the rural Sigfox dataset (over $25000$ rows), urban Sigfox dataset (over $14000$ rows) and the urban LoRaWAN dataset (over 123000 rows). A wide range of sensors and devices were used for sending and receiving messages through these proprietary Low-Power WAN channels in a variety of outdoor settings.

Readers are referred to table \ref{tab:PublicDataset} for more information on these datasets and other similar ones.

\section{Challenges and Opportunities}\label{sec:Challenges}
In this section we present a number of challenges for ML-based localization and interesting research directions. While many of them are common to classical solution techniques as well, our emphasis lies on the unique aspects of ML solutions. 
\subsection{Availability of Data}
\subsubsection{Discussion}
One major challenge of ML solutions is to acquire enough data to train and validate the solution. This can be manifested in different ways. When no labeled data are available, supervised and semi-supervised solutions cannot be used. Other solutions still rely on data, such as good representative non-labeled data for unsupervised learning. In fingerprinting solutions, as an example, the fingerprints should be collected when the system is initially deployed, e.g., through a drive-test. However, this is usually labor-intensive; the data must cover the entire service area and be recorded with suitable device(s). Furthermore, to combat the noise, and possible interference, several samples of the fingerprints should be recorded at each sample RP. %

{\bf Environment Drift:}
In realistic systems, the environment usually evolves, e.g., due to installation or relocation of APs or construction of new buildings that act as scatterers. Things become worse in highly dynamic environments, as the recorded data becomes quickly outdated, which degrades the performance of the solution. Data updates, either scheduled or triggered by environment change, might be needed. However, this is cumbersome and impractical over a long period.

{\bf Device Heterogeneity:}
In the exact same physical location, two different devices might observe the same feature differently. This is usually due to the effective RF characteristics of the devices. Different devices might have different antenna design, RF components, or packaging, which all impact the received signal. It is difficult to predict what device the user might use, and in particular it is quite possible that the user uses a different device from the one used in collecting the dataset to train the model.

Furthermore, different devices have different capabilities: while some devices can only read RSSI values, other might possibly be able to acquire CSI and other sensor readings. This is also rooted in the standards: different classes of devices are defined that may be equipped with different number of antennas or antennas architectures, and they might operate in different frequency bands. 

{\bf Body shadowing:}
Modern localization solutions are meant for mobile targets. Depending on the orientation of the user and the how the device is carried, the wireless signal could be occasionally shadowed by the user's body or hands. In fact, it is reported that body shadowing could result in as much as $>10$ dB loss in the received power; these values depend on body size, shape, distance and orientation and operation frequency \cite{tian2018human,chen2016modelling,cziezerski2019comparing,karedal2006shadowing}. This could alter or block certain features, and thus impact the range and distribution of the observed features. To demonstrate this with simple practical example: the constants calculated to normalize the input features, as part of preprocessing (see Sec. \ref{sec:prelWireless}), will be significantly impacted by the extreme variation of the features. 

{\bf User privacy:}
To collect large datasets, a number of works suggest the use of crowdsourcing \cite{8008794,magnetic_inertial,Chunjing2017WLAN,jung2015unsupervised_89,6805641}. In addition to the concern of the quality of the collected data and methods to label the them, the privacy of users is a major issue. Users generally avoid sharing their location and movement trends (if they are given the option). Furthermore, storing the location information in database for a relatively long time could also elevate that concern. With advent of data driven (ML) solutions, utilizing users data is both opportunity and threat to users privacy, which could lead to stringent regulations.
\subsubsection{Possible Solutions and Research Directions}

One solution to some of the aforementioned challenges is to collect the labeled data in brute-force manner, i.e., repeated collection of the feature-labels points, several times, by different administrated users, at different orientations, using different devices. This is clearly cost-prohibitive. Which motivated a number of interesting solutions that was covered in Sec. \ref{sec:LearnStrcut} (mainly, from subsection \ref{sub:semisuper} and onward), However, there many more interesting avenues to utilize the power of ML.
\begin{itemize}
\item New data augmentation techniques can be used to expand the dataset. Besides the classical techniques, e.g., adding noise or masking part of the features, the following can be good options. 
\begin{itemize}
    \item Augmenting sounding measurements with synthetic data.
    \item Utilize generative models to capture the possible mapping between the features and possible variations or different dimensions (in case of different features sets, e.g., from a set of possible CSI values from limited observations). A few papers have considered that, e.g., \cite{DataAugmentation,ZhangIndoor,xiao2018learning,eee112}. However, more research is still needed.
\end{itemize}
\item Employing TL could allow to use synthetic data or to train the solution in areas where access to a large number of points is relatively easy. As discussed in Sec. \ref{subsec:TL}, a number of works have considered TL, but both theoretical and experimental studies are needed to understand the best practices and limitation of TL for localization. For instance, motivated by different classes of channel models (e.g., micro, macro cells, urban, industrial etc.) or the dominant source of shadowing, TL may be easier between systems that are in similar environments.
\item Separable models, composed of both generic and device specific models. This could handle device heterogeneity and reduce the transfer of the explicit raw user data.
\item Utilizing measurement-validated channel characteristics, i.e., expert knowledge, in semi-supervised learning. Research is needed on how the few labeled data points can be used along with the unlabeled data, e.g., improved graphs based on modern channel models and environment structure. Furthermore, additional research can be done to improve where and when to update the annotated data points. 
%
\item Although a good number of works have investigated feature representation (e.g., ratio or difference of RSSIs), dimensionality reductions, and networks setups to combat channel variation and feature shifts, e.g., \cite{probabilistic_localization,LiUnsupervised,rezgui2017efficient,nguyen2017performance,adege2019mobility}, further research is still needed. In particular, the vast majority of the work in this area was dedicated to simple features, which may not be applicable to heterogeneous and complex features, which might be more robust to channel variations. As in illustrative example that was introduced in Fig. \ref{fig:MPCmanifold}, dimensionality reduction of channel multi-path multi-path components (that vary relatively slower than RSSI) could reveal underlying manifolds that may capture the structure in the environment. Furthermore, understanding the relation between the observed features and the proposed embedding and how the channel variations impact them could be of a value.
\item Use of FL. As discussed in Sec. \ref{subsec:Fed}, one goal of this technique is to preserve users' privacy. However, devices suffer from different local constrains and impairments. There are recent studies that investigate the different techniques to combine the data, and thus reduce the impact of device-specific issues. Additionally, the use of FL could encourage the user to engage in data sharing, as no raw data is being transferred. Given the constraints on location information and data sharing, localization driven advances in FL are desirable.
\end{itemize}
\subsection{Utilization of DL}
\subsubsection{Discussion}
\begin{figure}
\centering
\vspace{-0 mm}
 \includegraphics[width=\columnwidth, trim={5cm 2cm 5cm 2cm},clip]{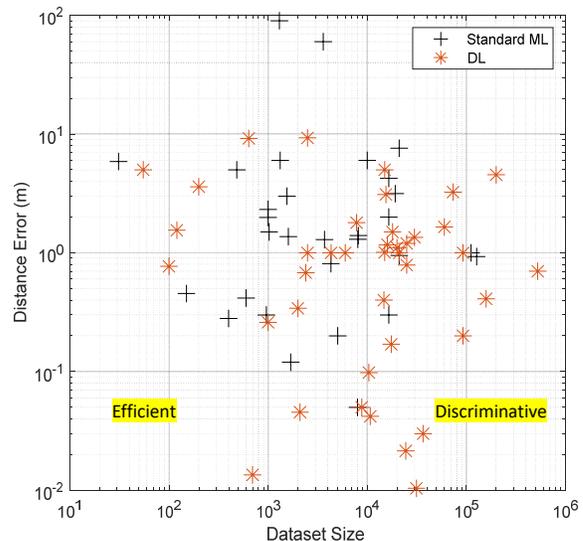}
\vspace{-10  mm}
\caption{\small Reported localization error in meter plotted vs the size of the used dataset. Solutions that use small datasets and achieve good results could indicate efficient performance. Small errors with larger datasets could suggest more discriminative power. Note these are not conclusive metrics as good performance in large datasets could be due to over sampling the area }
\label{fig:ErrVsSize}
\vspace{-0 mm}
\end{figure}

\begin{figure}[t]
\vspace{-0 mm}\centering
 \includegraphics[width=\columnwidth, trim={5cm 2cm 5cm 2cm},clip]{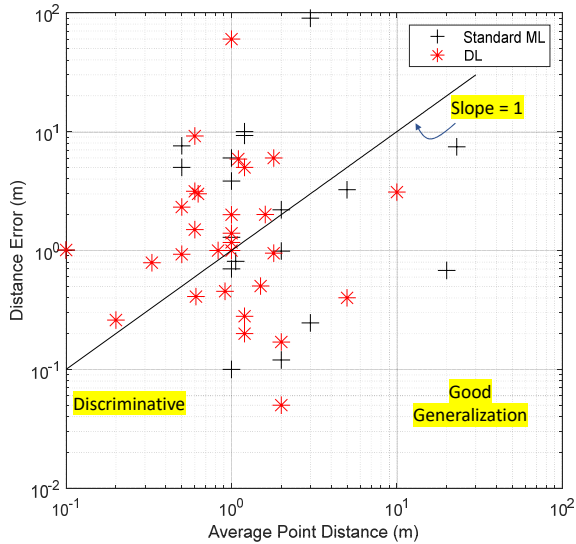}
\vspace{-10  mm}
\caption{\small Reported localization error in meters plotted vs the resolution of the points in the training dataset (in meters). Points below slope = 1 could suggest good regression and generalization abilities. The overall performance against density serve as an indicator of the cost of the solution (assuming similar set of features are used). Solutions that have small localization errors and use examples with small separation distance are preferred for practical applications.}
\label{fig:ErrVsDist}
\vspace{-0 mm}
\end{figure}

The number of published studies of DL-based localization is growing fast. Although DL offers a powerful solution that can utilize large datasets and is reported to easily generalize (at least in other fields), they are usually complex with a large number of trainable parameters (could be millions). This results in the following drawbacks:
\begin{itemize}
    \item Computational difficulty to train. They may need GPUs to speed the training process, which may not always be available or expensive to install.
    \item Deep models have a large number of operations; this limits their use in power-limited devices.
    \item While DL solutions flourish with large datasets, acquiring such data is difficult.
    \item Overfitting. While this is generally a concern in ML solution, the large number of parameters in DL can exacerbate the problem.
\end{itemize}
In the literature, the reported accuracy of different ML-based localization methods ranges from sub-centimeter to tens of meters. However, it is difficult to judge the solutions based on the reported accuracy only, as a number of factors play into the goodness of the solution and its practicality. In Fig. \ref{fig:ErrVsDist} and Fig. \ref{fig:ErrVsSize} we plot the reported localization error (in meter) vs dataset size and point resolution (in meter), respectively. For better comparability, we restricted the data to supervised learning in indoor environments. Usually, we are looking for solutions with small localization errors. However, the dataset plays a major role in the performance. Good solutions could achieve good accuracy with small examples. They could also generalize well even with large point separation. However, we do not see that for all the solutions. Rather, we notice that even with large datasets with DL solutions, large localization errors are observed in Fig. \ref{fig:ErrVsSize}. Similarly, in Fig. \ref{fig:ErrVsDist} we can observe that large errors are reported even at small separation distance. Additionally, in both figures, we notice that a number of DL solutions fall in the same region as the standard ML solutions. This should trigger the question of whether there is always a need for DL solutions. Nevertheless, we notice a number of DL solutions present in the right lower quadrant of Fig. \ref{fig:ErrVsSize} and the left lower quadrant of Fig. \ref{fig:ErrVsDist}, which confirms the {\em potential} advantages of DL when large datasets are available and better discriminative ability when points become close to one another. 

\subsubsection{Possible Solutions and Research Directions}

Thus it is always recommended to test the performance of DL-solutions against standard simple ML solutions. Moreover, research should consider the offered improvement against the model complexity. An urgent research item is developing a wide range of datasets that cover different possible features and environments, which will allow fair comparative studies of the proposed solution. For the same reason, researchers are encouraged to publish their trained models online.

\subsection{System Reliability} \label{subsec:SystReliab}
\subsubsection{Discussion}
In several proposed solutions, the device collects a large number of observations and sends them to the localization unit (e.g., BS or APs) for processing. Alternatively, the localization unit observes the features and processes them. With the expected increased demand for proximity and localization services in addition to the features complexity, the overhead of transmitting the observations and the processing shall increase, which may overwhelm the infrastructure. This has an impact on the localization and possibly the primary services, e.g., wireless communication.

\subsubsection{ Possible Solutions and Research Directions}
\begin{itemize}
    \item Systematic studies of overhead and processing of the solutions are required. The impact of spectrum efficiency in wireless systems can be also investigated. Since location aided wireless communication is becoming prevalent, metrics that capture the cost and gain of use of location information are also interesting directions.

    \item Joint localization can alleviate the possible degradation of location estimates, as target channel observations may be correlated. With efficient ML-based implementation, joint localization can reduce processing time and improve the location estimate. 
    
    \item Low dimensional feature extraction at the target end. To reduce the overhead, only compressed feature representations can be transmitted to the localization unit. 
    
    \item ML-based localization at the target end, which can minimize both overhead and processing overload at the localization unit, can also help preserving users' privacy. A relaxation of this is to implement part of the solution at the target end, which may be viewed as a generalization of the previous point. Few works have explicitly considered that, such as \cite{niitsoo2019deep_2}, which proposes "distributed CNN", extracting latent variables locally before the compressed representation of CIR is passed to the central localization unit; different from FL, the training is done end-to-end before the part of the solution is distributed. Still much work is needed in this important direction.
    
    \item Cooperative localization, where the devices can share information that may improve location estimates. This can be used along the line of the points above, where the targets can locally use their neighbours' estimates or features to improve their predictions. The large body of ML-based cooperative localization is based on range estimates, then applying MDS techniques (See Sec. \ref{sub:semisuper}); however, with the availability of rich features, solutions beyond range estimates can be proposed.
\end{itemize}
\subsection{Emerging applications and technologies:}
The demand for RF-based localization has increased dramatically due to the recent advances in wireless communication systems and the rise of new applications. These have raised the bar of the expected localization accuracy. A few emerging applications and technologies are listed below.
\begin{itemize}
    \item Accurate location information for vehicles; this is especially needed for smart and autonomous vehicles, where a few meters of localization errors may not be acceptable. What makes this application unique are (i) the environment is highly dynamic, (ii) communication is usually over low rate connections (e.g., IEEE 802.11p), (iii) availability of a wide range of side information types such as images and other sensors reading (iv) computational power may not be a constraint.
    \item Mission-critical applications in highly dynamic and harsh environments. This can be viewed as a generalization of the above point.  One example may be a factory, where wireless channels suffer from many (possibly dynamic) scatterers. Furthermore, excessive location errors can have catastrophic consequences. A few studies have considered ML based localization in industrial environment, e.g., \cite{IndustrialWSN,li2017measurement,wymeersch2012machine_72,krishnan2018improving_71,niitsoo2019deep_2,9272626}. However, ML-based methods to combat interference and provide estimates guarantees e.g., through adversarial networks could be studied \cite{goodfellow2016deep}.
    \item Localization in IoT, were the services cover applications with heterogeneous devices and wireless standards, This calls for flexible ML-based solutions that take the various constraints at the wide range of devices into account. Constraints include limitations on observable features, processing power, and acceptable localization error. An important goal is designing good embedding that captures the relation between the different devices, e.g., graph based solution through the graph neural networks \cite{wu2020comprehensive}.
    \item Massive MIMO. Although a number of approaches have already been proposed, there are a number of challenges that should be addressed. They include the excessive overhead for feature feedback (see also Sec. \ref{subsec:SystReliab}), or the needed beam-search along with possible beam outdatedness and misalignments. Massive MIMO is expected to be deployed in mmWave communication systems, which have multi-GHz bandwidth, method for efficient utilization of the large bandwidth or number of sub-carriers are needed.
    
    \item Utilizing RF-signals along with side information for localization related problems, such as Simultaneous Localization and Mapping (SLAM) and Radar. Both these areas have an overlap with localization problems. In SLAM the goal is to track a moving target (agent as referred to in robotics) along with the construction of the map of the environment. Typically, different types of data are used to solve this problem, such as Inertial Sensor readings and images, i.e., the side information listed in Sec. \ref{sec:SideInfo}. Solutions that implicitly construct the map (as in Sec. \ref{sub:semisuper}), and DL solutions that efficiently extract the RF-features especially in high bandwidth systems (e.g., UWB and high frequency ranges,i.e., mmWave and beyond), are good research directions in this field.
    In the Radar case, the goal is to detect, locate and identify the targets, which usually utilize different types of data as well. RF-based passive localization that we covered in this survey is a simpler version of that. Researchers can exploit some of the techniques used in the radar field to enhance the localization. 
    
    ML techniques have been used in both fields, for surveys see \cite{chen2020survey} and \cite{lang2020comprehensive}. With anticipated (DL driven) advances of RF-based localization, enhancements to both fields can be introduced.
    
\end{itemize}
\section{Conclusions}\label{sec:conc}
In this paper, we surveyed ML-based localization solutions. We focused on systems that use different attributes of the RF signals to localize their targets. The survey spans different aspects of ML-based localization: the system architectures, the used RF features, the ML methods, and the data acquisitions. Throughout the survey, we maintained structured reference lists for the relevant aspects of the solutions. To make the presentation accessible to readers from different backgrounds, we presented a concise review of the main aspects of ML and wireless channels. 

Based on the surveyed solutions, we identified different challenges and research directions when applying ML in localization. In particular, as data-driven solutions, enabling ML-based localization requires efficient methods to collect and maintain the datasets, which are usually subject to the environment's dynamics and the acquisition systems. One of the suggested solutions is based on augmentation techniques and deep generative models that may enrich the datasets, where the latter may learn the true distribution of the data and its evolution. Alternatively, methods that capture the fundamental properties of wireless channels and utilize TL and Semi-supervised learning can be used. The overhead of feature transfer and the complexity of DL are of concern, methods to distribute the learning (e.g., using FL or separable models) are of interest. Furthermore, datasets with rich features are needed to enable researchers to study and compare their solutions. Overall, ML-based localization has many interesting research avenues as new wireless systems, standards, and applications are being progressively introduced.
\\
{\bf Acknowledgements:} The authors thank Prof. Mahdi Soltonkotabi for helpful discussions and critical reading of the manuscript. 
 \begin{table}
  \caption{Used Acronyms}
 \begin{scriptsize}
\begin{tabularx}{\linewidth} { 
   >{\raggedright\arraybackslash}X 
   >{\centering\arraybackslash}X 
  }
  \hline
  \bf Acronym &  \bf Explanation \\
 \hline
 AE & Auto encoder\\
AP  & Access Point \\
BLE  & Bluetooth Low Energy \\
BS   & Base-station \\
  CFR & channel frequency response \\
 CIR  & channel impulse response \\
 CNN  & Convolution NN  \\
 CSI  & Channel State Information  \\
DBN &  Deep Belief Networks \\
DL  & Deep Learning\\
ELM &  Extreme Learning Machine \\
EM &  Expectation maximization\\
FL  & Federated Learning\\
GMM &   Gaussian mixture model\\
GP  & Gaussian Process   \\
GNNS  & Global Navigation Satellite Systems\\
GRU  & Gated Recurrent Units\\
 
HMM  & Hidden Markov Model \\
IoT  & Internet of Things \\
  KL divergence  & Kullback-Leibler divergence \\
KNN  & K-Nearest Neighbors  \\
LDA  &  Linear Discriminant Analysis \\
LOS  & Line of Sight \\

LoRaWAN & Long Range WAN (a protocol)\\ 
LSTM  & Long-Short-Term Memory \\
MDN &  multi density network\\
ML  & Machine Learning\\
mmWave  & millimeter-wave \\
MS &  Mobile station \\
NN  & Neural Network \\
PCA  & Principal Component Analysis \\
PDP  & Power Delay Profile \\
RBF  & Radial Basis Function  \\
RF  & Radio Frequency\\
RMSE  & Root Mean Square Error \\
RNN  & Recurrent NN \\
RP  & Reference Point \\
RSRP  & Reference Signal Received Power \\
RSRQ  & Reference Signal Received Quality  \\
RSS  & Received Signal Strength  \\
RSU  & Road Side Units \\
SINR  & Signal to interference and noise power ratio  \\
SVM  & Support Vector Machine  \\
TDOA & Time difference of arrival\\
TL  & Transfer Learning\\
ToA & Time of arrival\\
UAV   & Unmanned Aerial Vehicle \\
UE  & User Equipment \\
WAN & Wireless Area Network \\
WSN  & Wireless sensor network  \\
 \hline
\end{tabularx}
\label{tab:acry}
\end{scriptsize}
\end{table}

\bibliography{refsA, refs2, refsD, Dataset, DatasetsRef}
\bibliographystyle{IEEEtran}

\end{document}